

\hfuzz 70 pt

\magnification=\magstep1

\newbox\SlashedBox
\def\slashed#1{\setbox\SlashedBox=\hbox{#1}
\hbox to 0pt{\hbox to 1\wd\SlashedBox{\hfil/\hfil}\hss}#1}
\def\hboxtosizeof#1#2{\setbox\SlashedBox=\hbox{#1}
\hbox to 1\wd\SlashedBox{#2}}

\def\mathslashed#1{\setbox\SlashedBox=\hbox{$#1$}
\hbox to 0pt{\hbox to 1\wd\SlashedBox{\hfil/\hfil}\hss}#1}

\def\ifsmall{\iffalse}  
\def\titlepagefont{}  

\def\DefineTeXgraphics{%
\special{ps::[global] /TeXgraphics { } def}}  

\def\today{\ifcase\month\or January\or February\or March\or April\or May
\or June\or July\or August\or September\or October\or November\or
December\fi\space\number\day, \number\year}
\def\eatPrefix19{}
\def\Year{\expandafter\eatPrefix\the\year}
\newcount\hours \newcount\minutes
\def\monthname{\ifcase\month\or
January\or February\or March\or April\or May\or June\or July\or
August\or September\or October\or November\or December\fi}
\def\shortmonthname{\ifcase\month\or
Jan\or Feb\or Mar\or Apr\or May\or Jun\or Jul\or
Aug\or Sep\or Oct\or Nov\or Dec\fi}

\def\TimeStamp{\hours\the\time\divide\hours by60%
\minutes -\the\time\divide\minutes by60\multiply\minutes by60%
\advance\minutes by\the\time%
${\rm \shortmonthname}\cdot\if\day<10{}0\fi\the\day\cdot\the\year%
\qquad\the\hours:\if\minutes<10{}0\fi\the\minutes$}




\def\Title#1{%
\vskip 1in{\titlefont\centerline{#1}}\vskip .5in}



\newif\ifdraftmode
\newif\ifleftlabels  

\def\nolabels{\def\wrlabeL##1{}\def\eqlabeL##1{}\def\reflabeL##1{}}
\def\writelabels{\def\wrlabeL##1{\leavevmode\vadjust{\rlap{\smash%
{\line{{\escapechar=` \hfill\rlap{\sevenrm\hskip.03in\string##1}}}}}}}%
\def\eqlabeL##1{{\escapechar-1\rlap{\sevenrm\hskip.05in\string##1}}}%
\def\reflabeL##1{\noexpand\rlap{\noexpand\sevenrm[\string##1]}}}
\def\writeleftlabels{\def\wrlabeL##1{\leavevmode\vadjust{\rlap{\smash%
{\line{{\escapechar=` \hfill\rlap{\sevenrm\hskip.03in\string##1}}}}}}}%
\def\eqlabeL##1{{\escapechar-1%
\rlap{\sixrm\hskip.05in\string##1}%
\llap{\sevenrm\string##1\hskip.03in\hbox to \hsize{}}}}%
\def\reflabeL##1{\noexpand\rlap{\noexpand\sevenrm[\string##1]}}}
\nolabels

\newdimen\fullhsize
\newdimen\hstitle
\hstitle=\hsize 
\newdimen\hsbody
\hsbody=\hsize 
\newdimen\hbodyoffset
\hbodyoffset=\hoffset 
\newbox\leftpage
\def\abstract#1{#1}
\def\rotated{\special{ps: landscape}
\magnification=1000  
\baselineskip=14pt
\global\hstitle=9truein\global\hsbody=4.75truein
\global\vsize=7truein\global\voffset=-.31truein
\global\hoffset=-0.54in\global\hbodyoffset=-.54truein
\global\fullhsize=10truein
\def\DefineTeXgraphics{%
\special{ps::[global]
/TeXgraphics {currentpoint translate 0.7 0.7 scale
              -80 0.72 mul -1000 0.72 mul translate} def}}
\let\lr=L
\def\ifsmall{\iftrue}
\def\titlepagefont{\twelvepoint}
\trueseventeenpoint
\def\almostshipout##1{\if L\lr \count1=1
      \global\setbox\leftpage=##1 \global\let\lr=R
   \else \count1=2
      \shipout\vbox{\hbox to\fullhsize{\box\leftpage\hfil##1}}
      \global\let\lr=L\fi}

\output={\ifnum\count0=1 
 \shipout\vbox{\hbox to \fullhsize{\hfill\pagebody\hfill}}\advancepageno
 \else
 \almostshipout{\leftline{\vbox{\pagebody\makefootline}}}\advancepageno
 \fi}

\def\abstract##1{{\leftskip=1.5in\rightskip=1.5in ##1\par}} }

\def\linemessage#1{\immediate\write16{#1}}

\global\newcount\secno \global\secno=0
\global\newcount\appno \global\appno=0
\global\newcount\meqno \global\meqno=1
\global\newcount\subsecno \global\subsecno=0
\global\newcount\figno \global\figno=0

\newif\ifAnyCounterChanged
\let\terminator=\relax
\def\normalize#1{\ifx#1\terminator\let\next=\relax\else%
\if#1i\aftergroup i\else\if#1v\aftergroup v\else\if#1x\aftergroup x%
\else\if#1l\aftergroup l\else\if#1c\aftergroup c\else%
\if#1m\aftergroup m\else%
\if#1I\aftergroup I\else\if#1V\aftergroup V\else\if#1X\aftergroup X%
\else\if#1L\aftergroup L\else\if#1C\aftergroup C\else%
\if#1M\aftergroup M\else\aftergroup#1\fi\fi\fi\fi\fi\fi\fi\fi\fi\fi\fi\fi%
\let\next=\normalize\fi%
\next}
\def\makeNormal#1#2{\def\doNormalDef{\edef#1}\begingroup%
\aftergroup\doNormalDef\aftergroup{\normalize#2\terminator\aftergroup}%
\endgroup}

\def\warnIfChanged#1#2{%
\ifundef#1
\else\begingroup%
\edef\oldDefinitionOfCounter{#1}\edef\newDefinitionOfCounter{#2}%
\ifx\oldDefinitionOfCounter\newDefinitionOfCounter%
\else%
\linemessage{Warning: definition of \noexpand#1 has changed.}%
\global\AnyCounterChangedtrue\fi\endgroup\fi}

\def\Section#1{\global\advance\secno by1\relax\global\meqno=1%
\global\subsecno=0%
\bigbreak\bigskip
\centerline{\twelvepoint \bf %
\the\secno. #1}%
\par\nobreak\medskip\nobreak}
\def\tagsection#1{%
\warnIfChanged#1{\the\secno}%
\xdef#1{\the\secno}%
\ifWritingAuxFile\immediate\write\auxfile{\noexpand\xdef\noexpand#1{#1}}\fi%
}
\def\section{\Section}
\def\Subsection#1{\global\advance\subsecno by1\relax\medskip %
\leftline{\bf\the\secno.\the\subsecno\ #1}%
\par\nobreak\smallskip\nobreak}
\def\tagsubsection#1{%
\warnIfChanged#1{\the\secno.\the\subsecno}%
\xdef#1{\the\secno.\the\subsecno}%
\ifWritingAuxFile\immediate\write\auxfile{\noexpand\xdef\noexpand#1{#1}}\fi%
}

\def\subsection{\Subsection}

\def\romappno{\uppercase\expandafter{\romannumeral\appno}}
\def\makeNormalizedRomappno{%
\expandafter\makeNormal\expandafter\normalizedromappno%
\expandafter{\romannumeral\appno}%
\edef\normalizedromappno{\uppercase{\normalizedromappno}}}
\def\Appendix#1{\global\advance\appno by1\relax\global\meqno=1\global\secno=0
\bigbreak\bigskip
\centerline{\twelvepoint \bf Appendix %
\romappno. #1}%
\par\nobreak\medskip\nobreak}
\def\tagappendix#1{\makeNormalizedRomappno%
\warnIfChanged#1{\normalizedromappno}%
\xdef#1{\normalizedromappno}%
\ifWritingAuxFile\immediate\write\auxfile{\noexpand\xdef\noexpand#1{#1}}\fi%
}
\def\appendix{\Appendix}

\def\eqn#1{\makeNormalizedRomappno%
\ifnum\secno>0%
  \warnIfChanged#1{\the\secno.\the\meqno}%
  \eqno(\the\secno.\the\meqno)\xdef#1{\the\secno.\the\meqno}%
     \global\advance\meqno by1
\else\ifnum\appno>0%
  \warnIfChanged#1{\normalizedromappno.\the\meqno}%
  \eqno({\rm\romappno}.\the\meqno)%
      \xdef#1{\normalizedromappno.\the\meqno}%
     \global\advance\meqno by1
\else%
  \warnIfChanged#1{\the\meqno}%
  \eqno(\the\meqno)\xdef#1{\the\meqno}%
     \global\advance\meqno by1
\fi\fi%
\eqlabeL#1%
\ifWritingAuxFile\immediate\write\auxfile{\noexpand\xdef\noexpand#1{#1}}\fi%
}
\def\defeqn#1{\makeNormalizedRomappno%
\ifnum\secno>0%
  \warnIfChanged#1{\the\secno.\the\meqno}%
  \xdef#1{\the\secno.\the\meqno}%
     \global\advance\meqno by1
\else\ifnum\appno>0%
  \warnIfChanged#1{\normalizedromappno.\the\meqno}%
  \xdef#1{\normalizedromappno.\the\meqno}%
     \global\advance\meqno by1
\else%
  \warnIfChanged#1{\the\meqno}%
  \xdef#1{\the\meqno}%
     \global\advance\meqno by1
\fi\fi%
\eqlabeL#1%
\ifWritingAuxFile\immediate\write\auxfile{\noexpand\xdef\noexpand#1{#1}}\fi%
}
\def\anoneqn{\makeNormalizedRomappno%
\ifnum\secno>0
  \eqno(\the\secno.\the\meqno)%
     \global\advance\meqno by1
\else\ifnum\appno>0
  \eqno({\rm\normalizedromappno}.\the\meqno)%
     \global\advance\meqno by1
\else
  \eqno(\the\meqno)%
     \global\advance\meqno by1
\fi\fi%
}
\def\mfig#1#2{\global\advance\figno by1%
\relax#1\the\figno%
\warnIfChanged#2{\the\figno}%
\edef#2{\the\figno}%
\reflabeL#2%
\ifWritingAuxFile\immediate\write\auxfile{\noexpand\xdef\noexpand#2{#2}}\fi%
}

\catcode`@=11 

\font\ninerm=cmr9
\font\eightrm=cmr8
\font\sixrm=cmr6

\def\loadtrueseventeenpoint{
 \font\seventeenrm=cmr10 at 17.28truept
 \font\seventeeni=cmmi10 at 17.28truept
 \font\seventeenbf=cmbx10 at 17.28truept
 \font\seventeenit=cmti10 at 17.28truept
 \font\seventeensl=cmsl10 at 17.28truept
 \font\seventeensy=cmsy10 at 17.28truept
}
\def\loadfourteenpoint{
\font\fourteenrm=cmr10 at 14.4pt
\font\fourteeni=cmmi10 at 14.4pt
\font\fourteenit=cmti10 at 14.4pt
\font\fourteensl=cmsl10 at 14.4pt
\font\fourteensy=cmsy10 at 14.4pt
\font\fourteenbf=cmbx10 at 14.4pt
}
\def\loadtruetwelvepoint{
\font\twelverm=cmr10 at 12truept
\font\twelvei=cmmi10 at 12truept
\font\twelveit=cmti10 at 12truept
\font\twelvesl=cmsl10 at 12truept
\font\twelvesy=cmsy10 at 12truept
\font\twelvebf=cmbx10 at 12truept
}

\font\ninei=cmmi9
\font\eighti=cmmi8
\font\sixi=cmmi6
\skewchar\ninei='177 \skewchar\eighti='177 \skewchar\sixi='177

\font\ninesy=cmsy9
\font\eightsy=cmsy8
\font\sixsy=cmsy6
\skewchar\ninesy='60 \skewchar\eightsy='60 \skewchar\sixsy='60

\font\ninebf=cmbx9
\font\eightbf=cmbx8
\font\sixbf=cmbx6

\font\ninett=cmtt9
\font\eighttt=cmtt8

\hyphenchar\tentt=-1 
\hyphenchar\ninett=-1
\hyphenchar\eighttt=-1

\font\ninesl=cmsl9
\font\eightsl=cmsl8

\font\nineit=cmti9
\font\eightit=cmti8


\newskip\ttglue
\def\tenpoint{\def\rm{\fam0\tenrm}%
  \textfont0=\tenrm \scriptfont0=\sevenrm \scriptscriptfont0=\fiverm
  \textfont1=\teni \scriptfont1=\seveni \scriptscriptfont1=\fivei
  \textfont2=\tensy \scriptfont2=\sevensy \scriptscriptfont2=\fivesy
  \textfont3=\tenex \scriptfont3=\tenex \scriptscriptfont3=\tenex
  \def\it{\fam\itfam\tenit}\textfont\itfam=\tenit
  \def\sl{\fam\slfam\tensl}\textfont\slfam=\tensl
  \def\bf{\fam\bffam\tenbf}\textfont\bffam=\tenbf \scriptfont\bffam=\sevenbf
  \scriptscriptfont\bffam=\fivebf
  \normalbaselineskip=12pt
  \let\sc=\eightrm
  \let\big=\tenbig
  \setbox\strutbox=\hbox{\vrule height8.5pt depth3.5pt width\z@}%
  \normalbaselines\rm}

\def\twelvepoint{\def\rm{\fam0\twelverm}%
  \textfont0=\twelverm \scriptfont0=\ninerm \scriptscriptfont0=\sevenrm
  \textfont1=\twelvei \scriptfont1=\ninei \scriptscriptfont1=\seveni
  \textfont2=\twelvesy \scriptfont2=\ninesy \scriptscriptfont2=\sevensy
  \textfont3=\tenex \scriptfont3=\tenex \scriptscriptfont3=\tenex
  \def\it{\fam\itfam\twelveit}\textfont\itfam=\twelveit
  \def\sl{\fam\slfam\twelvesl}\textfont\slfam=\twelvesl
  \def\bf{\fam\bffam\twelvebf}\textfont\bffam=\twelvebf
\scriptfont\bffam=\ninebf
  \scriptscriptfont\bffam=\sevenbf
  \normalbaselineskip=12pt
  \let\sc=\eightrm
  \let\big=\tenbig
  \setbox\strutbox=\hbox{\vrule height8.5pt depth3.5pt width\z@}%
  \normalbaselines\rm}

\def\fourteenpoint{\def\rm{\fam0\fourteenrm}%
  \textfont0=\fourteenrm \scriptfont0=\tenrm \scriptscriptfont0=\sevenrm
  \textfont1=\fourteeni \scriptfont1=\teni \scriptscriptfont1=\seveni
  \textfont2=\fourteensy \scriptfont2=\tensy \scriptscriptfont2=\sevensy
  \textfont3=\tenex \scriptfont3=\tenex \scriptscriptfont3=\tenex
  \def\it{\fam\itfam\fourteenit}\textfont\itfam=\fourteenit
  \def\sl{\fam\slfam\fourteensl}\textfont\slfam=\fourteensl
  \def\bf{\fam\bffam\fourteenbf}\textfont\bffam=\fourteenbf%
  \scriptfont\bffam=\tenbf
  \scriptscriptfont\bffam=\sevenbf
  \normalbaselineskip=17pt
  \let\sc=\elevenrm
  \let\big=\tenbig
  \setbox\strutbox=\hbox{\vrule height8.5pt depth3.5pt width\z@}%
  \normalbaselines\rm}

\def\seventeenpoint{\def\rm{\fam0\seventeenrm}%
  \textfont0=\seventeenrm \scriptfont0=\fourteenrm \scriptscriptfont0=\tenrm
  \textfont1=\seventeeni \scriptfont1=\fourteeni \scriptscriptfont1=\teni
  \textfont2=\seventeensy \scriptfont2=\fourteensy \scriptscriptfont2=\tensy
  \textfont3=\tenex \scriptfont3=\tenex \scriptscriptfont3=\tenex
  \def\it{\fam\itfam\seventeenit}\textfont\itfam=\seventeenit
  \def\sl{\fam\slfam\seventeensl}\textfont\slfam=\seventeensl
  \def\bf{\fam\bffam\seventeenbf}\textfont\bffam=\seventeenbf%
  \scriptfont\bffam=\fourteenbf
  \scriptscriptfont\bffam=\twelvebf
  \normalbaselineskip=21pt
  \let\sc=\fourteenrm
  \let\big=\tenbig
  \setbox\strutbox=\hbox{\vrule height 12pt depth 6pt width\z@}%
  \normalbaselines\rm}

\def\ninepoint{\def\rm{\fam0\ninerm}%
  \textfont0=\ninerm \scriptfont0=\sixrm \scriptscriptfont0=\fiverm
  \textfont1=\ninei \scriptfont1=\sixi \scriptscriptfont1=\fivei
  \textfont2=\ninesy \scriptfont2=\sixsy \scriptscriptfont2=\fivesy
  \textfont3=\tenex \scriptfont3=\tenex \scriptscriptfont3=\tenex
  \def\it{\fam\itfam\nineit}\textfont\itfam=\nineit
  \def\sl{\fam\slfam\ninesl}\textfont\slfam=\ninesl
  \def\bf{\fam\bffam\ninebf}\textfont\bffam=\ninebf \scriptfont\bffam=\sixbf
  \scriptscriptfont\bffam=\fivebf
  \normalbaselineskip=11pt
  \let\sc=\sevenrm
  \let\big=\ninebig
  \setbox\strutbox=\hbox{\vrule height8pt depth3pt width\z@}%
  \normalbaselines\rm}

\def\eightpoint{\def\rm{\fam0\eightrm}%
  \textfont0=\eightrm \scriptfont0=\sixrm \scriptscriptfont0=\fiverm%
  \textfont1=\eighti \scriptfont1=\sixi \scriptscriptfont1=\fivei%
  \textfont2=\eightsy \scriptfont2=\sixsy \scriptscriptfont2=\fivesy%
  \textfont3=\tenex \scriptfont3=\tenex \scriptscriptfont3=\tenex%
  \def\it{\fam\itfam\eightit}\textfont\itfam=\eightit%
  \def\sl{\fam\slfam\eightsl}\textfont\slfam=\eightsl%
  \def\bf{\fam\bffam\eightbf}\textfont\bffam=\eightbf \scriptfont\bffam=\sixbf%
  \scriptscriptfont\bffam=\fivebf%
  \normalbaselineskip=9pt%
  \let\sc=\sixrm%
  \let\big=\eightbig%
  \setbox\strutbox=\hbox{\vrule height7pt depth2pt width\z@}%
  \normalbaselines\rm}

\def\tenbig#1{{\hbox{$\left#1\vbox to8.5pt{}\right.\n@space$}}}
\def\ninebig#1{{\hbox{$\textfont0=\tenrm\textfont2=\tensy
  \left#1\vbox to7.25pt{}\right.\n@space$}}}
\def\eightbig#1{{\hbox{$\textfont0=\ninerm\textfont2=\ninesy
  \left#1\vbox to6.5pt{}\right.\n@space$}}}

\def\footnote#1{\edef\@sf{\spacefactor\the\spacefactor}#1\@sf
      \insert\footins\bgroup\eightpoint
      \interlinepenalty100 \let\par=\endgraf
        \leftskip=\z@skip \rightskip=\z@skip
        \splittopskip=10pt plus 1pt minus 1pt \floatingpenalty=20000
        \smallskip\item{#1}\bgroup\strut\aftergroup\@foot\let\next}
\skip\footins=12pt plus 2pt minus 4pt 
\dimen\footins=30pc 

\newinsert\margin
\dimen\margin=\maxdimen
\def\titlefont{\seventeenpoint}
\loadtruetwelvepoint 
\loadtrueseventeenpoint
\catcode`\@=\active
\catcode`@=12  
\catcode`\"=\active

\def\eatOne#1{}
\def\ifundef#1{\expandafter\ifx%
\csname\expandafter\eatOne\string#1\endcsname\relax}
\def\notTrue{\iffalse}\def\isTrue{\iftrue}
\def\ifdef#1{{\ifundef#1%
\aftergroup\notTrue\else\aftergroup\isTrue\fi}}
\def\use#1{\ifundef#1\linemessage{Warning: \string#1 is undefined.}%
{\tt \string#1}\else#1\fi}


\global\newcount\refno \global\refno=1
\newwrite\rfile
\newlinechar=`\^^J
\def\ref#1#2{\the\refno\nref#1{#2}}
\def\nref#1#2{\xdef#1{\the\refno}%
\ifnum\refno=1\immediate\openout\rfile=\jobname.refs\fi%
\immediate\write\rfile{\noexpand\item{[\noexpand#1]\ }#2.}%
\global\advance\refno by1}
\def\lref#1#2{\the\refno\xdef#1{\the\refno}%
\ifnum\refno=1\immediate\openout\rfile=\jobname.refs\fi%
\immediate\write\rfile{\noexpand\item{[\noexpand#1]\ }#2\semi}%
\global\advance\refno by1}
\def\cref#1{\immediate\write\rfile{#1\semi}}

\def\semi{;\hfil\noexpand\break}

\def\listrefs{\vfill\eject\immediate\closeout\rfile
\centerline{{\bf References}}\bigskip\frenchspacing%
\input \jobname.refs\vfill\eject\nonfrenchspacing}

\def\inputAuxIfPresent#1{\immediate\openin1=#1
\ifeof1\message{No file \auxfileName; I'll create one.
}\else\closein1\relax\input\auxfileName\fi%
}
\def\NPB{Nucl.\ Phys.\ B}

\def\PLB{Phys.\ Lett.\ B}

\def\ZPC{Z.\ Phys.\ C}

\newif\ifWritingAuxFile
\newwrite\auxfile
\def\SetUpAuxFile{%
\xdef\auxfileName{\jobname.aux}%
\inputAuxIfPresent{\auxfileName}%
\WritingAuxFiletrue%
\immediate\openout\auxfile=\auxfileName}

\def\L{\left(}\def\R{\right)}

\def\LB{\left[}\def\RB{\right]}



\def\Tr{\mathop{\rm Tr}\nolimits}

\def\mod{\mathop{\rm mod}\nolimits}

\def\A#1{{\cal A}_{#1}}

\def\pol{\varepsilon}

\def\ksl{\slashed{k}}

\def\L{\left(}\def\R{\right)}

\def\spa#1.#2{\left\langle#1\,#2\right\rangle}
\def\spb#1.#2{\left[#1\,#2\right]}
\def\lor#1.#2{\left(#1\,#2\right)}
\def\sand#1.#2.#3{%
\left\langle\smash{#1}{\vphantom1}^{-}\right|{#2}%
\left|\smash{#3}{\vphantom1}^{-}\right\rangle}
\def\sandp#1.#2.#3{%
\left\langle\smash{#1}{\vphantom1}^{-}\right|{#2}%
\left|\smash{#3}{\vphantom1}^{+}\right\rangle}
\def\sandpp#1.#2.#3{%
\left\langle\smash{#1}{\vphantom1}^{+}\right|{#2}%
\left|\smash{#3}{\vphantom1}^{+}\right\rangle}
\catcode`@=11  
\def\meqalign#1{\,\vcenter{\openup1\jot\m@th
   \ialign{\strut\hfil$\displaystyle{##}$ && $\displaystyle{{}##}$\hfil
             \crcr#1\crcr}}\,}
\catcode`@=12  

\SetUpAuxFile
\loadfourteenpoint

\def\Li{\mathop{\rm Li}\nolimits}

\def\e{\epsilon}
\def\eps{\epsilon}
\def\del{\partial}
\def\hf{{\textstyle{1\over2}}}
\def\v{V}
\def\f{F}
\def\Atree{A^{\rm tree}}

\def\"#1{{\accent127 #1}}

\def\lr{\leftrightarrow}

\def\li{{\rm Li_2}}
\def\tr{{\rm tr}}

\def\tree{{\rm tree}}
\def\loop{{\rm 1-loop}}
\def\treemhv{{\rm tree\ MHV}}
\def\loopmhv{{\rm 1-loop\ MHV}}
\def\dlips{d{\rm LIPS}}
\def\lsl{\not{\hbox{\kern-2.3pt $\ell$}}}
\def\ksl{\not{\hbox{\kern-2.3pt $k$}}}
\def\Gr{{\rm Gr}}
\def\cg{c_\Gamma}
\def\rg{r_\Gamma}

\def\Split{\mathop{\rm Split}\nolimits}

\def\tn#1#2{t^{[#1]}_{#2}}

\def\Li{\mathop{\rm Li}\nolimits}
\def\Soft{{\cal S}}
%

\baselineskip 15pt
\overfullrule 0.5pt

\def\ref{\nref}

\ref\ManganoReview{M. Mangano and S.J. Parke, Phys.\ Rep.\ 200:301 (1991)}

\ref\ParkeTaylor{S.J. Parke and T.R. Taylor, Phys.\ Rev.\ Lett.\ 56:2459
(1986)}

\ref\RecursiveA{F.A. Berends and W.T. Giele, Nucl.\ Phys.\ B306:759 (1988)}

\ref\RecursiveB{D. A. Kosower, Nucl.\ Phys.\ B335:23 (1990)}

\ref\Susy{M.T.\ Grisaru, H.N.\ Pendleton and P.\ van Nieuwenhuizen,
Phys. Rev. {D15}:996 (1977)\semi
M.T. Grisaru and H.N. Pendleton, Nucl.\ Phys.\ B124:81 (1977)\semi
S.J. Parke and T. Taylor, Phys.\ Lett.\ B157:81 (1985)\semi
Z. Kunszt, Nucl.\ Phys.\ B271:333 (1986)}

\ref\BigTrees{S.J.\ Parke and T. Taylor, Nucl.\ Phys.\ B269:410 (1986)\semi
Z. Kunszt, Nucl.\ Phys.\ B271:333 (1986)\semi
J.F. Gunion and J. Kalinowski, Phys.\ Rev.\ D34:2119 (1986)\semi
F.A. Berends and W.T. Giele, Nucl.\ Phys.\ B294:700 (1987)}
\ref\BigTreesB{R. Kleiss and H. Kuijf, Nucl.\ Phys.\ B312:616 (1989)}

\ref\MPX{M.\ Mangano, S. Parke, and Z.\ Xu, Nucl.\ Phys.\ B298:653 (1988)}

\ref\Stirling{Z. Kunszt and W.J.\ Stirling, Phys.\ Rev.\ D37:2439 (1988)\semi
C.J.\ Maxwell, Phys.\ Lett.\ B192:190 (1987)}

\ref\Nair{ V.P. Nair, Phys.\ Lett.\ B214:215 (1988)}

\ref\Collinear{F.A. Berends and W.T. Giele, Nucl.\ Phys.\ B313:595 (1989)}

\ref\AltPar{G. Altarelli and G. Parisi, Nucl.\ Phys.\ B126:298 (1977)}

\ref\FiveGluon{Z. Bern, L. Dixon and D. A. Kosower, Phys.\ Rev. Lett.\
70:2677 (1993)}

\ref\Long{Z. Bern and D. A.\ Kosower, Nucl.\ Phys.\ B379:451 (1992)}

\ref\StringBased{
Z. Bern and D. A.\ Kosower, Phys.\ Rev.\ Lett.\ 66:1669 (1991)\semi
Z. Bern and D. A.\ Kosower, in {\it Proceedings of the PASCOS-91
Symposium}, eds.\ P. Nath and S. Reucroft (World Scientific, 1992)\semi
Z. Bern, Phys.\ Lett.\ 296B:85 (1992)\semi
Z. Bern, D.C. Dunbar and T. Shimada, Phys.\ Lett.\ B312:277 (1993)}

\ref\Mapping{Z. Bern and D.C. Dunbar,  Nucl.\ Phys.\ B379:562 (1992)}

\ref\Cutting{L.D. Landau, Nucl.\ Phys.\ 13:181 (1959)\semi
 S. Mandelstam, Phys.\ Rev.\ 112:1344 (1958), 115:1741 (1959)\semi
 R.E. Cutkosky, J.\ Math.\ Phys.\ 1:429 (1960)}

\ref\Siegel{W. Siegel, Phys.\ Lett.\ 84B:193 (1979)\semi
D.M.\ Capper, D.R.T.\ Jones and P. van Nieuwenhuizen, Nucl.\ Phys.\
B167:479 (1980)\semi
L.V.\ Avdeev and A.A.\ Vladimirov, Nucl.\ Phys.\ B219:262 (1983)}

\ref\Tasi{
Z. Bern, hep-ph/9304249, in {\it Proceedings of Theoretical
Advanced Study Institute in High Energy Physics (TASI 92)},
eds.\ J. Harvey and J. Polchinski (World Scientific, 1993)}

\ref\KST{Z. Kunszt, A. Signer and Z. Trocsanyi,
Nucl.\ Phys.\ B411:397 (1994)}

\ref\GSB{M. B.\ Green, J. H.\ Schwarz and L.\ Brink,
 Nucl.\ Phys.\ B198:472 (1982)}

\ref\Weak{Z.\ Bern and A.\ Morgan, hep-ph/9312218, to appear in Phys.
Rev. D}

\ref\Superspace{S.J. Gates, M.T. Grisaru, M. Rocek and W. Siegel,
 {\it Superspace}, (Benjamin/Cummings, 1983), pages 390-391}

\ref\PV{ L.M.\ Brown and R.P.\ Feynman, Phys.\ Rev.\ {85} (1952) 231\semi
G.\ Passarino and M.\ Veltman, Nucl.\ Phys.\ {B160} (1979) 151\semi
G. 't Hooft and M. Veltman, \NPB{153:365 (1979)}\semi
R. G. Stuart, Comp.\ Phys.\ Comm.\ 48:367 (1988)\semi
R. G. Stuart and A. Gongora, Comp.\ Phys.\ Comm.\ 56:337 (1990)}

\ref\MVNV{
D. B. Melrose, Il Nuovo Cimento 40A:181 (1965)\semi
W. van Neerven and J. A. M. Vermaseren, Phys.\ Lett.\ 137B:241 (1984)\semi
G. J. van Oldenborgh and J. A. M. Vermaseren, \ZPC{46:425 (1990)}\semi
G. J. van Oldenborgh, PhD thesis, University of Amsterdam (1990)\semi
A. Aeppli, PhD thesis, University of Zurich (1992)}

\ref\Integrals{Z. Bern, L. Dixon and D. A. Kosower,
Phys.\ Lett.\ B302:299 (1993); erratum B318:649 (1993);
Nucl.\ Phys.\ B412:751 (1994)}

\ref\BDKconf{Z. Bern, L. Dixon and D. A. Kosower, Proceedings of
Strings 1993,  May 24-29, Berkeley, CA, hep-th/9311026}

\ref\AllPlus{Z. Bern, G. Chalmers, L. Dixon and D. A. Kosower,
SLAC-PUB-6409, hep-ph/9312333}

\ref\MahlonB{G.D.\ Mahlon, preprint Fermilab-Pub-93/389-T,
hep-ph/9312276}

\ref\DP{L. Dixon and O. Puzyrko, in preparation}

\ref\MahlonA{G.D.\ Mahlon, preprint Fermilab-Pub-93/327-T, hep-ph/9311213}

\ref\Goldberg{H.\  Goldberg and R.\ Rosenfeld, hep-ph/9304238\semi
V.\ Del Duca, Phys.\ Rev.\ D48:5133 (1993) }

\ref\GieleGlover{W.T.\ Giele and E. W. N.\ Glover,
Phys.\ Rev.\ D46:1980 (1992)}

\ref\GGK{W.T.\ Giele, E. W. N.\ Glover and D. A. Kosower,
Nucl.\ Phys.\ B403:633 (1993)}

\ref\KunsztSoper{Z. Kunszt and D. Soper, Phys.\ Rev.\ D46:192 (1992)}

\ref\KunsztSingular{Z. Kunszt, A. Signer, and Z. Trocsanyi,
preprint ETH-TH--94--03, hep-ph/9401294}

\ref\SpinorHelicity{
F.\ A.\ Berends, R.\ Kleiss, P.\ De Causmaecker, R.\ Gastmans and T.\ T.\ Wu,
        Phys.\ Lett.\ 103B:124 (1981)\semi
P.\ De Causmaeker, R.\ Gastmans,  W.\ Troost and  T.\ T.\ Wu,
Nucl. Phys. B206:53 (1982)\semi
R.\ Kleiss and W.\ J.\ Stirling,
   Nucl.\ Phys.\ B262:235 (1985)\semi
   J.\ F.\ Gunion and Z.\ Kunszt, Phys.\ Lett.\ 161B:333 (1985)\semi
 R.\ Gastmans and T. T.\ Wu,
{\it The Ubiquitous Photon: Helicity Method for QED and QCD}
(Clarendon Press,1990)\semi
Z. Xu, D.-H.\ Zhang and L. Chang, Nucl.\ Phys.\ B291:392 (1987)}

\ref\Color{Z. Bern and D. A.\ Kosower, Nucl.\ Phys.\ B362:389 (1991)}

\ref\Subsequent{M.J.\ Strassler,  Nucl.\ Phys.\ B385:145 (1992)\semi
M. G.\ Schmidt and C. Schubert, Phys.\ Lett.\ B318:1993 438\semi
D. Fliegner, M.G.\ Schmidt and C. Schubert,
HD-THEP-93-44, hep-ph/9401221\semi
F. Bastianelli, USITP-93-17, hep-th/9308041}

\ref\Mandelstam{S. Mandelstam, Nucl.\ Phys.\ B213:149 (1983)}

\ref\ManganoParke{M. Mangano and S. J.\  Parke, Nucl.\ Phys.\ B299:673 (1988)}

\ref\Fermion{Z. Bern, L. Dixon and D. A. Kosower, in preparation}

\ref\QCDFact{J. Collins, G. Sterman and D. Soper in {\it Perturbative
Quantum Chromodynamics}, ed. A. Mueller (World Scientific, 1989)}

\ref\CFP{G. Curci, W. Furmanski and R. Petronzio, Nucl.\ Phys.\ B175:27
(1980)\semi
W. Furmanski and R. Petronzio, Phys.\ Lett.\ B97:437 (1980)}

\ref\Kalinowski{J. Kalinowski, K. Konishi, P.N. Scharbach and T.R.
Taylor, Nucl.\ Phys.\ B181:253 (1981)\semi
J. F. Gunion and J. Kalinowski, Phys.\ Rev.\ D32:2303 (1985)}

\ref\Lewin{ L.\ Lewin, {\it Dilogarithms and Associated Functions\/}
(Macdonald, 1958)}

\ref\ChanPaton{J. E.\ Paton and H. M.\ Chan, Nucl.\ Phys.\ B10:516 (1969)}

\ref\Ellis{R. K. Ellis and J. C. Sexton, Nucl.\ Phys.\ B269:445 (1986)}

\ref\FourMassBox{A. Denner, U. Nierste, and R. Scharf,
  \NPB{367:637 (1991)}\semi
N.I. Usyukina and A.I. Davydychev, \PLB{298:363 (1993)}; \PLB{305:136 (1993)}}


\def\FIGstrat{1}
\def\FIGcolinear{2}
\def\FIGcasea{3}
\def\FIGcaseb{4}
\def\FIGhexagon{5}
\def\FIGboxes{6}
\def\FIGboxtypes{7}
\def\FIGSubleadFigureA{8}
\def\FIGSubleadFigureB{9}


\nopagenumbers

\noindent

$\null$

\vskip -1.6 cm

hep-ph/9403226
\hfill SLAC-PUB-6415\hfil\break
\rightline{Saclay/SPhT--T94/20}
\rightline{UCLA/TEP/94/4}
\rightline{SWAT-94-17}

\vskip -2.4 cm

\baselineskip 12 pt
\Title{ One-Loop $n$-Point Gauge Theory Amplitudes, }
\vskip -3.8 cm
\Title{ Unitarity and Collinear Limits }

\vskip -.7 cm
\centerline{\ninerm ZVI BERN${}^{\sharp}$}
\baselineskip=13pt
\centerline{\nineit Department of Physics, UCLA, Los Angeles, CA 90024}
\vglue 0.3cm

\centerline{\ninerm LANCE DIXON${}^{\star}$}
\centerline{\nineit Stanford Linear Accelerator Center, Stanford University,
Stanford, CA 94309}
\vglue 0.3cm

\centerline{\ninerm DAVID C. DUNBAR${}^{\dagger\flat}$ }
\centerline{\nineit Department of Physics, UCLA, Los Angeles, CA 90024}

\vglue 0.2cm
\centerline{\ninerm and}
\vglue 0.2cm
\centerline{\ninerm DAVID A. KOSOWER${}^{\ddagger}$}
\baselineskip12truept
\centerline{\nineit Service de Physique Th\'eorique de Saclay,
 Centre d'Etudes de Saclay}
\centerline{\nineit F-91191 Gif-sur-Yvette cedex, France}

\vglue 0.7cm
\centerline{\tenrm ABSTRACT}
\vglue 0.3cm
{\rightskip=3pc
\leftskip=3pc
\tenrm\baselineskip=12pt
\noindent
We present a technique which utilizes unitarity and collinear limits
to construct ans\"atze for one-loop amplitudes in gauge
theory.  As an example, we obtain the one-loop
contribution to amplitudes for $n$ gluon scattering in $N=4$
supersymmetric Yang-Mills theory with the helicity configuration of
the Parke-Taylor tree amplitudes.  We prove that our $N=4$ ansatz is
correct using general properties of the relevant one-loop $n$-point
integrals.
We also give the ``splitting amplitudes'' which
govern the collinear behavior of one-loop helicity amplitudes
in gauge theories. }
\baselineskip 15 pt

\vglue 0.3cm

\vfil\vskip .2 cm
\noindent\hrule width 3.6in\hfil\break
${}^{\sharp}$Research supported in part by the US Department of Energy
under grant DE-FG03-91ER40662 and in part by the
Alfred P. Sloan Foundation under grant APBR-3222. \hfil\break
${}^{\star}$Research supported by the Department of
Energy under grant DE-AC03-76SF00515.\hfil\break
${}^{\dagger}$Address after Sept. 1, 1994: University of Swansea,
Swansea, SA2 8PP, UK. \hfil\break
${}^{\flat}$Research supported in part by the NSF under grant
PHY-9218990 and in part by the Department of Energy under grant
DE-FG03-91ER40662. \hfil\break
${}^{\ddagger}$Laboratory of the {\it Direction des Sciences de la Mati\`ere\/}
of the {\it Commissariat \`a l'Energie Atomique\/} of France.\hfil\break
\eject

\footline={\hss\tenrm\folio\hss}


\section{Introduction}

Although of fundamental interest in jet physics, perturbative QCD
amplitudes are notoriously difficult to calculate even at tree
level~[\use\ManganoReview].  It has nevertheless been possible to
derive a set of extremely simple formulae at tree level for
``maximally helicity-violating'' (MHV) amplitudes with an arbitrary
number of external gluons.  Parke and Taylor~[\use\ParkeTaylor]
formulated conjectures for these amplitudes in part by using an
analysis of collinear limits; they were later proven by Berends and
Giele~[\use\RecursiveA,\use\RecursiveB] using recursion relations.
The nonvanishing Parke-Taylor formulae are for amplitudes where two
gluons have a given helicity, and the remaining gluons all have the
opposite helicity, in the convention where all external particles are
treated as outgoing.  These amplitudes are called ``maximally
helicity-violating'' because amplitudes with all helicities identical,
or all but one identical, vanish at tree-level as a consequence of
supersymmetry Ward identities~[\use\Susy].  Analytic results are
available for tree amplitudes with all possible helicity
configurations and seven or fewer external
legs~[\use\BigTrees,\use\BigTreesB,\use\MPX], and
numerical implementation of the Berends-Giele recursion relations in
principle allows the computation of arbitrary helicity amplitudes with
an arbitrary number of external legs.  The Parke-Taylor formulae have
nonetheless proven useful both as exact results and in approximation
schemes [\use\Stirling].  Their simplicity also hints at possible
infinite-dimensional symmetries in four-dimensional gauge
theories~[\use\Nair].  At one-loop, results for all helicity
configurations are known for only up to five external legs; formulae
for one-loop amplitudes with special helicity choices but an arbitrary
number of external legs are thus perhaps even more desirable in
investigations of next-to-leading order QCD corrections to multi-jet
cross sections.

At loop level there are two important consistency constraints on
amplitudes, collinear behavior and unitarity, which can be used
as a guide in constructing ans\"atze for amplitudes.  The use of the
collinear limit at one loop is similar to that made by
Parke and Taylor at tree level.  As the momenta of two external legs
become collinear, an $n$-point amplitude reduces to a sum of
$(n-1)$-point amplitudes multiplied by known, singular functions ---
``splitting amplitudes''.
(The sum is over the helicity of the fused leg.)
At loop level, both tree and loop
$(n-1)$-point amplitudes appear, multiplied by loop and tree splitting
amplitudes, respectively.  The tree-level splitting
amplitudes~[\use\Collinear,\use\ManganoReview] are related to the
leading-order polarized Altarelli-Parisi coefficients~[\use\AltPar].
The loop-level splitting functions can be extracted from the collinear
behavior of one-loop five-parton amplitudes.
It is then possible to construct ans\"atze for $n$-point
one-loop amplitudes by a bootstrap approach,
in which the correct collinear behavior is imposed on the ansatz,
and the procedure is jump-started with known
one-loop lower-point results, such as the one-loop five-gluon
amplitudes~[\use\FiveGluon] calculated recently using string-based
methods~[\use\Long,\use\StringBased,\use\Mapping].

The second constraint is that of perturbative unitarity.  We apply the
Cutkosky rules~[\use\Cutting] at the amplitude level to determine the
absorptive parts (cuts) of the amplitude in all possible channels.
The cut amplitude can be written as a tree amplitude on one side of
the cut, multiplied by a tree amplitude on the other side of the cut,
with the loop integral replaced by an integral over the phase space of
the particles crossing the cut.  These cuts are generally much simpler
to evaluate than the full amplitude.  For example, for MHV one-loop
amplitudes it turns out that only the MHV tree amplitudes, given by
the Parke-Taylor formulae, are required to evaluate the cuts, so that
the procedure can be carried out for an arbitrary number of external
legs.  Furthermore, it is possible to calculate cuts in terms of the
imaginary parts of one-loop integrals that would have been encountered
in a direct calculation.  This makes it straightforward to write down
an analytic expression with the correct cuts in all channels, thus
avoiding the need to do a dispersion integral directly.  The unitarity
constraint leaves one with a potential ambiguity at the level of
additive polynomial terms in the amplitude.  (By `polynomial terms'
we actually mean any cut-free function of the kinematic invariants
and spinor products, that is any rational function of these variables.)
The collinear constraint
can be used to resolve much of this ambiguity.

We expect that the twin constraints of unitarity and the collinear
limits will have applicability in generating consistent ans\"atze for
a broad set of one-loop gauge amplitudes; both for general helicity
configurations but relatively few external legs, as are required for
next-to-leading-order QCD predictions for multi-jet processes at
hadron colliders; and also for special helicity configurations with an
arbitrary number of external legs.  In this paper we focus on the
latter application, and more specifically on one-loop MHV amplitudes
in $N=4$ $SU(N_c)$ super-Yang-Mills theory, where we will be able to
obtain results for an arbitrary number of external gluons.
Supersymmetry Ward identities~[\use\Susy] allow us to obtain
additional amplitudes, where certain of the gluons are replaced by
fermions or scalars in the $N=4$ theory.  (We use a
supersymmetry-preserving
regulator~[\use\Siegel,\use\Long,\use\Tasi,\use\KST].)  One-loop $N=4$
amplitudes for four external gluons were first calculated by Green,
Schwarz and Brink, as the low-energy limit of a superstring
amplitude~[\use\GSB].

The $N=4$ super-Yang-Mills results presented here can
be used as a part of the computation of the corresponding $n$-gluon
helicity amplitude in QCD, where gluons and quarks circulate in the
loop.  As is manifest in the string-based
rules~[\use\FiveGluon,\use\Tasi], one can
think of both the gluon and quark contributions to the $n$-gluon QCD
amplitude as different linear combinations of
(a) an $N=4$ supersymmetric amplitude,
(b) an $N=1$ supersymmetric amplitude,
and (c) a scalar in the loop.
The $N=4$ super-Yang-Mills results are thus one of the three
components of a QCD calculation organized in this manner.
There are two advantages of such an organization:
on the one hand, supersymmetry cancellations
(and explicit four- and five-point results)
suggest that the expressions for amplitudes (a) and
perhaps (b) should be
relatively simple; on the other hand, the remaining scalar loop
computation (c) is much easier than a direct gluon (or quark) loop
computation, because the scalar carries no spin
information around the loop.  This decomposition can also reveal
structure in gauge-theory amplitudes that would otherwise remain hidden
[\use\Weak].

For the case of the $N=4$ super-Yang-Mills theory there is a third
constraint which allows us to prove that the process outlined above
generates the correct amplitude.  One can perform a ``gedanken
calculation'' of the loop amplitude, using either superspace
techniques~[\use\Superspace] or a string-based
formalism~[\use\Long,\use\StringBased,\use\Weak].  In either approach,
one finds in each diagram a manifest cancellation in the numerator of
the integrand, such that the numerator loop-momentum polynomial has a
degree which is four less than in the pure gluonic case.  Standard
integral reduction formulae~[\use\PV,\use\MVNV] (or their equivalents
in the language of Feynman-parametrized integrals~[\use\Integrals])
allow one to evaluate loop integrals in terms of a linear combination
of box integrals, triangles and bubbles.  When the reduction formulae
are applied to the $N=4$ integrands, only {\it scalar\/} box integrals
survive (box integrals where the loop momentum polynomial in the
numerator is a constant).  Using this fact we can show that the cuts
uniquely determine the amplitude, proving the correctness of the
ansatz obtained via the unitary-collinear bootstrap.
The physical inputs we use to determine the result are summarized in
fig.~\FIGstrat .

In a non-supersymmetric theory, such as QCD, one-loop $n$-gluon
amplitudes with all gluon helicities identical, or all but one
helicity identical, do not have to vanish.  However, such amplitudes
are pure polynomials --- all cuts vanish.  Also, because these
amplitudes vanish in supersymmetric theories [\use\Susy], the
contributions from particles of different spin in the loop are equal
up to multiplicative constants.  For the identical-helicity case, the
collinear constraints described above have been used to construct an
ansatz~[\use\BDKconf,\use\AllPlus] which has been proven correct by a
recursive procedure~[\use\MahlonB,\use\DP].  Mahlon~[\use\MahlonB] has
also constructed an all-$n$ formula for the configuration with one leg
of opposite helicity from the rest.  Since tree-level amplitudes
vanish for these helicity configurations, these amplitudes do not
provide a vehicle for studying next-to-leading-order corrections to
multi-jet QCD cross-sections.

Other, related examples of $n$-point loop amplitudes that are
known for all $n$ include the $n$-photon massless QED
amplitudes where all photon helicities are identical,
or all but one are identical, which have recently been shown to vanish
for five or more legs by Mahlon~[\use\MahlonA].
The QED results can be generalized to amplitudes with external photons
and gluons, interacting via a massless quark loop.
Amplitudes with five or more legs,
where three or more legs are photons instead of gluons, have
been shown to vanish when all the helicities are
identical~[\use\AllPlus], and also when one of the photon helicities
is reversed~[\use\DP].

Recently, it has been suggested that the tree amplitudes for jet
production grow surprisingly fast for large numbers of external legs
[\use\Goldberg].  It may be interesting to apply our results to
determine whether the one-loop corrections modify this behavior.  To
do so one would need to cancel the infrared singularities using,
for example, the methods of refs.~[\use\GieleGlover,\use\GGK].
(For other methods, see refs.~[\use\KunsztSoper,\use\KunsztSingular].)

This paper is organized as follows: in section~\use\ReviewSection, we
review relevant previous results for tree and one-loop amplitudes.  In
section~\use\CollinearSection\ we describe the collinear behavior
required for a general amplitude and in section~\use\SusySection\ we
impose this collinear behavior in order to construct an ansatz for the
leading-color part of the $N=4$ super-Yang-Mills $n$-point MHV
amplitude.  In section~\use\UnitaritySection\ we describe how to
calculate the cuts for this amplitude, and show that the ansatz has
the correct cuts.  In section~\use\AmbiguitiesFixSection\ we prove
that the ansatz is correct using the structure of the loop integrals.
In section~\use\SubleadingSection\ we give a general formula which
expresses subleading-color contributions to $n$-gluon amplitudes in
terms of leading-color contributions for adjoint representation
particles in the loop.  Section~\use\ConclusionsSection\ contains our
conclusions, appendix~\use\IntegralsAppendix\ collects some needed
formulae for scalar box integrals,
appendix~\use\LoopSplittingAppendix\ contains the splitting amplitudes
appearing in the collinear limits,
and in appendix~\use\SubleadingAppendix\ we apply the general
formula of section~\use\SubleadingSection\ to
derive an explicit form for the subleading-color
$N=4$ supersymmetric amplitudes.


\section{Review of known results}
\tagsection\ReviewSection

Tree-level amplitudes for $U(N_c)$ or $SU(N_c)$ gauge theory
with $n$ external gluons can be decomposed
into color-ordered partial amplitudes, multiplied by
an associated color trace.
Summing over all non-cyclic permutations reconstructs the full
amplitude,
$$
\A{n}^\tree(\{k_i,\lambda_i,a_i\}) =
g^{n-2} \sum_{\sigma\in S_n/Z_n} \Tr(T^{a_{\sigma(1)}}
\cdots T^{a_{\sigma(n)}})
\ A_n^\tree(k_{\sigma(1)}^{\lambda_{\sigma(1)}},\ldots,
            k_{\sigma(n)}^{\lambda_{\sigma(n)}})\ ,
\eqn\TreeAmplitudeDecomposition
$$
where $k_i$, $\lambda_i$, and $a_i$ are respectively the momentum,
helicity ($\pm$), and color index of the $i$-th external
gluon, $g$ is the coupling constant, and $S_n/Z_n$ is the set of
non-cyclic permutations of $\{1,\ldots, n\}$.
The $U(N_c)$ ($SU(N_c)$) generators $T^a$ are the set of hermitian
(traceless hermitian) $N_c\times N_c$ matrices,
normalized so that $\Tr\L T^a T^b\R = \delta^{ab}$.
The color decomposition~(\use\TreeAmplitudeDecomposition)
can be derived in conventional field theory simply by using
$$
f^{abc} = -{i\over\sqrt2} \Tr\L \LB T^a, T^b\RB T^c\R,
\eqn\struct
$$
where the $T^a$ may by either $SU(N_c)$ matrices or $U(N_c)$ matrices.
The structure constants $f^{abc}$ vanish when any index belongs to the
$U(1)$, which is generated by the matrix
$T^{a_{U(1)}} \equiv {\bf 1}/\sqrt{N_c}$;
therefore the partial amplitudes satisfy the $U(1)$ decoupling
identities~[\use\MPX,\use\RecursiveA]
$$
\A{n}(\{k_i,\pol_i,a_i\}_{i=1}^{n-1};k_n,\pol_n,a_{U(1)}) = 0 \; .
\eqn\DecouplingTreeBase
$$
An advantage of using $U(N_c)$ matrices
in the color decomposition is that the $U(N_c)$ Fierz identities
\defeqn\Fierz
$$
\eqalignno{
\Tr(T^a X) \Tr(T^a Y) &= \Tr(X Y) &
(\Fierz{a}) \cr
\Tr(T^a X T^a Y) & = \Tr(X) \Tr(Y) &
(\Fierz{b}) \cr }
$$
are simpler than their $SU(N_c)$ counterparts.
This is useful both when performing a color decomposition on
Feynman diagrams, and when
squaring and summing over colors in order to obtain the cross section.

In a supersymmetric theory, amplitudes with all helicities identical,
or all but one identical, vanish due to supersymmetry Ward
identities~[\use\Susy].
Tree-level gluon amplitudes in super-Yang-Mills and in
purely gluonic Yang-Mills are identical
(fermions do not appear at this order), so that
$$
  A_n^\tree(1^{\pm},2^{+}, \ldots,n^+) = 0.
\anoneqn
$$
Parity may of course be used to simultaneously reverse all helicities
in a partial amplitude.
The non-vanishing Parke-Taylor formulae are for
maximally helicity-violating (MHV) partial amplitudes,
those with two negative helicities and the rest positive,
$$
\eqalign{
  A_{jk}^\treemhv(1,2,\ldots,n)\
&=\ i\, { {\spa{j}.{k}}^4 \over \spa1.2\spa2.3\cdots\spa{n}.1 }\ ,
  \cr}
\eqn\PT
$$
where we have introduced the notation
$$
\eqalign{
  A_{jk}^{\rm MHV}(1,2,\ldots,n)\ &\equiv\
  A_n(1^+,\ldots,j^-,\ldots,k^-,
                \ldots,n^+),  \cr}
\eqn\mhvdef
$$
for a partial amplitude where $j$ and $k$ are the only legs with
negative helicity. Our convention is that all legs are outgoing.
The result~(\PT) is written in terms of spinor inner-products,
$\spa{j}.{l} = \langle j^- | l^+ \rangle = \bar{u}_-(k_j) u_+(k_l)$ and
$\spb{j}.{l} = \langle j^+ | l^- \rangle = \bar{u}_+(k_j) u_-(k_l)$,
where $u_\pm(k)$ is a massless Weyl spinor with momentum $k$ and
chirality $\pm$~[\use\SpinorHelicity,\use\ManganoReview].

For one-loop amplitudes, one may perform a similar color decomposition
to the tree-level decomposition (\use\TreeAmplitudeDecomposition);
in this case, there are up to two traces
over color matrices [\use\Color],
and one must also sum over the different spins $J$ of the internal
particles circulating in the loop.
When all internal particles transform as color adjoints, as is the case
for $N=4$ supersymmetric Yang-Mills theory, the result takes the form
$$
{\cal A}_n\L \{k_i,\lambda_i,a_i\}\R =
  \sum_{J} n_J\,\sum_{c=1}^{\lfloor{n/2}\rfloor+1}
      \sum_{\sigma \in S_n/S_{n;c}}
     \Gr_{n;c}\L \sigma \R\,A_{n;c}^{[J]}(\sigma),
\eqn\ColorDecomposition$$
where ${\lfloor{x}\rfloor}$ is the largest integer less than or equal to $x$
and $n_J$ is the number of particles of spin $J$.
The leading color-structure factor
$$
\Gr_{n;1}(1) = N_c\ \Tr\L T^{a_1}\cdots T^{a_n}\R
\anoneqn
$$
is just $N_c$ times the tree color factor, and the subleading color
structures are given by
$$
\Gr_{n;c}(1) = \Tr\L T^{a_1}\cdots T^{a_{c-1}}\R\,
\Tr\L T^{a_c}\cdots T^{a_n}\R.
\anoneqn
$$
$S_n$ is the set of all permutations of $n$ objects,
and $S_{n;c}$ is the subset leaving $\Gr_{n;c}$ invariant.
Once again it is convenient to use $U(N_c)$ matrices; the extra $U(1)$
decouples from all final results [\use\Color].
(For internal particles in the fundamental ($N_c+\bar{N_c}$) representation,
only the single-trace color structure ($c=1$) would be present,
and the corresponding color factor would be smaller by a factor of $N_c$.
In this case the $U(1)$ gauge boson will {\it not} decouple from
the partial amplitude, so one should only sum over $SU(N_c)$ indices
when color-summing the cross-section.)
In each case the massless spin-$J$ particle is taken to have two
helicity states: gauge bosons, Weyl fermions, and complex scalars.

In the next-to-leading-order correction to the cross-section, summed
over colors, the leading contribution for large $N_c$ comes from
$A_{n;1}^{[J]}$; the subleading-in-$N_c$ corrections $A^{[J]}_{n;c}$
($c>1$) are down by a factor of $1/N_c^2$~[\use\Color].  In
section~\use\SubleadingSection\ we will show how to obtain
$A^{[J]}_{n;c}$ as a sum over permutations of the leading contribution
$A^{[J]}_{n;1}$.  Therefore, it is sufficient to focus on the
calculation of $A^{[J]}_{n;1}$.

Recently, a useful technique based on
string theory~[\use\Long,\use\StringBased] has
been developed for calculating one-loop
amplitudes explicitly, providing the first calculation of all one-loop
five-gluon helicity amplitudes~[\use\FiveGluon].  This method
is an alternative to the conventional Feynman diagram expansion,
but can be understood as a reorganization of field theory
[\use\Mapping] and is also useful in the calculation of effective actions
[\use\Subsequent].
With this method the gluon amplitudes are most naturally written in
a form~[\use\FiveGluon,\use\Tasi] which takes advantage
of the simplicity of contributions from supersymmetry multiplets,
$$
\eqalign{
A_{n;1}^{[0]} &= c_\Gamma  \L \v^s_n \Atree_n + i \f^s_n \R\,,\cr
A_{n;1}^{[1/2]} &= -c_\Gamma \L (\v^f_n+\v^s_n) \Atree_n
                             + i(\f^f_n+\f^s_n) \R\,
,\cr
A_{n;1}^{[1]} &= c_\Gamma \L (\v^g_n+4\v^f_n+\v^s_n) \Atree_n
+ i (\f^g_n + 4\f^f_n + \f^s_n) \R
  \,,\cr
}
\eqn\Totalamp
$$
where the prefactor is
$$
c_\Gamma\ =\ {(4 \pi)^\eps \over 16 \pi^2 }
{\Gamma(1+\eps)\Gamma^2(1-\eps)\over\Gamma(1-2\eps)}\ ,
\eqn\Prefactor
$$
with $\eps= (4-D)/2$ the dimensional regularization parameter.
The $\v_n$'s contain the singular parts of the amplitude (poles as
$\e\to0$), which must be proportional to the tree;
the $\f_n$'s are finite as $\e\to0$ and need not be proportional to
the tree.  (There is some freedom in assigning terms to either
$\v_n$ or $\f_n$.)

The organization~(\use\Totalamp) in terms of $g$, $f$ and $s$ pieces
amounts to calculating the fermion and gluon loop
contributions in terms of the scalar loop contributions plus the
contributions from supersymmetric multiplets.
For an $N=4$ super-Yang-Mills theory, summing over the contributions
from one gluon, four Weyl fermion and three complex (or six real)
scalars, all functions except $\v_n^g$ and $\f_n^g$ cancel from
eq.~(\use\Totalamp) and the amplitudes are
$$
A_{n;1}^{N=4}\ \equiv\
A_{n;1}^{[1]} + 4A_{n;1}^{[1/2]}+3 A_{n;1}^{[0]}\ =\
c_\Gamma \,  ( \v^g_n \; A_{n}^{\rm tree}   + i\f^g_n).
\eqn\Neqfoursum
$$
For an $N=1$ chiral multiplet, containing one scalar and one Weyl fermion,
only the functions $\v^f_n$ and $\f^f_n$ survive,
$$
A_{n;1}^{N=1\ {\rm chiral}}\ \equiv\
A_{n;1}^{[1/2]}\
+A_{n;1}^{[0]}
=\
-c_\Gamma\, \L \v^f_n \Atree_n + i\f^f_n \R.
\eqn\Neqonesum
$$

Organization in terms of $g$, $f$ and $s$ pieces ($N=4$, $N=1$ chiral, and
scalar contributions) is convenient because in the string-based method
or in a superspace approach
there are diagram-by-diagram cancellations within a supermultiplet
which lead to significant simplifications.
In the $N=1$ chiral contribution, the cancellation is easy to see.
The contribution of a complex scalar loop to the effective action is
$$
  \Gamma_{{\rm scalar}}[A]
  = \ln{\rm det^{-1}} (D_\mu D^\mu),
\eqn\scalarG
$$
where $D_\mu = \del_\mu + ig A_\mu$,
while the contribution of a Weyl fermion (coupled non-chirally),
in the second-order formalism motivated by the string-based method
[\use\Mapping], is
$$
\eqalign{
  \Gamma_{{\rm fermion}}[A] & = \ln \det{}_+ \slashed{D} \cr
&  = \ln{\rm det^{1/4}} (D_\mu D^\mu - \hf \sigma^{\mu\nu} F_{\mu\nu}), \cr}
\eqn\fermionG
$$
where $\sigma_{\mu\nu} = {i\over2}[\gamma_\mu,\gamma_\nu]$
is the spin-$\hf$ Lorentz generator.
Expanding the scalar operator $D_\mu D^\mu$ about the
free operator $\del_\mu\del^\mu$ generates derivative interactions,
$\bar\phi A^\mu \del_\mu \phi$, etc.
These lead to a loop-momentum polynomial of maximum degree $m$ for
an $m$-point contribution to the effective action for a scalar in the
loop.
For the $N=1$ chiral multiplet, however, the terms that come solely
from expanding $D_\mu D^\mu$ cancel between scalar and fermion
(since the Dirac trace $\Tr(1)=4$).
Surviving terms require at least
two insertions of $\sigma^{\mu\nu} F_{\mu\nu}$ in the fermion
loop, since $\Tr(\sigma_{\mu\nu})=0$.
Since $\sigma^{\mu\nu} F_{\mu\nu}$ contains no derivatives with
respect to the fermion field, the maximum degree of the loop-momentum
polynomial for the $N=1$ chiral multiplet is two smaller than
for a scalar, fermion, or gluon separately, namely $m-2$ for
an $m$-point contribution~[\use\Tasi,\use\Mapping,\use\Weak].

In an $N=4$ supersymmetric theory there are further cancellations.
In the string-based method (as mapped back from the language of
Feynman parameters to that of loop momenta~[\use\Mapping]),
two and three insertions of
the operator $\hf\sigma^{\mu\nu} F_{\mu\nu}$ for a fermion loop
cancel against two and three insertions of the corresponding operator
$\Sigma^{\mu\nu} F_{\mu\nu}$ for a gluon loop, where $\Sigma^{\mu\nu}$
is the spin-one Lorentz generator.
Thus the loop-momentum polynomials entering into the
calculation of $\v^g_n$ and $\f^g_n$ have a maximum degree four smaller
than for a scalar in the loop, namely $m-4$ for an $m$-point
contribution~[\use\Tasi,\use\Weak].

This $N=4$ cancellation may also be seen in superspace,
following Gates et~al.~[\use\Superspace].  In an $N=1$ superspace calculation
in supersymmetric background field gauge, there are three chiral
ghost superfields from fixing the gauge symmetry.
In the $N=4$ supersymmetric theory, there are three chiral matter
superfields which cancel the ghost contributions to the one-loop
effective action, leaving only the contribution of the vector
superfield $V$.  The vector superfield couples to external fields
via an interaction of the form
$$
  V \L W^\alpha D_\alpha + \bar{W}^{\dot\alpha} \bar{D}_{\dot\alpha}
 \ +\ (D_\alpha,\bar{D}_{\dot\alpha})\hbox{-independent terms} \R V,
\eqn\superspacecoupling
$$
where $W^\alpha$ is the supersymmetric background field strength and
$D_\alpha$ is the superspace covariant derivative.
At least four $D$'s are needed in a loop, so one needs to use
at least four insertions of $W^\alpha D_\alpha$ or
$\bar{W}^{\dot\alpha} \bar{D}_{\dot\alpha}$.
As in the earlier $N=1$ discussion,
each insertion costs a power of the
loop momentum, leading to a maximum degree four smaller than the
scalar case.
The $N=4$ result is related to the ultraviolet finiteness
of $N=4$ super-Yang-Mills~[\use\Mandelstam], since it forces the
potentially divergent two- and three-point contributions to the
effective action to vanish.

The simplicity of the $g,f,s$ organization, suggested by
supersymmetry, is confirmed by explicit four-gluon~[\use\Long,\use\KST]
and five-gluon~[\use\FiveGluon] results.
The separate contributions
$\v_n^s$, $\f_n^s$, $\v_n^f$, $\f_n^f$, $\v_n^g$ and $\f_n^g$
are simpler than the full answer for $n=4,5$.
The $N=4$ supersymmetric part is the simplest of all ---
$\f_4^g$ and $\f_5^g$ vanish, and $\v_4^g$ and $\v_5^g$ are
universal functions, independent of the particular helicity
configuration.
For the four-point amplitude,
$$
\v_4^g =
 -{ 2 \over \eps^2 }\LB
 \Bigl({ \mu^2  \over -s_{12}} \Bigr)^{\eps}
+ \Bigl({ \mu^2  \over -s_{23}} \Bigr)^{\eps}\RB
+ \ln^2 \L {-s_{12} \over -s_{23}} \R
+ \pi^2\ ,
\eqn\fourptVg
$$
where $s_{12}=(k_1+k_2)^2$ and $s_{23}=(k_2+k_3)^2$
are the usual Mandelstam variables, $\mu$ is the renormalization
scale,
and we have used a supersymmetry-preserving regulator
such as dimensional reduction~[\use\Siegel,\use\KST] or
the `four-dimensional helicity scheme'~[\use\Long,\use\Tasi].
This four-point $N=4$ super Yang-Mills amplitude was first calculated
using string theory~[\use\GSB].
For the five-point amplitude~[\use\FiveGluon],
$$
\v^g_5 =
\sum_{i=1}^{5} -{ 1 \over \eps^2 } \Bigl(
{ \mu^2  \over -s_{i,i+1} } \Bigr)^{\eps}
           +\sum_{i=1}^5 \ln\L{-s_{i,i+1} \over-s_{i+1,i+2} }\R\,
                     \ln\L{-s_{i+2,i+3}\over-s_{i-2,i-1}}\R+{5\over6}\pi^2
 \; .
\eqn\fiveptVg
$$
As usual,  $s_{i,i+1}=(k_i+ k_{i+1})^2$.

The remarkable simplicity of these results suggests that it may
be possible to find
a closed-form expression for certain $N=4$ supersymmetric
amplitudes for all $n$.
Since in a supersymmetric theory $A_n (1^\pm,2^+,3^+,\ldots, n^+)$
vanishes~[\use\Susy], we consider in this paper
the maximally helicity-violating one-loop amplitudes that do not
vanish, $A_{jk}^{N=4\ {\rm MHV}}(1,2,\ldots,n)$.


\section{Collinear singularities}
\tagsection\CollinearSection

In the next section we shall construct an ansatz for the one-loop
helicity amplitude
$A_{jk}^{N=4\ {\rm MHV}}(1,2,\ldots,n)$.
The ansatz is based on examining collinear singularities of amplitudes,
starting from the known four- and five-point functions in
eqs.~(\use\fourptVg) and (\use\fiveptVg).
In this section we describe the collinear behavior of
the partial amplitudes $A_{n;1}$ which multiply the single trace
structure $N_c\Tr(T^{a_1} T^{a_2} \cdots T^{a_n})$.
(The partial amplitudes $A_{n;c}$, relevant at subleading order in
$N_c$, are
determined explicitly
in section~\use\SubleadingSection\ in terms of $A_{n;1}$,
so we need not examine their collinear behavior separately.)
We envisage broader applications of these techniques than just to
$N=4$ supersymmetric theories, so in
appendix~\use\LoopSplittingAppendix\ we give the more general one-loop
collinear behavior of (nonsupersymmetric) gauge theory
amplitudes with external quarks as well as gluons.

Consider first the $n$-point tree-level partial amplitude
$A_n(1,2,\ldots,n)$ with an arbitrary helicity configuration.
The external legs may be fermions or gluons.
There is an implicit color ordering of the vertices $1,2,\ldots,n$,
so that collinear singularities arise only from {\it neighboring\/}
 legs $a$ and $b$
becoming collinear~[\use\Collinear,\use\ManganoReview].  These
singularities have the form
$$
A_{n}^{\rm tree}\ \mathop{\longrightarrow}^{a \parallel b}\
\sum_{\lambda=\pm}
 \Split^{\rm tree}_{-\lambda}(a^{\lambda_a},b^{\lambda_b})\,
      A_{n-1}^{\rm tree}(\ldots(a+b)^\lambda\ldots)\ ,
\eqn\treesplit
$$
where the non-vanishing splitting amplitudes diverge as
$1/\sqrt{s_{ab}}$ in the collinear limit $s_{ab}=(k_a+k_b)^2\rightarrow0$.
In the collinear limit $k_a = z\,P$, $k_b = (1-z)\,P$, where $P$ is
the sum of the collinear momenta; $\lambda$ is the helicity of the
intermediate state with momentum $P$.
The tree splitting amplitudes
$\Split^{\rm tree}_{-\lambda}(a^{\lambda_a},b^{\lambda_b})$
may be found in
refs.~[\use\ParkeTaylor,\use\ManganoParke,\use\RecursiveA,\use\ManganoReview]
and are included in appendix~\use\LoopSplittingAppendix.

The collinear limits of the (color-ordered) one-loop partial
amplitudes are expected to have the following form:
$$
\eqalign{
A_{n;1}^{\rm loop}\ \mathop{\longrightarrow}^{a \parallel b}\
\sum_{\lambda=\pm}  \biggl(
  \Split^{\rm tree}_{-\lambda}(a^{\lambda_a},b^{\lambda_b})\,
&
      A_{n-1;1}^{\rm loop}(\ldots(a+b)^\lambda\ldots)
\cr
&  +\Split^{\rm loop}_{-\lambda}(a^{\lambda_a},b^{\lambda_b})\,
      A_{n-1}^{\rm tree}(\ldots(a+b)^\lambda\ldots) \biggr) ,
\cr}
\eqn\loopsplit
$$
as shown diagrammatically in fig.~\FIGcolinear .
The splitting amplitudes
$\Split^{\rm tree}_{-\lambda}(a^{\lambda_a},b^{\lambda_b})$ and
$\Split^{\rm loop}_{-\lambda}(a^{\lambda_a},b^{\lambda_b})$
are universal: they depend only on the two legs becoming
collinear, and not upon the specific amplitude under consideration.
Intuitively, the collinear splitting amplitude describes infrared or
long-distance behavior,
which is not sensitive to the short-distance details
such as the helicities and momenta of the other, hard, external legs.
We expect this universal behavior to hold for all one-loop amplitudes,
with external (massless) fermions as well as gluons;
all one-loop amplitudes that we have inspected do indeed obey
eq.~(\use\loopsplit).  (A similar equation is expected to govern the
limit of one-loop partial amplitudes as one external
gluon momentum becomes
soft.)
The explicit $\Split^{\rm loop}_{-\lambda}(a^{\lambda_a},b^{\lambda_b})$
have been determined from the known four- and five-point
one-loop amplitudes~[\use\FiveGluon,\use\Fermion], and are collected in
appendix~\use\LoopSplittingAppendix.
An outline of a direct proof of the universality of the splitting
amplitudes for the scalar-loop contributions to amplitudes with
external gluons was presented in ref.~[\use\AllPlus];
a more detailed discussion of collinear limits will be given elsewhere.
One can also give an indirect universality argument using
QCD factorization theorems~[\use\QCDFact].
{}From this point of view, the one-loop splitting amplitudes presented here
should be related to the spin-polarized versions of virtual corrections
to evolution of parton distribution functions~[\use\CFP] and of
jet calculus~[\use\Kalinowski].

We may extract the splitting amplitudes
$\Split^{\rm loop}_{-\lambda}(a^{\lambda_a},b^{\lambda_b})$
for $g\to gg$ in an $N=4$ supersymmetric theory rather easily by
inspecting the collinear limits of the
expressions~(\use\Neqfoursum), (\use\fourptVg) and
(\use\fiveptVg) for the four- and five-point amplitudes.
Since these $N=4$ scattering amplitudes are proportional to the
corresponding tree amplitudes ($\f_4^g=\f_5^g=0$),
the supersymmetric loop splitting amplitudes must be proportional
to the tree splitting amplitudes,
$$
 \Split^{\rm loop}_{-\lambda}(a^{\lambda_a},b^{\lambda_b})
   \ =\ {c_\Gamma }
  \times \Split^{\rm tree}_{-\lambda}(a^{\lambda_a},b^{\lambda_b})
  \times r_S^{\rm SUSY}(z,s_{ab}),
\eqn\rSdef
$$
where $s_{ab}=(k_a+k_b)^2$, and
the ratio $r_S^{\rm SUSY}(z,s_{ab})$ is independent of the
helicities.  Equations~(\use\treesplit), (\use\loopsplit)
further require the ratio to obey
$$
 V^g_5\ \mathop{\longrightarrow}^{a \parallel b}\
 V^g_{4} \hskip 0.2  truecm  + r_S^{\rm SUSY}(z,s_{ab} )\ ,
\eqn\fivetofourcoll
$$
implying that
$$
\eqalign{
 r_S^{\rm SUSY}(z,s)
  \ &=
\ - {1\over\e^2}\L{\mu^2\over z(1-z)(-s)}\R^{\e}
        + 2 \ln z\,\ln(1-z) - {\pi^2\over6}\ .\cr}
\eqn\rSanswer
$$
Equations~(\use\rSdef), (\use\rSanswer) turn out to give
the collinear behavior of amplitudes in $N=1$ super Yang-Mills
(with no matter fields), as well as in the $N=4$ theory,
and they hold for external gluinos as well as gluons.
(See appendix~\use\LoopSplittingAppendix.)

The collinear behavior places tight constraints on the possible
form of amplitudes.  However, one must be aware that there do exist
non-vanishing functions which may appear in amplitudes
but do not have singular collinear behavior in any channel.
The simplest non-trivial
example of such a function is the five-point function
$$
{\varepsilon(1,2,3,4) \over \spa1.2 \spa2.3 \spa3.4 \spa4.5 \spa5.1}
\ ,
\eqn\fivepointambiguity
$$
since the contracted antisymmetric tensor
$\varepsilon(1,2,3,4) \equiv  4i \varepsilon_{\mu\nu\rho\sigma}
 k_1^\mu k_2^\nu k_3^\rho k_4^\sigma$
vanishes when any two of the five vectors
$k_i$ become parallel ($\sum_{i=1}^5 k_i = 0$).
Another example is the six-point function
$$
\sum_{P(1,\ldots,5)}
{ \ln(-s_{12}) + \ln(-s_{23}) + \cdots + \ln(-s_{61})
\over \spa1.2 \spa2.3 \spa3.4 \spa4.5 \spa5.6 \spa6.1 }
\eqn\sixpointambiguity
$$
where the
summation is over all 120 permutations of legs 1 through 5.
Without additional information, functions
such as~(\use\fivepointambiguity) and (\use\sixpointambiguity)
represent additive ambiguities in the collinear bootstrap.
Functions of the type~(\use\fivepointambiguity) are somewhat more
insidious because they cannot be detected by unitarity cuts.

In the following section we apply the constraints on collinear
behavior to
construct an ansatz for $N=4$ MHV amplitudes with an arbitrary number of
external legs.  In spite of the potential ambiguities mentioned above,
we will prove that
at least for these MHV amplitudes, the collinear
bootstrap leads naturally to the correct result.


\section{ $N=4$ supersymmetric amplitudes }
\tagsection\SusySection

In this section we use the collinear limits to construct an ansatz
for the one-loop leading-color MHV partial amplitudes in $N=4$
super-Yang-Mills theory.
Our starting point will be the assumption that the $n$-point
amplitude has the same structure as that found at the four- and
five-point level,
$$
 A_{n;1}^{N=4\ {\rm MHV\ loop}}\ =\
   c_\Gamma \times A_{n}^{\rm tree} \times  V_n^g\ ,
\eqn\Ansatz
$$
where $V_n^g$ has no collinear poles but contains the
logarithms and dilogarithms found in a loop amplitude.
With this assumption, and demanding that this amplitude have
the expected collinear limit~(\use\loopsplit),
we find that the function $V^g_n$ must satisfy the generalization
of equation~(\use\fivetofourcoll),
$$
V^g_n\ \mathop{\longrightarrow}^{a \parallel b}\
V^g_{n-1} \hskip 0.2  truecm  +r_S^{\rm SUSY}(z,s_{ab} ).
\eqn\collcondition
$$

In constructing the ansatz it is useful to know the set of integrals
that can appear as a result of explicit diagrammatic calculation,
since this information restricts the form of the possible
logarithms and dilogarithms in the result.
As we shall discuss later,
this set is precisely the set of scalar box integrals, which are
given in appendix~\use\IntegralsAppendix.
Using the five-point function $\v_5^g$~(\use\fiveptVg)
to jump-start the ansatz,
and experimenting with functional forms suggested by
these integrals for small $n$,
we can find a function with the expected
collinear limit (\collcondition) for all $n\geq 5$,
$$
V^g_n=
\sum_{i=1}^{n}  -{ 1 \over \eps^2 } \biggl(
{ \mu^2  \over -\tn2{i} } \biggr)^{\eps}
-\sum_{r=2}^{\lfloor n/2\rfloor -1}
\sum_{i=1}^n
  \ln \biggl({ -\tn{r}{i}\over -\tn{r+1}{i} }\biggr)
  \ln \biggl({ -\tn{r}{i+1}\over -\tn{r+1}{i} }\biggr) +
D_n + L_n +{ n \pi^2 \over 6 }\ ,
\eqn\UniversalFunc
$$
where
$$\eqalign{
\tn{r}{i} &= (k_i+\cdots +k_{i+r-1})^2
}\anoneqn$$
are the momentum invariants, so that $\tn1i = 0$ and $\tn2i = s_{i,i+1}$.
(All indices are understood to be $\mod n$.)
The form of $D_n$ and $L_n$ depends upon whether $n$ is odd or even.
For $n=2m+1$,
$$
D_{2m+1}= -\sum_{r=2}^{m-1} \Biggl( \sum_{i=1}^{n}
\li \biggl[ 1- { \tn{r}{i} \tn{r+2}{i-1}
\over \tn{r+1}{i} \tn{r+1}{i-1} } \biggr]  \Biggr)\ ,
\anoneqn
$$
$$
L_{2m+1}= -{ 1\over 2} \sum_{i=1}^n
  \ln \biggl({ -\tn{m}{i}\over -\tn{m}{i+m+1}  } \biggr)
  \ln \biggl({ -\tn{m}{i+1}\over -\tn{m}{i+m} } \biggr)\ ,
\anoneqn
$$
whereas for $n=2m$,
$$
D_{2m}= -\sum_{r=2}^{m-2} \Biggl( \sum_{i=1}^{n}
\li \biggl[ 1- { \tn{r}{i} \tn{r+2}{i-1}
\over \tn{r+1}{i} \tn{r+1}{i-1} }  \biggr]  \Biggr)
-\sum_{i=1}^{n/2} \li \biggl[ 1- { \tn{m-1}{i} \tn{m+1}{i-1}
\over \tn{m}{i} \tn{m}{i-1}} \biggr]\ ,
\anoneqn
$$
$$
L_{2m}=-{1\over 4} \sum_{i=1}^n
  \ln \biggl({ -\tn{m}{i}\over -\tn{m}{i+m+1}  } \biggr)
  \ln \biggl({ -\tn{m}{i+1}\over -\tn{m}{i+m} } \biggr)\ .
\hskip 4 truecm
\anoneqn
$$
For $n=6$ we have verified this result explicitly
by direct calculation
using the string-based method.

The functions $D_n$ containing the dilogarithms first appear in the
six-point amplitude.
One must use various dilogarithm
identities to show that the function $\v^g_n$ has the expected collinear
behavior. Note that the explicit cyclic symmetry of $\v^g_n$ allows
us to verify the behavior~(\use\collcondition) for just one collinear
pair of legs $a,b$.
For example, consider the six-point function where
$$
D_6\ =\
-\li \biggl[1-{ \tn{2}{1} \tn{4}{6}\over
\tn{3}{1} \tn{3}{6} } \biggr]
-\li \biggl[1-{ \tn{2}{2} \tn{4}{1}\over
\tn{3}{2} \tn{3}{1} } \biggr]
-\li \biggl[1-{ \tn{2}{3} \tn{4}{2}\over
\tn{3}{3} \tn{3}{2} } \biggr]\ ,
\anoneqn
$$
and let legs $5$ and $6$ become collinear:
$k_5 \rightarrow z k_5'$ and $k_6 \rightarrow (1-z) k_5'$. We find
$$
D_6\ \rightarrow\
-\li \biggl[ 1- {z\tn{2}{1} \over z\tn{2}{1}+(1-z)\tn{2}{3}}
   \biggr]
-{\pi^2 \over 6 }
-\li \biggl[1-{(1-z)\tn{2}{3} \over z\tn{2}{1}+(1-z)\tn{2}{3}}
   \biggr]\ ,
\anoneqn
$$
where we have used momentum conservation appropriate for the
five point function.
Using the dilogarithm identity~[\use\Lewin]
$$
 \li(x) + \li(1-x) = {\pi^2\over6} - \ln(x)\ln(1-x),
\eqn\easydilog
$$
this reduces to
$$
-{\pi^2 \over 3 }
+\ln\Biggl( { z \tn{2}{1} \over z\tn{2}{1} +(1-z) \tn{2}{3} } \Biggr)
\ln\Biggl( { (1-z) \tn{2}{3} \over z\tn{2}{1} +(1-z) \tn{2}{3} } \Biggr)
\ .
\anoneqn
$$
These logarithms combine with those arising from other parts of the
amplitude to give the correct collinear limit.

For the collinear limit of a seven-point function the dilogarithms
become more complicated.
Take the dilogarithms in the seven-point function
$$
\eqalign{
D_7 \ &=\
-\li \biggl[1-{ \tn{2}{1} \tn{4}{7}\over
\tn{3}{1} \tn{3}{7} } \biggr]
-\li \biggl[1-{ \tn{2}{2} \tn{4}{1}\over
\tn{3}{2} \tn{3}{1} } \biggr]
-\li \biggl[1-{ \tn{2}{3} \tn{4}{2}\over
\tn{3}{3} \tn{3}{2} } \biggr]
-\li \biggl[1-{ \tn{2}{4} \tn{4}{3}\over
\tn{3}{4} \tn{3}{3} } \biggr]
\cr
&
-\li \biggl[1-{ \tn{2}{5} \tn{4}{4}\over
\tn{3}{5} \tn{3}{4} } \biggr]
-\li \biggl[1-{ \tn{2}{6} \tn{4}{5}\over
\tn{3}{6} \tn{3}{5} } \biggr]
-\li \biggl[1-{ \tn{2}{7} \tn{4}{6}\over
\tn{3}{7} \tn{3}{6} } \biggr]
\cr}
\anoneqn
$$
and consider the limit where legs $6$ and $7$ become collinear,
$k_6 \rightarrow z k_6'$, $k_7  \rightarrow (1-z) k_6'$.
Then
$$
\eqalign{
D_7 \rightarrow &
-\li \biggl[1-{ \tn{2}{1} ( (1-z)\tn{4}{6} +z\tn{3}{1} ) \over
\tn{3}{1} ( (1-z)\tn{3}{6} +z\tn{2}{1} ) } \biggr]
-\li \biggl[1-{ \tn{2}{2} \tn{4}{1}\over
\tn{3}{2} \tn{3}{1} } \biggr]
-\li \biggl[1-{ \tn{2}{3} \tn{4}{2}\over
\tn{3}{3} \tn{3}{2} } \biggr]
\cr &
-\li \biggl[1-{ \tn{2}{4} ((1-z)\tn{3}{3} +z\tn{4}{3} ) \over
  \tn{3}{3} ( (1-z)\tn{2}{4} +z\tn{3}{4} )  } \biggr]
-\li \biggl[1-{ z\tn{3}{4} \over (1-z)\tn{2}{4} +z\tn{3}{4} } \biggr]
-{\pi^2 \over 6 }
\cr &
-\li \biggl[1-{(1-z)\tn{3}{6} \over (1-z)\tn{3}{6}+z\tn{2}{1}}\biggr]
\; . \cr}
\anoneqn
$$
This expression can be rewritten,
using momentum conservation appropriate for the six-point
amplitude, as
$$
\eqalign{
D_6
&+\li \biggl[1-{ \tn{2}{1} \tn{2}{4}\over
\tn{3}{1} \tn{3}{3} } \biggr]
-\li \biggl[1-{ \tn{2}{1} ( (1-z)\tn{2}{4} +z\tn{3}{1} ) \over
\tn{3}{1} ( (1-z)\tn{3}{3} +z\tn{2}{1} ) } \biggr]
-\li \biggl[1-{ \tn{2}{4} ((1-z)\tn{3}{3} +z\tn{2}{1} ) ) \over
  \tn{3}{3} ( (1-z)\tn{2}{4} +z\tn{3}{1} )  } \biggr]
\cr &
-\li\biggl[1-{  z \tn{3}{1} \over (1-z)\tn{2}{4} +z\tn{3}{1} }\biggr]
-{\pi^2 \over 6}
-\li\biggl[1-{(1-z)\tn{3}{3} \over (1-z)\tn{3}{3} +z\tn{2}{1}}\biggr]
\; .\cr}
\anoneqn
$$
If we define variables $X=(1-z)\tn{2}{4} / ( (1-z)\tn{2}{4} +z\tn{3}{1} ) $
and $Y=z\tn{2}{1}/( (1-z)\tn{3}{3} +z\tn{2}{1} )$ then this collinear limit
of $D_7$ can be expressed in terms of $D_6$ along with
a function of $X$ and $Y$,
$$
\eqalign{
&D_6
-{\pi^2 \over 6 }
+\li \Bigl[1-{ X Y \over (1-X)(1-Y) }\Bigr]
-\li \Bigl[1-{Y\over 1-X }\Bigr]
\cr &
-\li \Bigl[1-{X \over 1-Y }\Bigr]
-\li \Bigl[1-{(1-X)} \Bigr]
-\li \Bigl[1-{(1-Y)} \Bigr]\ .
\cr}
\anoneqn
$$
The dilogarithms in this expression can be eliminated using
the following dilogarithm identity,
$$
\eqalign{
\li \Bigl[1&-{ xy\over (1-x)(1-y) }\Bigr] =
\li \Bigl[1-{(1-x)} \Bigr]
+\li \Bigl[1-{(1-y)} \Bigr]
+\li \Bigl[1-{x \over 1-y }\Bigr]
\cr
&
+\li \Bigl[1-{y\over 1-x }\Bigr]
+{\rm ln} (x){\rm ln}(1-x) +{\rm ln} (y){\rm ln}(1-y)
-{\rm ln} (1-x){\rm ln}(1-y) -{\pi^2 \over 6}\ ,
\cr}
\eqn\AbelId
$$
which is equivalent via~(\use\easydilog)
to a rearrangement of Abel's identity~[\use\Lewin].
With this identity we have, in the collinear limit
$$
D_7\ \rightarrow\
D_6 +{\rm logarithms} -{\pi^2 \over 3}\ .
\anoneqn
$$
The logarithms combine with those arising from
other terms in $V_7^g$ to ensure that
the ansatz has the correct collinear limit.

\def\Fn{n}
\def\Fs#1#2{F^{{#1}}_{\Fn:#2}}
\def\Fone{\Fs{\rm 1m}}
\def\Feasy{\Fs{{\rm 2m}\,e}}
\def\Fhard{\Fs{{\rm 2m}\,h}}
\def\Fthree{\Fs{\rm 3m}}
\def\Ffour{\Fs{\rm 4m}}

An alternate way to write the function $V_n^g$ is
in terms of functions related to the scalar box integrals:
$$
\eqalign{
(\mu^2)^{-\eps} V_{2m+1}^g\ &=\  \sum_{r=2}^{m-1} \sum_{i=1}^{n}
  \Feasy{r;i}
+\sum_{i=1}^{n} \Fone{i}\ , \cr
(\mu^2)^{-\eps} V_{2m}^g\ &=\
\sum_{r=2}^{m-2} \sum_{i=1}^{n}   \Feasy{r;i}
+ \sum_{i=1}^{n} \Fone{i} +\sum_{i=1}^{n/2} \Feasy{m-1;i}\ , \cr}
\eqn\CoeffDefine
$$
where the box functions $F$ are defined in
appendix~\use\IntegralsAppendix\ (and are the box integrals
multiplied by the dimensionful denominator and a constant).
Only certain special types of box functions appear
in the expression~(\use\CoeffDefine).
The ``easy two-mass'' box function $\Feasy{r;i}$
has two diagonally-opposite massless legs,
with momenta equal to the external momenta $k_{i-1}$ and $k_{i+r}$;
the remaining two diagonally-opposite
legs are massive, as they contain the sum of $r$ adjacent external
momenta between $i-1$ and $i+r$,
and of $n-r-2$ adjacent momenta between $i+r$ and $i-1$.
The one-mass box function $\Fone{i}$ is a special
case of $\Feasy{r;i-2}$ with $r=1$.
Box functions with two adjacent
external masses, or with three or four external masses,
do not appear.
This property can be understood from the structure of the unitarity
cuts to be calculated directly in the next section;
the representation~(\use\CoeffDefine) is a convenient one for
comparing with the results of that calculation.

This representation is also convenient
for proving that the ansatz possesses the required collinear
limit~(\use\collcondition) for arbitrary $n$; consider, for example the limit
$k_{n-1} \parallel k_n$ for $n$ odd, and split up the sums in
equation~(\use\CoeffDefine) as follows,
$$\hskip -20pt\eqalign{
(\mu^2)^{-\eps} V_{2m+1}^g\ &=\
\sum_{r=3}^{m-1} \sum_{i=n-r+1}^{n-1} \Feasy{r;i}
+\sum_{r=2}^{m-2} \sum_{i=2}^{n-r-2} \Feasy{r;i}
+\sum_{i=2}^m \Feasy{m-1;i}
+\sum_{i=4}^{n-1} \Fone{i}\cr
&+ \sum_{r=3}^{m-1} \LB\Feasy{r;n}+\Feasy{r-1;1}\RB
+\sum_{r=3}^{m-1}\LB\Feasy{r;n-r}+\Feasy{r-1;n-r}\RB
+\LB \Feasy{m-1;1}+\Feasy{m-1;m+1}\RB\cr
&+ \LB \Feasy{2;n}+\Fone{3}\RB
+\LB\Feasy{2;n-2} + \Fone{n}\RB
+\LB\Feasy{2;n-1} + \Fone{1}+\Fone{2}\RB\;.\cr
}\anoneqn$$
Using the collinear limits of the $F$ functions detailed in
appendix~\use\IntegralsAppendix, these sums reduce to
$$\eqalign{
&\sum_{r=2}^{m-2} \sum_{i=n-r}^{n-1} {\def\Fn{n-1}\Feasy{r;i}}
+\sum_{r=2}^{m-2} \sum_{i=2}^{n-r-2} {\def\Fn{n-1}\Feasy{r;i}}
+\sum_{i=2}^m {\def\Fn{n-1}\Feasy{m-1;i}}
+\sum_{i=4}^{n-1} {\def\Fn{n-1}\Fone{i}}\cr
&+ \sum_{r=2}^{m-2} \LB{\def\Fn{n-1}\Feasy{r;1}}\RB
+\sum_{r=2}^{m-2} \LB{\def\Fn{n-1}\Feasy{r;n-r-1}}\RB
+\LB{\def\Fn{n-1}\Feasy{m-1;1}}\RB\cr
&+ \LB{\def\Fn{n-1}\Fone{3}}\RB + \LB{\def\Fn{n-1}\Fone{1}}\RB
+\LB{\def\Fn{n-1}\Fone{2}}+(\mu^2)^{-\e}r_S^{\rm SUSY}(z,s_{n-1,n})\RB\cr
&= (\mu^2)^{-\eps} V_{2m}^g+(\mu^2)^{-\e}r_S^{\rm SUSY}(z,s_{n-1,n})\;.
}\anoneqn$$
The proof is similar for $n$ even.


\section{ Unitarity constraints}
\tagsection\UnitaritySection

We now show that the unitarity cuts in the expression~(\use\CoeffDefine)
obtained from the collinear limits are correct for the maximally
helicity-violating (MHV) configurations~(\use\mhvdef),
$A_{jk}^{{\rm MHV}}(1,2,\ldots,n)$.
We do this by calculating the cuts directly from the Cutkosky
rules~[\use\Cutting], which turn out to require only the
Parke-Taylor (MHV) tree amplitudes~[\use\ParkeTaylor]  for their evaluation.
The simple structure of the Parke-Taylor amplitudes~(\use\PT)
allows us to evaluate the cuts for all $n$ in terms of a single
hexagon integral, or four box integrals.
The cuts determine all terms in the amplitude apart
from additive polynomial terms without logarithms or dilogarithms.
We will fix the remaining polynomial ambiguity in the next section.

We compute the cuts of the MHV amplitudes in all possible channels.  We
consider the amplitude, not in a physical kinematic configuration, but
in a region where exactly one of the momentum invariants is taken to
be positive (time-like), and the rest are negative (space-like).  In
this way we isolate cuts in a single momentum channel.  We apply the
Cutkosky rules at the amplitude level, rather than at the diagram
level.  That is, we write the sum of all cut diagrams as the sum of
all tree diagrams on one side of the cut, multiplied by the sum of all
tree diagrams on the other side of the cut.  Thus the cut in the
one-loop amplitude is given by the integral over a two-body
phase-space of the product of two tree amplitudes, which is then
summed over each intermediate helicity configuration that contributes.
Since we use the helicity convention that all particles are outgoing, the
helicity of each of the two intermediate particles is reversed upon
crossing the cut.

There are two distinct cases to consider: (a) the two negative
helicity gluons are on the same side of the cut, and (b) the two
negative helicity gluons are on opposite sides of the cut.  Case (a)
is easier, so we consider it first. This cut amplitude is shown in
fig.~\FIGcasea .  First, we note that the contributions to the cut
from intermediate fermions vanish in this case since there is no way
to assign intermediate helicity configurations so that fermion
helicity is conserved and both the tree amplitudes are non-zero.  For
example, the intermediate helicity assignment in fig.~\FIGcasea\ does
not conserve fermion helicity.  Since tree amplitudes with two
external fermions vanish if the gluons all carry the same
helicity~[\use\ManganoReview],
the helicity assignments which conserve fermion helicity also vanish.
The contributions to the cut from intermediate
complex scalars vanish by an analogous argument where the `helicity'
of a complex scalar refers to particle or antiparticle assignment rather
than genuine helicity.

Thus in case (a) the only contribution is from intermediate gluons
with the helicity assignment shown in fig.~\FIGcasea .
The tree amplitudes on either side of the cut are
pure-glue MHV tree amplitudes, given in eq.~(\use\PT).
Consider the cut in the channel
$(k_{m_1}+k_{m_1+1}+\cdots+k_{m_2-1}+k_{m_2})^2$ for the
loop amplitude $A_{jk}^\loopmhv(1,2,\ldots,n)$, where $m_1 \leq j < k
\leq m_2$.  The cut for this channel is given by
$$
\eqalign{
  &\int \dlips(-\ell_1,\ell_2)
  \ A^\treemhv_{jk}(-\ell_1,m_1,\ldots,m_2,\ell_2)
  \ A^\treemhv_{(-\ell_2)\ell_1}(-\ell_2,m_2+1,\ldots,m_1-1,\ell_1) \cr
  &=\  -i
{ A_{jk}^\treemhv(1,2,\ldots,n)}
\int \dlips(-\ell_1,\ell_2)
   { \spa{(m_1-1)}.{m_1} {\spa{\ell_1}.{\ell_2}}^2 \spa{m_2}.{(m_2+1)}
    \over \spa{(m_1-1)}.{\ell_1} \spa{\ell_1}.{m_1} \spa{m_2}.{\ell_2}
    \spa{\ell_2}.{(m_2+1)} }
   \ , \cr
}\eqn\caseacut
$$
where the spinor inner products are labelled by either loop
momenta or external particle labels.
We have removed minus signs from the spinor inner products by
cancelling constant phases.
Here $\ell_1$ is the running loop momentum between vertices $(m_1-1)$
and $m_1$, and $\ell_2$ is the running loop momentum between vertices
$m_2$ and $(m_2+1)$.  The arguments in the second amplitude
($m_2+1,\ldots,m_1-1$) should be understood to
mean~($m_2+1,\ldots,n,1,\ldots,m_1-1$).
The $(4-2\e)$-dimensional
Lorentz-invariant phase space measure is denoted by
$\dlips(-\ell_1,\ell_2)$.
(The factor of $(\mu^2)^\eps$ has been
suppressed here and in subsequent formulae.)
In equation~(\caseacut) the integration
momenta $\ell_1$ and $\ell_2$ appear in only
a few of the factors, even though we are considering an
arbitrary number of external legs.  This simplicity
is due to the simple form of the
Parke-Taylor amplitudes (\use\PT).  Observe
that the integral~(\caseacut)
does not depend on the locations $j,k$ of the negative helicity gluons.
This result does not require
$N=4$ supersymmetry, but only the assumption that the two negative
helicity external states in the MHV amplitude are on the same side of the
cut.
We will see that the same independence holds
in case~(b), for the special case of an $N=4$ super-Yang-Mills theory.

Consider case (b) now. This amplitude is shown in fig.~\FIGcaseb\
with the possible intermediate helicity assignments.
As shown, there are two sets of possible helicity
configurations across this cut.
The cut has contributions from all possible intermediate states,
the scalars, the fermions and the gluons.
However, the $N=4$ supersymmetric sum turns out to be very
simple, thanks to the supersymmetric Ward
identities~[\use\Susy,\use\ManganoReview]
for the nonvanishing MHV tree amplitudes.
These identities relate amplitudes
with all external gluons, $g$, to amplitudes where a pair of gluons is
replaced by a pair of gluinos, $\Lambda$, or a pair of complex scalars,
$\phi$.
Specifically, if we have the pure gluon MHV tree amplitude
$\Atree( g_1^-, g_2^+ , \ldots, g_j^-, \ldots g_n^+)$,
given explicitly in eq.~(\use\PT), then the amplitudes
with gluinos and scalars are
$$
\eqalign{
\Atree(\Lambda_1^-, g_2^+, \ldots , g_j^-, \ldots, \Lambda_n^+)
\ &=\ { \spa{j}.{n} \over \spa{j}.{1} }
\ \Atree( g_1^-, g_2^+ , \ldots, g_j^-, \ldots, g_n^+),
\cr
\Atree(\phi_1^-, g_2^+, \ldots ,g_j^-, \ldots, \phi_n^+)
\ &=\ { {\spa{j}.{n}}^2 \over {\spa{j}.{1}}^2 }
\ \Atree( g_1^-, g_2^+ , \ldots, g_j^-, \ldots, g_n^+),
\cr}
\eqn\scalarContrib
$$
where `$\ldots$' denotes positive-helicity gluons,
and the helicity assignments on $\phi$ refer to particle or antiparticle
assignments rather than genuine helicity.
These and the corresponding relations for the other MHV amplitudes
allow us to sum the contributions due to intermediate states.
If we consider the contribution due to a single scalar state
we have
$$
\eqalign{
\  -i
& { { A_{jk}^\treemhv  (1,2,\ldots,n)} \over
{\spa{j}.{k}}^4 }
\int \dlips(-\ell_1,\ell_2)
\cr & \hskip 3.5 truecm
\times
   { \spa{(m_1-1)}.{m_1} \spa{m_2}.{(m_2+1)}
{\spa{\ell_1}.{j}}^2{\spa{\ell_2}.{j}}^2
{\spa{\ell_1}.{k}}^2{\spa{\ell_2}.{k}}^2
    \over
{\spa{\ell_1}.{\ell_2}}^2
\spa{(m_1-1)}.{\ell_1} \spa{\ell_1}.{m_1} \spa{m_2}.{\ell_2}
    \spa{\ell_2}.{(m_2+1)} }\ . \cr}
\eqn\casebbcut
$$
The contributions for the other states can be obtained using
eq.~(\use\scalarContrib).

In the $N=4$ multiplet we have one gluon, four Weyl fermions
(both with plus and minus helicities) and three complex
scalars (or six real scalars).
If we use~(\use\scalarContrib) to write the different contributions
to the integrand
as multiples of the single scalar state contribution~(\use\casebbcut),
then the overall factor for the $N=4$ multiplet is
$$
 \rho^2\ \equiv\ \left({ \spa{\ell_1}.{j} \spa{\ell_2}.{k} \over
         \spa{\ell_2}.{j} \spa{\ell_1}.{k} } \right)^2
 - 4 \left({ \spa{\ell_1}.{j} \spa{\ell_2}.{k} \over
         \spa{\ell_2}.{j} \spa{\ell_1}.{k} } \right)
 + 6 - 4 \left({ \spa{\ell_1}.{j} \spa{\ell_2}.{k} \over
         \spa{\ell_2}.{j} \spa{\ell_1}.{k} } \right)^{-1}
 +  \left({ \spa{\ell_1}.{j} \spa{\ell_2}.{k} \over
         \spa{\ell_2}.{j} \spa{\ell_1}.{k} } \right)^{-2}\ ,
\eqn\neqfourfactor
$$
where the central term is the contributions from the three
complex scalars (the two `helicity' assignments give equal
contributions for complex scalars),
the terms flanking the central term are the fermion contributions
(one for each possible helicity configuration),
and the remaining terms are the contributions of the two gluon
helicities.  Using the Schouten identity,
$$
\spa a.b\spa c.d = \spa a.d\spa c.b + \spa a. c\spa b.d\; ,
\eqn\Schouten
$$
and some rearrangements we have
$$
\eqalign{
 \rho^2\ &=\ { ( \spa{\ell_1}.{j} \spa{\ell_2}.{k}
               - \spa{\ell_2}.{j} \spa{\ell_1}.{k} )^4
 \over {\spa{\ell_1}.{j}}^2 {\spa{\ell_2}.{k}}^2
       {\spa{\ell_2}.{j}}^2 {\spa{\ell_1}.{k}}^2 }
  =\  { {\spa{\ell_1}.{\ell_2}}^4 {\spa{j}.{k}}^4
 \over {\spa{\ell_1}.{j}}^2 {\spa{\ell_2}.{j}}^2
       {\spa{\ell_1}.{k}}^2 {\spa{\ell_2}.{k}}^2 } \ .\cr}
\eqn\newneqfourfactor
$$
Using the form~(\newneqfourfactor) for $\rho^2$,
we see that the product of $\rho^2$ with the
integrand of the scalar loop contribution~(\use\casebbcut)
is identical to the integrand for the cut in case~(a),
equation~(\caseacut).
Thus in both cases (a) and (b), the cut reduces to eq.~(\caseacut).
We now evaluate this integral.

\def\Slash#1{\slash\hskip -0.17 cm #1}
The integral~(\caseacut) can be viewed
as the cut hexagon loop integral shown in fig.~\FIGhexagon.
To see this, one may use the on-shell condition $\ell_1^2=\ell_2^2=0$
to rewrite the four spinor product denominators in~(\caseacut)
as scalar propagators, multiplied by a numerator factor, for
example
$$
     {1 \over \spa{\ell_1}.{m_1}}
  = {\spb {m_1}.{\ell_1} \over \spa{\ell_1}.{m_1} \spb {m_1}.{\ell_1}}
  = {\spb {m_1}.{\ell_1} \over 2 \ell_1 \cdot k_{m_1}}
  = {-\spb {m_1}.{\ell_1} \over (\ell_1 - k_{m_1})^2}\ .
\eqn\propagators
$$
In addition to these four propagators, there are two cut propagators
implicit in the phase-space integral $\int \dlips(-\ell_1,\ell_2)$.
(For a four- or five-point loop amplitude there
are obviously not enough external momenta to make it a genuine hexagon;
in this case one can take some of the external momenta to be null.)

Rather than evaluate the cut hexagon integral directly,
we can perform a ``partial fraction'' decomposition of the integrand
in order to reduce the number of spinor product factors in the
denominator of each term, which will break up the integral
into a sum of cut box integrals.
(This is equivalent to a Passarino-Veltman reduction~[\use\PV].)
Using the Schouten identity~(\use\Schouten) we rewrite the integrand
of~(\use\caseacut) as
$$
\eqalign{
 {\cal I}\ &=\
 - \biggl( { \spa{(m_1-1)}.{m_1} {\spa{\ell_1}.{\ell_2}}
 \over \spa{(m_1-1)}.{\ell_1} \spa{\ell_1}.{m_1} } \biggr)
  \biggl( { \spa{m_2}.{(m_2+1)} {\spa{\ell_2}.{\ell_1}}
 \over \spa{m_2}.{\ell_2} \spa{\ell_2}.{(m_2+1)} } \biggr) \cr
 &=\
  -\biggl( { \spa{(m_1-1)}.{\ell_2} \over \spa{(m_1-1)}.{\ell_1} }
         - { \spa{m_1}.{\ell_2} \over \spa{m_1}.{\ell_1} } \biggr)
   \biggl( { \spa{m_2}.{\ell_1} \over \spa{m_2}.{\ell_2} }
         - { \spa{(m_2+1)}.{\ell_1} \over \spa{(m_2+1)}.{\ell_2} }
   \biggr) \cr
 &=\ {\spa{m_1}.{\ell_2}\spa{m_2}.{\ell_1} \over
      \spa{m_1}.{\ell_1}\spa{m_2}.{\ell_2}}
   \pm \Bigl[\ m_1 \lr (m_1-1)\ ,\ m_2 \lr (m_2+1)\ \Bigr]\ , \cr
}
\eqn\firstpartialfr
$$
antisymmetrizing in each exchange.  In terms of propagators,
$$
 \eqalign{
  {\cal I}\ &=\
   -{\spb{\ell_1}.{m_1}\spa{m_1}.{\ell_2}
     \spb{\ell_2}.{m_2}\spa{m_2}.{\ell_1}
    \over (\ell_1-k_{m_1})^2 (\ell_2+k_{m_2})^2}
    \pm \Bigl[\ m_1 \lr (m_1-1)\ ,\ m_2 \lr (m_2+1)\ \Bigr] \cr
 &=\ -{\tr_-(\lsl_1 \ksl_{m_1} \lsl_2 \ksl_{m_2})
     \over (\ell_1-k_{m_1})^2 (\ell_2+k_{m_2})^2}
     -{\tr_-(\lsl_1 \ksl_{m_1-1} \lsl_2 \ksl_{m_2})
      \over (\ell_1+k_{m_1-1})^2 (\ell_2+k_{m_2})^2} \cr
&\qquad
     -{\tr_-(\lsl_1 \ksl_{m_1} \lsl_2 \ksl_{m_2+1})
     \over (\ell_1-k_{m_1})^2 (\ell_2-k_{m_2+1})^2}
     -{\tr_-(\lsl_1 \ksl_{m_1-1} \lsl_2 \ksl_{m_2+1})
      \over (\ell_1+k_{m_1-1})^2 (\ell_2-k_{m_2+1})^2}\ ,\cr
}
\eqn\traceterms
$$
where the $\tr_-$
indicates that a $(1-\gamma_5)/2$ projector has been inserted
into the trace.
Thus the integral of ${\cal I}$ is the sum of four cut box
integrals, shown in fig.~\FIGboxes, corresponding to cancelling
different pairs of propagators in the cut hexagon integral in
fig.~\FIGhexagon.

We use a Passarino-Veltman reduction~[\use\PV] to evaluate the
box integrals in terms of scalar boxes, triangles and bubbles.
First we evaluate the traces in~(\use\traceterms) using
$$
{\rm tr_{\pm}}( \Slash{a} \; \Slash{b} \; \Slash{c} \; \Slash{d} )
\ =\ 2 ( a\cdot b \; c\cdot d - a\cdot c\; b\cdot d+a\cdot d\; b\cdot c)
\mp \hf \varepsilon(a,b,c,d)\ ,
\anoneqn
$$
where
$\varepsilon(a,b,c,d)=
4 i \varepsilon_{\mu\nu\rho\sigma} a^{\mu}b^{\nu}c^{\rho}d^{\sigma}$
with $\varepsilon_{\mu\nu\rho\sigma}$ the totally antisymmetric tensor.
First consider the $\varepsilon$ terms.
The two momenta $\ell_1$ and $\ell_2$ are related to
each other by
$\ell_2=\ell_1-P-k_{m_1}- k_{m_2}$
with $P=\sum_{i=m_1+1}^{m_2-1} k_i$.
Since $\varepsilon(a, b , c , d )$ is antisymmetric the
$\varepsilon$
terms in the traces reduce to terms linear in $\ell_1$.
For example, the first term in eq.~(\use\traceterms), corresponding
to fig.~\FIGboxes a,\ has an $\varepsilon$ term
$$
\varepsilon( \ell_1, k_{m_1}, \ell_2 , k_{m_2}  )
=-\varepsilon( \ell_1, k_{m_1}, P ,  k_{m_2} ).
\anoneqn
$$
Since the box only contains only three independent momenta,
which can be taken to be $k_{m_1}$, $ k_{m_2}$ and $P$,
evaluation of the box momentum integral must give zero for the
$\varepsilon$ term.
The $\gamma_5$ contribution drops out of the remaining terms
in analogous fashion, and so we can replace $\tr_- \to \hf\tr$
in eq.(\traceterms).

Thus the trace for fig.~\FIGboxes a\ is
$$
-\hf\tr ( \lsl_1 \ksl_{m_1} \lsl_2 \ksl_{m_2} ) =
- 2 (\ell_1\cdot k_{m_1}) (\ell_2\cdot k_{m_2})
- 2 (\ell_1\cdot k_{m_2}) (\ell_2\cdot k_{m_1})
+ 2 (k_{m_1}\cdot k_{m_2})(\ell_1\cdot \ell_2)
\anoneqn
$$
which may be rewritten as
$$
\eqalign{
&\qquad\Bigl( 2 P\cdot k_{m_1} P\cdot k_{m_2} -P^2 k_{m_1}\cdot k_{m_2}
\Bigr)  \cr
&+\ (P+k_{m_1})\cdot k_{m_2}\ (\ell_1-k_{m_1})^2\
 +\ (P+k_{m_2})\cdot k_{m_1}\ (\ell_2+k_{m_2})^2 \cr
&+\ (\ell_1-k_{m_1} )^2\ (\ell_2+k_{m_2})^2\; . \cr}
\eqn\numeratorrewrite
$$
The first term in~(\use\numeratorrewrite) yields a scalar box integral
with the coefficient
$(2 P\cdot k_{m_1} P\cdot k_{m_2} -P^2 k_{m_1}\cdot k_{m_2})$,
the next two terms cancel propagators to give triangle integrals,
and the last term gives a bubble (two-point) integral.
Each of the four cut boxes in~(\use\traceterms) is reduced in this
way into a cut scalar box, two cut scalar triangles and a cut scalar
bubble.
However, when the four contributions are added together the
triangles and bubbles cancel leaving just the cut scalar boxes.
This cancellation is expected since, as discussed in
sections~\use\ReviewSection\ and \use\AmbiguitiesFixSection,
the $N=4$ supersymmetric result can be expressed as a sum of
scalar boxes with no triangles or bubbles.
Note that (up to a factor of 2),
the coefficient of the scalar box is precisely its denominator
(see eqs.~(\use\boxes):
$$
( 2 P\cdot k_{m_1} P\cdot k_{m_2} -P^2 k_{m_1}\cdot k_{m_2} )
= \Bigl({1\over 2}\Bigr)
\Bigl( (P+  k_{m_1})^2 ( P+k_{m_2})^2 -P^2 ( P+k_{m_1}+ k_{m_2} )^2 \Bigr)
\anoneqn
$$
so that the cuts will be given in terms of the scalar box functions
$F$ defined in appendix~\use\IntegralsAppendix.

Thus the cuts in the amplitude are given simply by the cuts in the
scalar box functions $\Feasy{r;i}$ (including the limiting case
$\Fone{i+2}$ for $r=1$) with the coefficients
$$
\eqalign{   \cg \, (\mu^2)^\e
& { A_{jk}^\treemhv(1,2,\ldots,n)}
\cr}
\anoneqn
$$
where $i=m_1+1$ and $r=m_2-m_1-1$.
The coefficients are precisely those given in eqs.~(\use\CoeffDefine),
thus confirming that the cuts in the ansatz (\UniversalFunc)
are correct.
Although suggestive, this agreement does not yet prove that the
ansatz is correct, because the cuts do not necessarily fix
possible polynomial terms.
In the following section we shall use $N=4$ supersymmetry
to show that no ambiguities are present.

For amplitudes other than the MHV amplitudes the evaluation of the
cuts is more involved.  Furthermore, all-$n$ tree formulas for general
non-MHV amplitudes are unknown.  For the $N=4$ supersymmetric non-MHV
amplitudes the ansatz~(\use\Ansatz), which expresses $A_{n;1}^{N=4\
{\rm loop}}$ as a product of the tree amplitude and the universal
function $\v_n^g$, still defines a function with perfectly well
behaved two-particle collinear limits.  However, explicit calculation
for the six-point amplitude,
$A_{6;1}^{N=4\ {\rm
loop}}(1^-,2^-,3^-,4^+,5^+,6^+)$,
by the string-based method shows that the
ansatz~(\Ansatz) does {\it not} hold for the general non-MHV case.  We
have also verified this by evaluating the cut in
the channel
$(k_1+k_2+k_3)^2$ for this amplitude.


\section{Fixing remaining ambiguities}
\tagsection\AmbiguitiesFixSection

The collinear limits do not necessarily provide a tight enough
constraint to allow us to prove the uniqueness of our ansatz;
unitarity, on the other hand, determines the cuts uniquely
--- and hence the dilogarithms and logarithms
--- but does not provide any
information about polynomial terms in the amplitude.  The $N=4$
supersymmetric case is however special: a knowledge of the cuts
completely determines the amplitude.  The set of functions which can
appear in the amplitude are all known since they arise from one-loop
integrals which are all known [\use\PV,\use\MVNV,\use\Integrals].
For the $N=4$ theory the set of functions is a restricted set.
As we will demonstrate, for this restricted set the cuts uniquely
determine the coefficients of all functions
that may appear in the amplitude, including polynomials.

The key to this result is the property discussed in
section~\use\ReviewSection, that
for the $N=4$ supersymmetric theory the loop-momentum
polynomials encountered have a degree that is four less
than the purely gluonic case, namely $m-4$ for an $m$-point integral.
Now, when calculating a general amplitude one may evaluate the
tensor integrals (integrals with nontrivial loop-momentum polynomials
in the numerator) by Passarino-Veltman reduction~[\use\PV]
to lower-point integrals.
Passarino-Veltman reduction takes an $m$-point
($m\ge 4$) tensor integral of degree $d$ ($d\ge 1$) and reduces it to a
sum of $m$- and $(m-1)$-point integrals of degree $d-1$.
Any scalar $m$-point integral
(an integral with a numerator independent of the loop momentum)
can also be reduced to a sum of scalar $(m-1)$-point integrals
for $m\ge 5$~[\use\MVNV].
In this way any one-loop $m$-point integral can be reduced to a
combination of tensor box integrals of degree up to $d+4-m$.
In a gauge theory the maximum degree of the polynomial in an
$m$-point integral is $m$, and when one iterates the reduction one
arrives at a combination of tensor box integrals of degree four.
(To evaluate the tensor boxes explicitly one often performs
another iteration to arrive at a combination of known
scalar box and tensor triangle integrals.)
Since in $N=4$ super-Yang-Mills the polynomials for the $m$-point
integrals have a maximum degree of $m-4$, this process will express the
$m$-point integrals in terms of a sum of {\it scalar\/} boxes.
Thus the full $N=4$ amplitude can be written simply
as a sum over scalar boxes
$$
A^{N=4} = \sum_i  c_i \, I^{\rm box }_i
\anoneqn
$$
without any triangles or bubbles.
(This is why the triangles and bubbles
cancelled in the previous calculation of
the cuts.)
The set of possible scalar box integrals, with massless internal lines,
but all possible combinations of external masses, is given explicitly
in appendix~\use\IntegralsAppendix.
The coefficients $c_i$ may be polynomial functions of the
momentum invariants and spinor helicity factors but may not contain
logarithms and dilogarithms.

Is this decomposition in terms of scalar boxes determined uniquely,
given the cuts?
For uniqueness to hold, the equation
$$
 \sum_i   c_i I^{\rm box }_i
= {\rm polynomial }
\eqn\LinInd
$$
must have only the trivial solution $c_i=0$.  The set of
integrals $I^{\rm box}_i$ include the cases where one, two, three, or
four legs may be off-shell (massive)
as depicted in fig.~\FIGboxtypes .  All these
integrals contain logarithms or dilogarithms, which produce
cuts that are independent of those produced by other integrals.
For example, consider the coefficient of the three-mass box
$I_4^{3{\rm m},r,r',i}$
(\use\boxes{$d$}) appearing in fig.~\FIGboxtypes.
This box integral is a function of five kinematic invariants;
two of the invariants, $\tn{r}{i}$ and $\tn{n-r-r'-1}{i+r+r'}$,
appear together only in this one box, in the term
$2\,\Li_2[ 1 -
(\tn{r}{i}\tn{n-r-r'-1}{i+r+r'})/(\tn{r+1}{i-1}\tn{r+r'}{i}) ]$.
Consider the cut in the $\tn{r}{i}$ channel.
The cut for this term is proportional to
$\ln(-\tn{r}{i})+\ln(-\tn{n-r-r'-1}{i+r+r'})
-\ln(-\tn{r+1}{i-1})-\ln(-\tn{r+r'}{i})$.
The $\ln(-\tn{n-r-r'-1}{i+r+r'})$ part of the cut can only arise from
this box, $I_4^{3{\rm m},r,r',i}$, and thus the coefficient
of this three mass box in eq.~(\LinInd) must vanish.
One can continue in this way to
show that eq.~(\LinInd) has only the trivial solution $c_i=0$ for
all coefficients.
Thus, for the $N=4$ supersymmetric case, the coefficients of the
scalar boxes in eq.~(\use\CoeffDefine) are uniquely determined
by the cuts, and we have proven that the ansatz is correct.

Since we have a supersymmetric theory, we can use
supersymmetry Ward identities to generate amplitudes with external
fermions and scalars from $n$-gluon amplitudes.
For supersymmetric MHV amplitudes the Ward
identity~(\use\scalarContrib)
holds for loop amplitudes as well as tree amplitudes,
and we obtain the amplitudes with two
external fermions trivially from the gluon amplitudes~(\use\Ansatz),
(\use\UniversalFunc), (\use\CoeffDefine).

In general, for theories other than $N=4$ super-Yang-Mills, the cuts
may not uniquely determine the full amplitude.  As a simple example,
the five-point
helicity amplitudes $A_{5;1}(1^-,2^+,\ldots,5^+)$ and
$A_{5;1}(1^+,2^+,\ldots,5^+)$ each have no cuts but are not equal.
One cannot reconstruct the full amplitude from the cuts in this case
because the amplitude is a more general sum of boxes, triangles,
and bubbles, including combinations without branch cuts.
In such cases the collinear limits provide restrictions
on the form of rational functions that may appear in the
amplitudes~[\use\AllPlus].


\section{Remaining partial amplitudes}
\tagsection\SubleadingSection

In this section we show that the one-loop
partial amplitudes relevant at subleading orders in $N_c$,
$A_{n;c>1}$, can be obtained by performing appropriate sums over
permutations of the leading-color partial amplitudes $A_{n;1}$.
The result obtained holds not just in $N=4$ supersymmetry,
but also for any one-loop gauge theory amplitude where the external
particles and the particles circulating around the loop are both in the
adjoint representation of $SU(N_c)$.
(In the case of e.g. a quark loop where the internal particle is
in the fundamental representation, there are no $A_{n;c>1}$
contributions.)
In string theory, the color
decomposition is manifest~[\use\Color,\use\Long]; it thus provides a natural
framework for discussing this result.
We shall also provide the outline
of a conventional field theory argument leading to
the same result.

As we shall explain, the coefficients of the subleading color structures
$$
\Gr_{n;c}(1) = \Tr\L T^{a_1}\cdots T^{a_{c-1}}\R\,
\Tr\L T^{a_{c}}\cdots T^{a_n}\R
\eqn\sublcolorstr
$$
are
$$
 A_{n;c}(1,2,\ldots,c-1;c,c+1,\ldots,n)\ =\
 (-1)^{c-1} \sum_{\sigma\in COP\{\alpha\}\{\beta\}} A_{n;1}(\sigma)
\eqn\sublanswer
$$
where $\alpha_i \in \{\alpha\} \equiv \{c-1,c-2,\ldots,2,1\}$,
$\beta_i \in \{\beta\} \equiv \{c,c+1,\ldots,n-1,n\}$,
and $COP\{\alpha\}\{\beta\}$ is the set of all
permutations of $\{1,2,\ldots,n\}$ with $n$ held fixed
that preserve the cyclic
ordering of the $\alpha_i$ within $\{\alpha\}$ and of the $\beta_i$
within $\{\beta\}$, while allowing for all possible relative orderings
of the $\alpha_i$ with respect to the $\beta_i$.
For example if $\{\alpha\} = \{2,1\}$ and
$\{\beta\} = \{3,4,5\}$, then $COP\{\alpha\}\{\beta\}$
contains the twelve elements
$$
\eqalign{
 &(2,1,3,4,5),\quad (2,3,1,4,5),\quad (2,3,4,1,5),\quad
  (3,2,1,4,5),\quad (3,2,4,1,5),\quad (3,4,2,1,5), \cr
 &(1,2,3,4,5),\quad (1,3,2,4,5),\quad (1,3,4,2,5),\quad
  (3,1,2,4,5),\quad (3,1,4,2,5),\quad (3,4,1,2,5) \cr}
\anoneqn
$$
(cyclic ordering for a two-element set is meaningless).
Note that the ordering of the first sets of indices is reversed
with respect to the second.

In an open string theory where the trace structures are just Chan-Paton
factors~[\use\ChanPaton] it is easy to see that a formula
like~(\use\sublanswer) should be expected.
Open string vertex operators corresponding to the two different traces
are attached to the two different boundaries of
the open string annulus, and the relative orderings of
operators from opposite boundaries are summed over in the world-sheet
path integral, while the ordering of operators on the same boundary
is preserved.  (Indeed, for the $N=4$ super-Yang-Mills result, it would
suffice to consider an open superstring in the infinite-tension limit,
as its trivial compactification
to four dimensions yields precisely this theory.)

More formally, string-rules for generating the subleading-in-$N_c$
partial amplitudes with external gluons are given in ref.~[\use\Long].
For proving eq.~(\use\sublcolorstr) we
require only a few salient features of the string rules.
The rules are in terms of diagrams with only three-point vertices.
The tree and loop
parts of diagrams are evaluated by a set of substitution rules on a
`master' kinematic expression described in the reference.
The right-hand-side of eq.~(\sublanswer) contains all the
leading-in-$N_c$ diagrams, with legs permuted over
$COP\{\alpha\}\{\beta\}$.
For the left-hand-side of eq.~(\sublanswer) the string-based rules for
subleading-in-$N_c$ amplitudes give exactly the same set of diagrams
except that two classes are explicitly excluded.  These two sets are:
\par\noindent
1) diagrams where indices from {\it both} sets
$\{\alpha\}$ and $\{\beta\}$ label leaves of the same tree
(attached to the loop), and
\par\noindent
2) diagrams where a single tree contains {\it all} elements of
either $\{\alpha\}$ or $\{\beta \}$.
\par\noindent
In fig.~\use\FIGSubleadFigureA\ two examples of
diagrams which are excluded from the left-hand-side of
eq.~(\sublanswer) are given.
If we can prove that the two classes of excluded
diagrams, when summed over the permutations in
$COP\{\alpha\}\{\beta\}$, have vanishing contribution, then
eq.~(\use\sublanswer) follows; this is not difficult to do
using the string-based rules.

For diagrams of the first type where the tree legs are labelled by
the indices from both sets $\{\alpha\}$ and $\{\beta\}$, the
diagrams can be arranged so that diagrams cancel pairwise.  This follows
from the anti-symmetry of the tree substitution rules under the
interchange of the ordering of two outer legs of a vertex.  (This
anti-symmetry is completely analogous to the anti-symmetry of the
color-decomposed three-gluon
self-interaction~[\use\ManganoReview,\use\Tasi] of field theory.)
For example, the pairs of diagrams in figs.~\FIGSubleadFigureA{a} and
\FIGSubleadFigureA{b} cancel
in the string based-rules.
Figure~\FIGSubleadFigureA{c} is an example of the pairwise cancellation
for diagrams of
the second type where a single tree contains all elements of
$\{\beta\}=\{\beta_1,\beta_2,\beta_3\}$.
For all trees in either class we
can similarly arrange the diagrams to cancel pairwise
when one sums over the permutations in $COP\{\alpha\}\{\beta\}$.
This completes the proof of eq.~(\use\sublanswer).

The corresponding analysis in field theory also uses a representation
of graphs in terms of trees attached to the loop.  Using the trace (or
double-line) representation~(\use\struct) of the structure constants
$f^{abc}$, and also eq.~(\Fierz), it is easy to see that the set of
all Feynman diagrams (in Feynman gauge) which have only three-point
vertices, and no non-trivial trees, feed into both $A_{n;1}$ and
$A_{n;c>1}$ in the correct way so that eq.~(\use\sublanswer) is
satisfied for this class of diagrams.  For diagrams containing
non-trivial trees that are attached to the loop, one again needs to
show (as in the string-based proof), that the permutation sum over
different orderings, $COP\{\alpha\}\{\beta\}$, cancels out
right-hand-side contributions to eq.~(\use\sublanswer) of the types~1
and~2 above, since these contributions can be seen to be absent from
the left-hand-side.  For diagrams with only three-point vertices, the
permutation sum drops out from the antisymmetry of the vertices as in
the string-based argument.  For diagrams with trees containing
four-point vertices, the cancellations occur in triplets such as
those shown
in fig.~\FIGSubleadFigureB .  Diagrams with
four-point vertices attached to
the loop
can be color decomposed into the same color structures encountered above.
In this way one can construct a purely field-theoretic proof of
eq.~(\use\sublanswer).

Instead of relying on an explicit representation
of the vertices to prove the legitimacy of omitting contributions
of type~1 and~2,
one can use ``$U(P)\times U(N_c-P)$ decoupling'' of tree amplitudes
(even when one leg is off-shell --- this is equivalent to the
additional decoupling properties of the Berends-Giele
current~[\Collinear,\BigTreesB]).
The sum over orderings in $COP\{\alpha\}\{\beta\}$, where
some $\alpha_i$ and some $\beta_i$ belong to the same tree,
amounts to computing the color-ordered tree amplitude for those
$\alpha_i$ belonging to (say) $U(P)$ for some $P$, and the $\beta_i$
belonging to $U(N_c-P)$; but this quantity vanishes.  The same
is true for the sum over all given orderings when all indices from
a given set label leaves on a single tree.
The field theory analysis really only relies on the properties of the
$U(N_c)$ structure constants $f^{abc}$, and so it applies to any amplitude
containing only such vertices, for example to $N=4$ or pure $N=1$
super-Yang-Mills amplitudes with an arbitrary number of external gluinos
as well as gluons.

In the $N=4$ supersymmetric case, the special form of
$A^{\rm MHV}_{n;1}$ in
equations~(\use\Ansatz) and (\use\CoeffDefine) allows
us to simplify the expression for $A_{n;c}$ by explicitly carrying out
the permutation sums in eq.~(\use\sublanswer).
The computation
and results
are given in appendix~\use\SubleadingAppendix .

\section{Conclusions}
\tagsection\ConclusionsSection

Although they are important to the analysis of experimental jet data,
few one-loop QCD amplitudes have been calculated. Only four-
[\use\Ellis,\use\Long,\use\KST] and five-point [\use\FiveGluon]
amplitudes relevant for next-to-leading order corrections are known,
the latter already made possible only by the development of new
techniques.  In this paper we have introduced a technique, based on
unitarity and collinear limits, which allows one to compute amplitudes
without performing explicit diagrammatic calculations. Unitarity, in
the form of the Cutkosky rules, fixes the form of the cuts in the
amplitudes without ambiguity but imposes no direct constraints on the
polynomials in the kinematic variables.  The constraints imposed by
the collinear limits allow one to guess extrapolations of known
results to higher-point functions, and in particular constrain the
form of rational functions (lacking cuts) of the invariants and spinor
products if present. We presented all one-loop splitting
amplitudes for external gluons and fermions required for the collinear
bootstrap.  Collinear singularities are also very useful in checking
results obtained by other means.

We have applied this technique to produce an all-$n$ formula for the
non-vanishing maximally helicity-violating amplitudes in an $N=4$
supersymmetric theory.  We fixed the remaining ambiguity in this
amplitude by noting that the set of integrals that would appear in a
string-based or superspace calculation is only a subset of the usual
set of tensor integrals.  Within this restricted set of integrals, the
cuts uniquely determine the amplitudes, thereby proving that our
amplitude is the unique solution to the constraints imposed by
unitarity.

For the $N=4$ supersymmetry case that we have presented here the
collinear limits are not actually needed.  In contrast, for the
all-plus helicity amplitudes the cuts are trivial (they all vanish)
and it is the collinear limits that allow one to give an ansatz for
the amplitude~[\use\BDKconf,\use\AllPlus], which has subsequently been
proven correct [\use\MahlonB,\use\DP].  In general, the restrictions
imposed by collinear behavior and unitarity complement each other.

In the string based-method, it is convenient to organize QCD
amplitudes with external gluons into contributions which correspond to
an $N=4$ supersymmetric piece, an $N=1$ chiral piece, and a scalar
piece~[\use\FiveGluon,\use\Tasi,\use\Weak].
With this type of organization
the $N=4$ supersymmetric amplitude is one of three pieces needed
for the QCD loop amplitude.  The other pieces are also amenable
to the methods described here and will be discussed elsewhere.

Using a ``unitary-collinear bootstrap'' we have thus constructed
a class of one-loop amplitudes with an arbitrary number of
external gluons.  These amplitudes are one-loop analogs of the
Parke-Taylor tree amplitudes.
We expect that this method can be used
to generate further fixed-$n$ and all-$n$ amplitudes, while
bypassing the algebraic barrier usually present in explicit computations.

\vskip0.3in
\par\noindent
{\bf Acknowledgements}
\vskip0.1in

We thank G. Chalmers for useful discussions regarding the uniqueness
of the collinear limits and Z. Kunszt for useful discussions regarding
splitting amplitudes.  We are grateful for the support
of NATO Collaborative Research Grants CRG--921322 (L.D. and D.A.K.)
and CRG--910285 (Z.B. and D.C.D).

\vfill\eject


\appendix{Scalar Box Integrals}
\tagappendix\IntegralsAppendix

The scalar box integrals that can arise in principle in the $N=4$
super-Yang-Mills computation (or QCD computation) have vanishing
internal masses, but may have one, two, three or four nonvanishing
external masses, and there are two types of two-mass boxes.
These integrals are defined and given in ref.~[\use\Integrals]
(the four-mass boxes are from ref.~[\use\FourMassBox])
and are shown in fig.~\FIGboxtypes.

The scalar box integral is
$$
I_4 = -i \L4\pi\R^{2-\e} \,\int {d^{4-2\e}p\over \L2\pi\R^{4-2\e}}
\;{1\over p^2 \L p-K_1\R^2 \L p-K_1-K_2\R^2 \L p+K_4\R^2}\;.
\anoneqn$$
It is convenient to define the scalar box function
$$
F(K_1,K_2,K_3,K_4) = -{2\sqrt{\mathop{\rm det} S}\over\rg} \,I_4
\eqn\GeneralFdefn$$
where
$$
\rg =  {\Gamma(1+\e)\Gamma^2(1-\e)\over\Gamma(1-2\e)}
\anoneqn$$
and where the symmetric $4\times4$
matrix $S$ has components ($i$, $j$ are $\mod 4$)
$$
S_{ij} = -{1\over2}\L K_i+\cdots+ K_{j-1}\R^2, \quad i\neq j;
\hskip 1truecm S_{ii} = 0.
\anoneqn$$
The external momentum arguments $K_{1\ldots4}$ in equation~(\use\GeneralFdefn)
are sums
of external momenta $k_i$ that are the arguments of
the $n$-point amplitude.

With the labelling of legs shown in fig.~\FIGboxtypes\
(that is re-expressing the functions in terms of the invariants
$\tn{r}i$ of the $n$-point amplitude),
the scalar box function $F$ expanded through order ${\cal O}(\e^0)$
for the different cases reduces to
\defeqn\Fboxes
$$
\eqalign{
  \Fone{i}
&=\ -{1\over\e^2} \Bigl[ (-\tn{2}{i-3})^{-\e} +
(-\tn{2}{i-2} )^{-\e} - (-\tn{n-3}{i})^{-\e} \Bigr] \cr
 &\ + \Li_2\left(1-{\tn{n-3}{i} \over \tn{2}{i-3}}\right)
  \ + \ \Li_2\left(1-{\tn{n-3}{i} \over \tn{2}{i-2}}\right)
  \ +{1\over 2} \ln^2\left({ \tn{2}{i-3} \over \tn{2}{i-2}}\right)\
+\ {\pi^2\over6}\ ,
\cr}
\eqno({\rm\Fboxes{a}})
$$
$$
\eqalign{
  \Feasy{r;i}
&=\  - {1\over\e^2} \Bigl[ (-\tn{r+1}{i-1})^{-\e} + (-\tn{r+1}{i})^{-\e}
              - (-\tn{r}{i} )^{-\e} - (-\tn{n-r-2}{i+r+1} )^{-\e} \Bigr] \cr
&\ +\ \Li_2\left(1-{\tn{r}{i} \over \tn{r+1}{i-1} }\right)
 \ +\ \Li_2\left(1-{\tn{r}{i} \over \tn{r+1}{i}}\right)
 \ +\ \Li_2\left(1-{\tn{n-r-2}{i+r+1} \over \tn{r+1}{i-1} }\right)
\cr
&\
 \ +\ \Li_2\left(1-{\tn{n-r-2}{i+r+1} \over \tn{r+1}{i}}\right)
-\ \Li_2\left(1-{\tn{r}{i} \tn{n-r-2}{i+r+1}
\over \tn{r+1}{i-1} \tn{r+1}{i}}\right)
   \ +\ {1\over 2} \ln^2\left({\tn{r+1}{i-1} \over \tn{r+1}{i}}\right)\ , \cr }
\eqno({\rm\Fboxes{b}})
$$
$$
\eqalign{ \hskip -1.4 truecm
  \Fhard{r;i}
&=\ -{1\over\e^2} \Bigl[ (- \tn{2}{i-2})^{-\e} + (-\tn{r+1}{i-1})^{-\e}
              - (-\tn{r}{i} )^{-\e} - (-\tn{n-r-2}{i+r})^{-\e} \Bigr]
\cr &
  \ -\ {1\over2\e^2}
    { (-\tn{r}{i})^{-\e}(-\tn{n-r-2}{i+r})^{-\e}
     \over (- \tn{2}{i-2})^{-\e} }
  \ +\ {1\over 2} \ln^2\left({ \tn{2}{i-2}\over \tn{r+1}{i-1} }\right)
\cr &
  \ +\ \Li_2\left(1-{ \tn{r}{i} \over \tn{r+1}{i-1}}\right)
  \ +\ \Li_2\left(1-{\tn{n-r-2}{i+r}\over \tn{r+1}{i-1} }\right)
  \ , \cr}
\eqno({\rm\Fboxes{c}})
$$
$$
\eqalign{ \hskip -1.4 truecm
  \Fthree{r,r';i}
&=\ -{1\over\e^2} \Bigl[ (-\tn{r+1}{i-1} )^{-\e} + (-\tn{r+r'}{i})^{-\e}
     - (-\tn{r}{i})^{-\e}
     - (-\tn{r'}{i+r} )^{-\e}
     - (-\tn{n-r-r'-1}{i+r+r'} )^{-\e} \Bigr] \cr
  &  \ -\ {1\over2\e^2}
   { (-\tn{r}{i})^{-\e}(-\tn{r'}{i+r} )^{-\e} \over(-\tn{r+r'}{i})^{-\e} }
  \ -\ {1\over2\e^2}
    {(-\tn{r'}{i+r})^{-\e}(-\tn{n-r-r'-1}{i+r+r'})^{-\e}
           \over (-\tn{r+1}{i-1})^{-\e} }
      \ +\ {1\over2}\ln^2\left({\tn{r+1}{i-1} \over \tn{r+r'}i}\right)
\cr
  &\ +\ \Li_2\left(1-{\tn{r}{i}\over \tn{r+1}{i-1} }\right)
   \ +\ \Li_2\left(1-{\tn{n-r-r'-1}{i+r+r'} \over \tn{r+r'}i}\right)
  \ -\  \Li_2
\left(1-{\tn{r}{i} \tn{n-r-r'-1}{i+r+r'}\over \tn{r+1}{i-1}\tn{r+r'}i }\right)
\ ,  \cr }
\eqno({\rm\Fboxes{d}})
$$
$$
\eqalign{
 \Ffour{r, r', r'';i} = &
{1\over 2} \biggl(  \Li_2\left(\hf(1-\lambda_1+\lambda_2+\rho)\right)
  \ +\ \Li_2\left(\hf(1-\lambda_1+\lambda_2-\rho)\right) \cr
 &\ +\ \Li_2\left(
   \textstyle-{1\over2\lambda_1}(1-\lambda_1-\lambda_2-\rho)\right)
 \ +\ \Li_2\left(\textstyle-{1\over2\lambda_1}(1-\lambda_1-
    \lambda_2+\rho)\right) \cr
  &\ -\ {1\over2}\ln\left({\lambda_1\over\lambda_2^2}\right)
   \ln\left({ 1+\lambda_1-\lambda_2+\rho \over 1+\lambda_1
        -\lambda_2-\rho }\right) \biggr)\ ,
   \cr}
\eqno({\rm\Fboxes{e}})
$$
where
$$
 \rho\ \equiv\ \sqrt{1 - 2\lambda_1 - 2\lambda_2
+ \lambda_1^2 - 2\lambda_1\lambda_2 + \lambda_2^2}\ ,
\eqn\rdefinition
$$
and
$$
\lambda_1 = {t_i^{[r]} \; t_{i+r +r'}^{[r'']} \over t_i^{[r+ r']} \;
t_{i+r}^{[r'+r'']} } \; , \hskip 1.5 cm
\lambda_2 = {t_{i+r}^{[r']}\; t_{i+r+r'+r''}^{[n-r-r'-r'']} \over
t_i^{[r+ r']}\; t_{i+r}^{[r'+r'']} }\ .
\anoneqn
$$

In terms of these variables, the relations between the scalar box
functions and scalar box integrals are given by
\defeqn\boxes
$$
\eqalignno{
  I_{4:i}^{1{\rm m}} &=\ -2 \rg {\Fone{i} \over \tn{2}{i-3} \tn{2}{i-2} }
        \, , \hskip 4 cm & ({\rm\boxes{a}}) \cr
 I_{4:r;i}^{2{\rm m}e}
&=\ -2 \rg {\Feasy{r;i}
      \over \tn{r+1}{i-1}\tn{r+1}{i} -\tn{r}{i}\tn{n-r-2}{i+r+1} }\,, &
     ({\rm\boxes{b}}) \cr
  I_{4:r;i}^{2{\rm m}h}
&=\ -2 \rg {\Fhard{r;i} \over \tn{2}{i-2} \tn{r+1}{i-1} } \,, &
 ({\rm\boxes{c}}) \cr
  I_{4:r,r',i}^{3{\rm m}}
&=\ -2 \rg {\Fthree{r,r';i}
     \over \tn{r+1}{i-1} \tn{r+r'}i -\tn{r}{i} \tn{n-r-r'-1}{i+r+r'} }\,,
 & ({\rm\boxes{d}}) \cr
I_{4: r, r', r'', i}^{4{\rm m}} & =
-2 {\Ffour{r, r', r'';i}\over t_i^{[r+ r']}\; t_{i+r}^{[r'+r'']}\;\rho}\, .
 & ({\rm\boxes{e}})   \cr}
$$

We also record here the limits of the functions appearing in
$V^g_n$ as two
adjacent external momenta, say $k_c$ and $k_{c+1}$, become collinear.
We denote the momentum fraction within the fused leg by $z$
($k_c = z k_P$ and $k_{c+1}=(1-z) k_P$), replace $P\to c$,
and shift the labels of legs $c+2,\ldots,n$
down by one. (The indices on the right-hand sides of the following
equations are to be understood $\mod\ (n-1)$ rather than $n$.)
If both $k_c$ and $k_{c+1}$ are part of the sum forming one of the external
masses, say the one labelled $r$ in fig.~7, and $r>2$, then the $F$ functions
simply reduce to the corresponding ones with $n\rightarrow n-1$,
$$\eqalign{
\Fone{i} &\rightarrow {\def\Fn{n-1} \Fone{i}}\;,\cr
\Feasy{r;i} &\rightarrow {\def\Fn{n-1} \Feasy{r-1;i}}\;;\cr
}\anoneqn$$
the behavior is analogous in the case that both external momenta
are part of the sum forming a different external mass, so long
as three or more external momenta make up that mass.
In the case that $r=2$, the two-mass box $\Feasy{r;c}$ reduces to
a one-mass box,
$$
\Feasy{r;c} \rightarrow
{\def\Fn{n-1} \Fone{c+2}} +{1\over\e^2} \L-\tn2c\R^{-\e} \;.
\anoneqn$$

Using the collinear limit
$$
\tn{r+1}{c+1} \rightarrow z \tn{r}{c+1} + (1-z)\tn{r+1}{c}\;,
\anoneqn$$
and Abel's identity [\use\Lewin], one can show that
$$\eqalign{
\Feasy{r;c+1} + \Feasy{r-1;c+2} &\rightarrow {\def\Fn{n-1} \Feasy{r-1;c+1}}\;,
  \cr
\Feasy{r;c-r+1} + \Feasy{r-1;c-r+1} &\rightarrow
    {\def\Fn{n-1} \Feasy{r-1;c-r+1}}\;.  \cr
}\anoneqn$$

For the one-mass box function, there are four additional cases to
consider, corresponding to $c=i-4,\ldots,i-1$.  The latter two are
equivalent to the former two under a reflection; the first combines
with an $\Feasy{2;c+1}$ as follows,
$$
\Feasy{2;c+1} + \Fone{c+4} \rightarrow {\def\Fn{n-1} \Fone{c+3}}
\anoneqn$$
while the second reduces to
$$\Fone{c+3} \rightarrow
-{1\over\e^2}\L -(1-z) \tn{2}{c}\R^{-\e}-\Li_2(z)\;.
\anoneqn$$
Combining the limits of
$$
\Feasy{2;c}+ \Fone{c+3} + \Fone{c+2}
\anoneqn$$
yields $(\mu^2)^{-\e}r_S^{\rm SUSY}(z,\tn2c)$ in the form also obtained from
gluino amplitudes in the following appendix,
$$\eqalign{
\Feasy{2;c}+ &\Fone{c+3} + \Fone{c+2}
\longrightarrow
{\def\Fn{n-1} \Fone{c+2}}\cr &+{1\over\e^2} \L-\tn2c\R^{-\e}
-{1\over\e^2}\L -(1-z) \tn{2}{c}\R^{-\e}-\Li_2(z)
-{1\over\e^2}\L -z \tn{2}{c}\R^{-\e}-\Li_2(1-z)\;.\cr
}\anoneqn$$


\appendix{Loop Splitting Amplitudes}
\tagappendix\LoopSplittingAppendix

In this appendix we collect the various splitting amplitudes which are
useful for bootstrapping higher-point one-loop amplitudes from known
amplitudes.  The one-loop splitting amplitudes are obtained by taking
the collinear limit of known five parton
amplitudes~[\use\FiveGluon,\use\GieleGlover,\use\Fermion].
The universality of the ($g\to gg$) splitting amplitudes for arbitrary
numbers of legs has been shown for scalar
contributions~[\use\AllPlus],
but it is likely to be true in general.
All known one-loop amplitudes satisfy eq.~(\use\loopsplit),
including amplitudes with external fermions.

In discussing gauge theory amplitudes with external fermions,
we distinguish two cases: external fermions in the adjoint
representation (gluinos, $\tilde{g}$),
and external fermions in the fundamental
$N_c$ and $\bar{N_c}$ representations
(quarks, $q$, and antiquarks, $\bar{q}$).
The color decomposition of scattering amplitudes with external
gluons and gluinos are identical to the $n$-gluon color decompositions
(\use\TreeAmplitudeDecomposition) and (\use\ColorDecomposition).
The tree and loop splitting amplitudes for $g\to \tilde{g}\tilde{g}$
and $\tilde{g}\to\tilde{g}g$ may therefore be defined
via the same equations~(\use\treesplit), (\use\loopsplit) used
to define $g\to gg$.

The color decomposition of amplitudes with external quarks as well as
gluons is somewhat different, but the collinear behavior
(\use\treesplit), (\use\loopsplit) again holds for the tree
partial amplitudes $A_n$, and is expected to hold for the
leading-in-$N_c$ one-loop partial amplitudes $A_{n;1}$, with
an appropriate definition of $A_n$ and $A_{n;1}$.
(For amplitudes with four or more external quarks, one must restrict to
the leading-in-$N_c$ contributions even at tree level, in order
to obtain the simple color-adjacent collinear
behavior~(\use\treesplit).)
For example, tree amplitudes with two external quarks and $n-2$ gluons
have the decomposition
$$
 {\cal A}_n^\tree(1_{\bar{q}},2_q,3,\ldots,n)
 \ =\ \sum_{\sigma\in S_{n-2}}
   (T^{a_{\sigma(3)}}\ldots T^{a_{\sigma(n)}})_{i_2}^{~\bar i_1}\
    A_n^\tree(1_{\bar{q}},2_q;\sigma(3),\ldots,\sigma(n)).
\eqn\gentreeqqdecomp
$$
The $A_n$ obey~(\use\treesplit) for all four possibilities:
$g\to gg$, $q\to qg$, $\bar{q}\to g\bar{q}$ and $g\to \bar{q}q$.
The $g\to \bar{q}q$ collinear limit produces an $(n-1)$-gluon
partial amplitude on the right-hand-side of~(\use\treesplit),
whereas the remaining limits produce
two-quark-$(n-3)$-gluon partial amplitudes.
The one-loop amplitudes can be decomposed as follows,
$$
 {\cal A}_n^\loop(1_{\bar{q}},2_q,3,\ldots,n)
 \ =\ \sum_{j=1}^{n-1} \sum_{\sigma\in S_{n-2}/S_{n;j}}
    \Gr_{n;j}^{(\bar{q}q)}(\sigma(3,\ldots,n))\
  A_{n;j}(1_{\bar{q}},2_q;\sigma(3,\ldots,n))\ ,
\eqn\genqqdecomp
$$
where
$$
\eqalign{
 \Gr_{n;1}^{(\bar{q}q)}(3,\ldots,n)
  \ &=\ N_c\ (T^{a_3}\ldots T^{a_n})_{i_2}^{~\bar i_1}\ ,\cr
 \Gr_{n;2}^{(\bar{q}q)}(3;4,\ldots,n)
  \ &=\ 0\ ,\cr
 \Gr_{n;j}^{(\bar{q}q)}(3,\ldots,j+1;j+2,\ldots,n)
 \ &=\ \Tr(T^{a_3}\ldots T^{a_{j+1}})\ \
   (T^{a_{j+2}}\ldots T^{a_n})_{i_2}^{~\bar i_1}\ ,
  \quad j=3,\ldots,n-2, \cr
 \Gr_{n;n-1}^{(\bar{q}q)}(3,\ldots,n)\ &=\ \Tr(T^{a_3}\ldots T^{a_n})\ \
     \delta_{i_2}^{\bar i_1} \ ,\cr}
\eqn\grqq
$$
and $S_{n;j}$ is the subset of the permutation group $S_{n-2}$ that
leaves $\Gr_{n;j}^{(\bar{q}q)}$ invariant.
The color-ordered partial amplitudes $A_{n;1}$ give the
leading-in-$N_c$ one-loop contribution to the color-summed
cross-section, and for $n=5$ ($\bar{q}qggg$)
the $A_{n;1}$ have the expected one-loop collinear
behavior~(\use\loopsplit) in all channels~[\use\Fermion],
with the splitting amplitudes given below.

Before presenting the explicit loop splitting amplitudes,
we first review the tree-level splitting amplitudes, since they enter
into the collinear behavior of loop amplitudes as well.
Also, most of the one-loop splitting amplitudes are
proportional to the tree-level ones.

The splitting amplitudes for the process $g\to gg$, when the gluon
momenta $k_a$ and $k_b$ become collinear
are~[\use\ParkeTaylor,\use\ManganoParke,\use\RecursiveA,\use\ManganoReview]
$$
\eqalign{
\Split^{\rm tree}_{-}(a^{-},b^{-})\ &=\ 0,\cr
\Split^{\rm tree}_{-}(a^{+},b^{+})
           \ &=\ {1\over \sqrt{z (1-z)}\spa{a}.b},\cr
\Split^{\rm tree}_{-}(a^{+},b^{-})
          \ &=\ -{z^2\over \sqrt{z (1-z)}\spb{a}.b},\cr
\Split^{\rm tree}_{-}(a^{-},b^{+})
          \ &=\ -{(1-z)^2\over \sqrt{z (1-z)}\spb{a}.b},\cr}
\eqn\gggtree
$$
where $k_a = z P$ and $k_b = (1-z) P$ with $P = (k_a + k_b)$.
The remaining
$\Split^{\rm tree}_{+}$ can be obtained from these by parity.

The $g \to \bar{q}q$ splitting amplitudes are
$$
\eqalign{
\Split^{\rm tree}_{+}(\bar{q}^{+},q^{-})
          \ &=\ {z^{1/2}(1-z)^{3/2}
          \over \sqrt{z(1-z)}\spa{\bar{q}}.q},\cr
\Split^{\rm tree}_{+}(\bar{q}^{-},q^{+})
          \ &=\ -{z^{3/2}(1-z)^{1/2}
          \over \sqrt{z(1-z)}\spa{\bar{q}}.q},\cr
\Split^{\rm tree}_{-}(\bar{q}^{+},q^{-})
          \ &=\ {z^{3/2}(1-z)^{1/2}
          \over \sqrt{z(1-z)}\spb{\bar{q}}.q},\cr
\Split^{\rm tree}_{-}(\bar{q}^{-},q^{+})
          \ &=\ -{z^{1/2}(1-z)^{3/2}
          \over \sqrt{z(1-z)}\spb{\bar{q}}.q},\cr}
\eqn\barqqgtree
$$
and the $q \to qg$ and $\bar{q} \to g\bar{q}$ splitting amplitudes are
$$
\eqalign{
\Split^{\rm tree}_{-}(q^{+},a^{+})
          \ &=\ {z^{1/2}\over \sqrt{z(1-z)}\spa{q}.a},\cr
\Split^{\rm tree}_{-}(q^{+},a^{-})
          \ &=\ -{z^{3/2}\over \sqrt{z(1-z)}\spb{q}.a},\cr
\Split^{\rm tree}_{-}(a^{+},\bar{q}^{+})
          \ &=\ {(1-z)^{1/2}\over \sqrt{z(1-z)}\spa{a}.{\bar{q}}},\cr
\Split^{\rm tree}_{-}(a^{-},\bar{q}^{+})
          \ &=\ -{(1-z)^{3/2}\over \sqrt{z(1-z)}\spb{a}.{\bar{q}}}.\cr}
\eqn\qgqtree
$$
Again the remaining ones can be obtained by parity.
The tree-level splitting amplitudes with gluinos are identical to
those with quarks; simply replace $q\to\tilde{g}$,
$\bar{q}\to\tilde{g}$ in the above expressions.

The loop splitting functions have a structure similar to the
tree splitting amplitudes, so it is useful to express them in terms
of a proportionality constant $r_S$ defined by
$$
 \Split^{\rm loop}_{-\lambda}(a^{\lambda_a},b^{\lambda_b})
 \ =\ \cg
 \times \Split^{\rm \tree}_{-\lambda}(a^{\lambda_a},b^{\lambda_b})
 \times r_S(-\lambda,a^{\lambda_a},b^{\lambda_b})
\eqn\genrsdef
$$
for general partons $a$ and $b$.
The only exception to eq.~(\use\genrsdef) is for $g^-\to g^+g^+$
(and its parity conjugate $g^+\to g^-g^-$), where $\Split^{\rm tree}$
vanishes but $\Split^{\rm loop}$ does not.
In general $r_S(-\lambda,a^{\lambda_a},b^{\lambda_b})$ depends
on the parton helicities, although in a supersymmetric theory
it turns out to be helicity-independent.

The loop splitting amplitudes also depend on the particles circulating in
the loop.  The contribution to the $g\to gg$ loop splitting amplitudes,
$\Split^{\rm loop}(a,b)$, for an adjoint spin-$J$ particle (with two
helicity states) are denoted by $\Split^{[J]}$, and the corresponding
proportionality constant by $r_S^{[J]}$.
The splitting amplitudes with internal particles in the fundamental
representation are the same but multiplied by $1/N_c$.
The splitting amplitudes we present describe the collinear behavior
before subtraction of the ultraviolet pole, that is
of unrenormalized amplitudes.
These are slightly simpler than the corresponding ones for physical
(`renormalized') amplitudes, but it is easy to convert from the
former to the latter; in the $\overline{MS}$ subtraction scheme one simply adds
to $r_S$ the helicity-independent term
$-{1\over2\e}\hat\beta_0$, where
$\hat\beta_0={11\over3}-{2\over3}{n_f\over N_c}$ in QCD with $n_f$ quark
flavors.

The $g \to gg$ splitting amplitudes may be directly obtained from
the four- [\use\Long,\use\KST] and five-point [\use\FiveGluon] helicity
amplitudes.
The $\Split^{[J]}_{+}(a^{+},b^{+})$ obey the supersymmetry relation
$\Split^{[1]} = -\Split^{[1/2]} = \Split^{[0]}$, where
$$
\Split^{[1]}_{+}(a^{+},b^{+})\ =\
-{1\over 48\pi^2}
  \sqrt{z(1-z)}\ {\spb{a}.b  \over {\spa{a}.b}^2}\ .
\eqn\finitegggloop
$$
We present the remaining $g\to gg$ loop splitting amplitudes
in terms of $r_S$:
$$\eqalign{
r_S^{[1]}(-,a^{+},b^{+})\ &=\
 - {1\over\e^2}\L{\mu^2\over z(1-z)(-s_{ab})}\R^{\e}
 + 2 \ln z\,\ln(1-z) + {1\over3} z (1-z) - {\pi^2\over6}\ , \cr
r_S^{[1/2]}(-,a^{+},b^{+})\ &=\ -{1\over 3} z (1-z)\ ,\cr
r_S^{[0]}(-,a^{+},b^{+})\ &=\ +{1\over 3} z (1-z)\ ,\cr
r_S^{[1]}(+,a^\pm,b^\mp)\ &=\
 - {1\over\e^2}\L{\mu^2\over z(1-z)(-s_{ab})}\R^{\e}
 + 2 \ln z\,\ln(1-z) - {\pi^2\over6}\ ,\cr
r_S^{[1/2]}(+,a^\pm,b^\mp)\ &=\ 0\ ,\cr
r_S^{[0]}(+,a^\pm,b^\mp)\ &=\ 0\ .\cr}
\eqn\gggrenorm
$$

One can extract the loop-level $q \to qg$ and $\bar{q}\to \bar{q}g$
splitting amplitudes from two independent sources:
Giele and Glover's expressions for the one-loop
$(\gamma^*,Z)\to q\bar{q}$ and $(\gamma^*,Z)\to qg\bar{q}$
helicity amplitudes [\use\GieleGlover],
and a calculation of $\bar{q}q \to ggg$ at one-loop [\use\Fermion].
Both methods agree, and we find
$$\eqalign{
r_S(q^{-},a^{+})\ &=\ r_S(q^{+},a^{-})
  \ =\ f(1-z,s_{qa})\ ,\cr
r_S(q^{-},a^{-})\ &=\ r_S(q^{+},a^{+})
  \ =\  f(1-z,s_{qa}) + \left(1+{1\over N^2}\right) {1-z\over2}\ ,\cr
r_S(a^{+},\bar{q}^{+})\ &=\ r_S(a^{-},\bar{q}^{-})
  \ =\ f(z,s_{a\bar{q}}) + \left(1+{1\over N^2}\right) {z\over2}\ ,\cr
r_S(a^{-},\bar{q}^{+})\ &=\ r_S(a^{+},\bar{q}^{-})
  \ =\ f(z,s_{a\bar{q}})\ ,\cr}
\eqn\qgqrenorm
$$
where the function $f(z,s)$ is
$$
\eqalign{
  f(z,s)\ &=\ - {1\over\e^2}\L{\mu^2\over z(-s)}\R^{\e} - \Li_2(1-z) \cr
  &\quad  -{1\over N_c^2} \left[
     - {1\over\e^2}\L{\mu^2\over(1-z)(-s)}\R^{\e}
     + {1\over\e^2}\L{\mu^2\over(-s)}\R^{\e} - \Li_2(z) \right]\ . \cr}
\eqn\fdef
$$

We have also extracted the factor $r_S^{[J]}$ for the
loop-level $g \to \bar{q}q$ splitting amplitudes from the one-loop
$\bar{q}q \to ggg$ helicity amplitudes [\use\Fermion].
By symmetry, $r_S^{[J]}$ is the same here for every helicity
configuration.  The results are:
$$\eqalign{
r_S^{[1]}(\pm,\bar{q}^{\mp},q^{\pm})\ &=\
  -{1\over\e^2} \LB \L{\mu^2\over z(-s)}\R^{\e}
                  + \L{\mu^2\over (1-z)(-s)}\R^{\e}
                  - 2\L{\mu^2\over (-s)}\R^{\e} \RB \cr
&\qquad
  + {13\over6\e} \L{\mu^2\over (-s)}\R^{\e}
  + \ln(z)\,\ln(1-z) - {\pi^2\over6} + {83\over18}
                                     - {\delta_R\over6} \cr
&\qquad   -{1\over N_c^2} \LB
    -{1\over\e^2} \L{\mu^2\over (-s)}\R^{\e}
    -{3\over2\e} \L{\mu^2\over (-s)}\R^{\e} - {7\over2}
                                     - {\delta_R\over2} \RB\ ,\cr
r_S^{[1/2]}(\pm,\bar{q}^{\mp},q^{\pm})\ &=\
   -{2\over3\e} \L{\mu^2 \over -s_{\bar{q}q}}\R^\e - {10\over9}\ ,\cr
r_S^{[0]}(\pm,\bar{q}^{\mp},q^{\pm})\ &=\
   -{1\over3\e} \L{\mu^2 \over -s_{\bar{q}q}}\R^\e - {8\over9}\ .\cr}
\eqn\qqgrenorm
$$
Here $\delta_R$ is a parameter controlling the variant of dimensional
regularization used.  In a supersymmetric scheme such as dimensional
reduction~[\use\Siegel,\use\KST] or four-dimensional
helicity~[\use\Long] with 2 physical gluon helicity states,
$\delta_R=0$;
in a ``conventional'' scheme~[\use\Ellis]
with $2-2\e$ physical gluon helicity states, $\delta_R=1$.

To convert the external quark results~(\use\qgqrenorm), (\use\fdef) and
(\use\qqgrenorm) into external gluino results, one must correct
for the different $SU(N_c)$ group theory factors; however,
this simply amounts to replacing $1/N_c^2 \to -1$ in the expressions.
Making these replacements, and setting $\delta_R=0$,
we find that in $N=1$ super-Yang-Mills theory,
for every possible helicity configuration and every choice of
adjoint external states
($g\to gg$, $\tilde{g}\to\tilde{g}g$ or $g\to\tilde{g}\tilde{g}$),
the proportionality constant $r_S$ is given by
$$
\eqalign{
  r_S^{\rm SUSY}(z,s)
  \ &=\ - {1\over\e^2}\L{\mu^2\over z(-s)}\R^{\e}
        - {1\over\e^2}\L{\mu^2\over(1-z)(-s)}\R^{\e}
        + {1\over\e^2}\L{\mu^2\over(-s)}\R^{\e}
        - \Li_2(1-z) - \Li_2(z) \cr
  \ &=\ - {1\over\e^2}\L{\mu^2\over z(1-z)(-s)}\R^{\e}
        + \ln z\,\ln(1-z) - \Li_2(1-z) - \Li_2(z) \cr
  \ &=\ - {1\over\e^2}\L{\mu^2\over z(1-z)(-s)}\R^{\e}
        + 2 \ln z\,\ln(1-z) - {\pi^2\over6}\ .\cr}
\eqn\rSsusy
$$
The independence of $r_S^{\rm SUSY}$
from the external states is consistent with the supersymmetry Ward
identities~[\use\Susy],
when the regulator is consistent with supersymmetry ($\delta_R=0$)
[\use\Siegel,\use\Long,\use\Tasi,\use\KST].

Due to supersymmetry, the same result~(\use\rSsusy) should also hold
in $N=4$ super-Yang-Mills theory.
This is easiest to verify directly for the $g\to gg$ splitting amplitudes,
because the difference between the $N=4$ and $N=1$ contributions
to $n$-gluon amplitudes is just the contribution of three chiral
multiplets (fermions plus scalars).  From the supersymmetry relation
for $\Split^{[J]}_{+}(a^+,b^+)$, and equation~(\use\gggrenorm)
for $r_S^{[J]}$, we see that the chiral multiplet contribution
($[1/2]+[0]$) to the loop splitting amplitudes vanishes for every
helicity configuration.

The $\tilde{g}\to\tilde{g}g$ and $g\to\tilde{g}\tilde{g}$
splitting amplitudes are slightly subtler, because
amplitudes with external fermions in $N=4$ super-Yang-Mills theory
have contributions with virtual scalars coupling to fermion lines
via Yukawa interactions, in addition to the gauge interactions
assumed in the above results.
We have calculated these additional Yukawa contributions to five-point
amplitudes with two external fermions (they are a natural intermediate
step in the string-based gauge-theory calculation),
and have extracted the
corresponding extra terms in the loop splitting amplitudes.
They produce an extra term $r_S^{\rm Yukawa}$ which should be added
to the proportionality constant $r_S$ in the case of external fermions.

For $q\to qg$ and $\bar{q}\to g\bar{q}$, with the virtual scalar
in the adjoint representation, the correction is
$$
\eqalign{
  r_S^{\rm Yukawa}(q^-,a^{+})\ &=\ r_S^{\rm Yukawa}(q^+,a^{-})
   \ =\ r_S^{\rm Yukawa}(a^{-},\bar{q}^{+})
   \ =\ r_S^{\rm Yukawa}(a^{+},\bar{q}^{-})\ =\ 0, \cr
  r_S^{\rm Yukawa}(q^-,a^{-})\ &=\ r_S^{\rm Yukawa}(q^+,a^{+})
   \ =\ \left(1+{1\over N_c^2}\right){1-z\over2}\ , \cr
   r_S^{\rm Yukawa}(a^{+},\bar{q}^{+})
   \ &=\ r_S^{\rm Yukawa}(a^{-},\bar{q}^{-})
   \ =\ \left(1+{1\over N_c^2}\right){z\over2}\ . \cr}
\eqn\qgqscalarrenorm
$$
The corresponding result where quarks are replaced by gluinos can
be found by letting $1/N_c^2 \to -1$; it vanishes in every case.
Thus we see that $r_S^{\rm SUSY}$ controls the $\tilde{g}\to\tilde{g}g$
behavior for $N=4$ as well as $N=1$ super-Yang-Mills.

Finally, the Yukawa contribution to $g\to \bar{q}q$ is the same
for all helicities,
$$
r_S^{\rm Yukawa}(\pm,\bar{q}^{\mp},q^{\pm})\ =\
  {1\over2\e} \L{\mu^2 \over -s_{\bar{q}q}}\R^\e + {3\over2}
  \ -\ {1\over N_c^2} \left[
  {1\over2\e} \L{\mu^2 \over -s_{\bar{q}q}}\R^\e + {1\over2}
  \right]\ .
\eqn\qqgscalarrenorm
$$
Again making the substitution $1/N_c^2 \to -1$ in order
to get the gluino result,
we find that the Yukawa contribution $r_S^{\rm Yukawa}$
cancels the extra chiral loop contribution $r_S^{[1/2]}+r_S^{[0]}$
in~(\use\qqgrenorm), verifying that $r_S^{\rm SUSY}$ also
governs $g\to\tilde{g}\tilde{g}$ in $N=4$ super-Yang-Mills.


\appendix{Explicit Summation of Subleading-Color Terms}
\tagappendix\SubleadingAppendix

In this appendix we carry out the sum over permutations in
eq.~(\use\sublanswer) to obtain the explicit form of the
subleading-in-$N_c$ partial amplitudes $A_{n;c}$.
We calculate the coefficient of the subleading color structure
$$
\Gr_{n;c}(1) = \Tr\L T^{a_1}\cdots T^{a_{c-1}}\R\,
\Tr\L T^{a_{c}}\cdots T^{a_n}\R
\; .
\anoneqn
$$
and define
$\alpha_i \in \{\alpha\} \equiv \{ c-1,c-2,\ldots,2,1 \}$,
$\beta_i \in \{\beta\} \equiv \{c,c+1, \ldots, n-1,n \}$,

The partial amplitude $A_{n;1}$ in eqs.~(\use\Ansatz),
(\use\CoeffDefine) is a sum of the
one-mass and easy two-mass box integral functions
with the appropriate cyclic ordering.
Since $A_{n;c}$ is given by a sum of permuted $A_{n;1}$
it will also be a sum of these integrals but now with
all the orderings specified by $COP\{\alpha\}\{\beta\}$.
Because many orderings of the external momenta appear,
in this section we use the more explicit notation for
the arguments of the scalar box function $F$.
The arguments $k_{i_1},P_1,k_{i_2},P_2$ denote the four external
momenta of the box diagram, in cyclic order around the box; in
the cases we shall encounter here,
$k_{i_1}$ and $k_{i_2}$ are massless external
momenta, while $P_1$ and $P_2$ are in general massive vectors
(sums of external momenta).
In $A_{n;1}$, $P_1$ and $P_2$ were always sums of cyclicly consecutive
momenta, but in $A_{n;c}$ this is no longer the case.
If either $P_1$ or $P_2$ consists of a single
external momentum, then $F$ reduces to a rescaled one-mass box.

In terms of these rescaled boxes,
the function $V_n^g$ of eqs.~(\use\CoeffDefine) is
$$
  V_n^g\ =\ \mu^{2\e} \sum_{i_1,i_2}
  F(k_{i_1},P_{i_1+1,i_2-1},k_{i_2} ,P_{i_2+1,i_1-1}),
\eqn\newVndef
$$
where we define the sum of consecutive momenta
$$
  P_{i,j}\ \equiv\ k_{i}+k_{i+1}+\cdots+k_{j}.
\eqn\Pijdef
$$
The sum in equation~(\use\newVndef)
(and analogous sums in future equations)
runs over all distinct $i_1,i_2$ such that
$P_{i_1+1,i_2-1}$ and $P_{i_2+1,i_1-1}$ are nonzero;
the indices in~(\use\newVndef) are all treated mod $n$.

Now we sum the expression $A_{n;1} = c_\Gamma A^\treemhv_{jk} \v^g_n$
over the permutations in $COP\{\alpha\}\{\beta\}$, to obtain $A_{n;c}$.
The $\alpha_i$ indices are all treated mod $c-1$ (i.e. cyclicly in
$\{c-1,c-2,\ldots,2,1\}$), while the $\beta_i$ indices are treated mod
$n-(c-1)$ (cyclicly in $\{c,c+1,\ldots,n\}$).
Focus first on the (rescaled) box integral where
$\beta_1$ and $\beta_2$ have been ``pulled out'',
that is, where the two massless legs have labels $\beta_1$ and
$\beta_2$,
$$
 F(k_{\beta_1} , P_{\beta_1+1,\beta_2-1}+P_{\alpha_2,\alpha_1-1},
   k_{\beta_2} , P_{\beta_2+1,\beta_1-1}+P_{\alpha_1,\alpha_2-1}).
\eqn\firstF
$$
Here $P_{\alpha_i,\alpha_j}$ and $P_{\beta_i,\beta_j}$ are sums of
consecutive momenta within the respective $\{\alpha\}$ and $\{\beta\}$ sets,
$$
\eqalign{
  P_{\alpha_1,\alpha_2-1}\ &\equiv\
  k_{\alpha_1}+k_{\alpha_1+1}+\cdots+k_{\alpha_2-1},\qquad
  \hbox{(indices mod $c-1$)}, \cr
  P_{\beta_1+1,\beta_2-1}\ &\equiv\
  k_{\beta_1+1}+k_{\beta_1+2}+\cdots+k_{\beta_2-1},\qquad
  \hbox{(indices mod $n-(c-1)$)}, \cr}
\eqn\Palphadef
$$
etc.
Note that both the $\{\alpha\}$ and $\{\beta\}$ sets have been partitioned
in two in a specific way in the integral~(\use\firstF).
Denote the coefficient which multiplies this rescaled box in $A_{n;c}$ by
$(\mu^2)^\e c(\alpha_1,\alpha_2;\beta_1,\beta_2)$.

In general, quite a few permutations in $COP\{\alpha\}\{\beta\}$
contribute to $c(\alpha_1,\alpha_2;\beta_1,\beta_2)$, with different
spinor product denominators from permuting $A_{jk}^\treemhv$.
Define $M(\beta_1,\beta_2;\alpha_1-1,\alpha_2)$ to be the
set of all {\it mergings} of the two ordered sets
$\{\beta_1+1,\beta_1+2,\ldots,\beta_2-1\}$ and
$\{\alpha_1-1,\alpha_1-2,\ldots,\alpha_2\}$ within the range
$[\beta_1,\beta_2]$.
The coefficient $c(\alpha_1,\alpha_2;\beta_1,\beta_2)$ is obtained
by summing over mergings in $M(\beta_1,\beta_2;\alpha_1-1,\alpha_2)$
on one side of the pulled-out legs $\beta_1$ and $\beta_2$,
and over mergings in $M(\beta_2,\beta_1;\alpha_2-1,\alpha_1)$
on the other side:
$$
\eqalign{
  c(\alpha_1,\alpha_2;\beta_1,\beta_2)
  \ =\ (-1)^{c-1}\, \cg \, i \, {\spa{j}.{k}}^4 \,
  & \sum_{\{i_1,i_2,\ldots,i_{m}\}
        \in M(\beta_1,\beta_2;\alpha_1-1,\alpha_2)}
  {1 \over \spa{\beta_1}.{i_1} \spa{i_1}.{i_2} \cdots
    \spa{i_{m}}.{\beta_2}}  \cr
 \times
  & \sum_{\{i_1,i_2,\ldots,i_{m^\prime}\}
        \in M(\beta_2,\beta_1;\alpha_2-1,\alpha_1)}
  {1 \over \spa{\beta_2}.{i_1} \spa{i_1}.{i_2} \cdots
    \spa{i_{m^\prime}}.{\beta_1}}\ . \cr}
\eqn\cabsum
$$

Spinor product identities can be used to simplify the permutation sum.
The standard ``eikonal'' identity reads
$$
  \sum_{j=j_1+1}^{j_2} \Soft_n(j-1,j)\ =\ \Soft_n(j_1,j_2),
\eqn\eikonalid
$$
where the soft factor $\Soft_n(i,j)$ is
$$
 \Soft_n(i,j)\ =\ {\spa{i}.{j}\over\spa{i}.{n}\spa{n}.{j}}\ .
\eqn\softdef
$$
With the help of~(\use\eikonalid), one can show that
$$
\eqalign{
 & \sum_{\{i_1,i_2,\ldots,i_{m}\}
        \in M(\beta_1,\beta_2;\alpha_1-1,\alpha_2)}
  {1 \over \spa{\beta_1}.{i_1} \spa{i_1}.{i_2} \cdots
    \spa{i_{m}}.{\beta_2}}  \cr
 \ &=\
 {1 \over \spa{\alpha_1-1}.{\alpha_1-2} \cdots \spa{\alpha_2+1}.{\alpha_2}}
 {1 \over \spa{\beta_1}.{\beta_1+1} \cdots \spa{\beta_2-1}.{\beta_2}}
  \L {\spa{\beta_1}.{\beta_2} \over
    \spa{\beta_1}.{\alpha_1-1} \spa{\alpha_2}.{\beta_2}} \R\ .\cr}
\eqn\mainspinorid
$$
The proof is by induction on $\alpha_1$, i.e. on the number of $\alpha$'s.
For each merging in $M(\beta_1,\beta_2;\alpha_1-1,\alpha_2)$ there
are several mergings in $M(\beta_1,\beta_2;\alpha_1,\alpha_2)$,
from inserting $\alpha_1$ all possible places between $\beta_1$ and
$\alpha_1-1$.  Due to~(\use\eikonalid) they all produce the same
multiplicative factor, namely
$$
\eqalign{
&  \Soft_{\alpha_1}(\beta_1,\beta_1+1)
+ \Soft_{\alpha_1}(\beta_1+1,\beta_1+2)
+ \cdots + \Soft_{\alpha_1}(\beta_s,\alpha_1-1) \cr
\ &=\ \Soft_{\alpha_1}(\beta_1,\alpha_1-1)
\ =\ {\spa{\beta_1}.{\alpha_1-1}
   \over\spa{\beta_1}.{\alpha_1}\spa{\alpha_1}.{\alpha_1-1}}\ , \cr}
\eqn\multfactor
$$
which is just the factor needed to go from $\alpha_1-1$ to $\alpha_1$ on
the right-hand side of equation~(\use\mainspinorid).
It is also easy to see from~(\use\eikonalid)
that the induction starts correctly when there is only one $\alpha$.

Applying equation~(\use\mainspinorid) twice, the coefficient
$c(\alpha_1,\alpha_2;\beta_1,\beta_2)$ is given by the product
$$
\eqalign{
 & c(\alpha_1,\alpha_2;\beta_1,\beta_2)\cr
 &=\
  { (-1)^{c-1} \, \cg \, i \, {\spa{j}.{k}}^4 \over
   \spa{\alpha_1-1}.{\alpha_1-2} \cdots  \spa{\alpha_2+1}.{\alpha_2}}
  {1 \over \spa{\beta_1}.{\beta_1+1} \cdots \spa{\beta_2-1}.{\beta_2}}
  \L {\spa{\beta_1}.{\beta_2} \over
    \spa{\beta_1}.{\alpha_1-1} \spa{\alpha_2}.{\beta_2}} \R \cr
  &\hskip 0.5 truecm
 \times   {1 \over \spa{\beta_2}.{\beta_2+1} \cdots
    \spa{\beta_1-1}.{\beta_1}}
    {1 \over \spa{\alpha_2-1}.{\alpha_2-2} \cdots
    \spa{\alpha_1+1}.{\alpha_1}}
  \L {\spa{\beta_2}.{\beta_1} \over
    \spa{\beta_2}.{\alpha_2-1} \spa{\alpha_1}.{\beta_1}} \R \cr
  &=\ \cg
   {  i \, {\spa{j}.{k}}^4 \over \spa1.2\cdots\spa{c-1,}.{1}
       \spa{c}.{,c+1}\cdots \spa{n}.{c} }
     (-1) {\spa{\beta_1}.{\beta_2}}^2
     \Soft_{\beta_1}(\alpha_1-1,\alpha_1)
     \Soft_{\beta_2}(\alpha_2-1,\alpha_2)\ . \cr}
\eqn\ccoeff
$$
The same analysis works also for the integrals where $\alpha_1,\beta_2$,
etc., are ``pulled out''.  Also, if $\beta_1,\beta_2$ are pulled out,
and all the $\alpha$ variables are on one side, then we only get one
factor of the type~(\use\mainspinorid) instead of two; this generates
the terms in $S_{n;c}$ in eq.~(\use\GandS).

Altogether, we find that $A_{n;c}$ for the MHV $N=4$ supersymmetric
amplitudes becomes
for $c\geq3$,
$$
\eqalign{
  A_{n;c}\ &=\ \cg\  (\mu^2)^\e i \,
   {{\spa{j}.{k}}^4 \over \spa1.2\cdots\spa{c-1,}.{1}
   \ \spa{c,}.{c+1}\cdots\spa{n}.{c}}
  \L G_{n;c} + S_{n;c} \R\ ,  \cr }
\eqn\Ancansatz
$$
where,
$$
\eqalign{
  G_{n;c}\ =\
&  \sum_{{\alpha_1,\alpha_2=1 \atop \alpha_1\neq\alpha_2}}^{c-1}
   \sum_{{\beta_1,\beta_2=c \atop \beta_1 < \beta_2}}^{n}
 \Bigl[
 -{\spa{\beta_1}.{\beta_2}}^2 \Soft_{\beta_1}(\alpha_1-1,\alpha_1)
     \Soft_{\beta_2}(\alpha_2-1,\alpha_2) \cr
 &\quad\times
   F(k_{\beta_1} , P_{\beta_1+1,\beta_2-1}+P_{\alpha_2,\alpha_1-1},
    k_{\beta_2} , P_{\beta_2+1,\beta_1-1}+P_{\alpha_1,\alpha_2-1})
    \Bigr]   \cr
+
&  \sum_{{\alpha_1,\alpha_2=1 \atop \alpha_1 < \alpha_2}}^{c-1}
   \sum_{{\beta_1,\beta_2=c \atop \beta_1 \neq \beta_2}}^{n}
 \Bigl[
 -{\spa{\alpha_1}.{\alpha_2}}^2 \Soft_{\alpha_1}(\beta_1-1,\beta_1)
     \Soft_{\alpha_2}(\beta_2-1,\beta_2) \cr
 &\quad\times
   F(k_{\alpha_1} , P_{\alpha_1+1,\alpha_2-1}+P_{\beta_2,\beta_1-1},
    k_{\alpha_2} , P_{\alpha_2+1,\alpha_1-1}+P_{\beta_1,\beta_2-1})
    \Bigr]   \cr
+
&  \sum_{\alpha_1,\alpha_2=1}^{c-1}
   \sum_{\beta_1,\beta_2=c}^{n}
 \Bigl[
 +{\spa{\alpha_1}.{\beta_2}}^2 \Soft_{\alpha_1}(\beta_1-1,\beta_1)
     \Soft_{\beta_2}(\alpha_2-1,\alpha_2) \cr
 &\quad\times
   F(k_{\alpha_1} , P_{\beta_1,\beta_2-1}+P_{\alpha_2,\alpha_1-1},
     k_{\beta_2} , P_{\beta_2+1,\beta_1-1}+P_{\alpha_1+1,\alpha_2-1})
    \Bigr]\ ,   \cr
  S_{n;c}\ =\
&- \sum_{\alpha_1=1}^{c-1}
   \sum_{{\beta_1,\beta_2=c \atop \beta_1 \neq \beta_2}}^{n}
    {\spa{\beta_1}.{\beta_2} \spa{\alpha_1-1}.{\alpha_1}
      \over \spa{\beta_1}.{\alpha_1} \spa{\beta_2}.{\alpha_1-1}}
   \ F( k_{\beta_1} , P_{\beta_1+1,\beta_2-1},
     k_{\beta_2} ,  P_{\beta_2+1,\beta_1-1}+P_{\{\alpha\}})   \cr
&
 - \sum_{{\alpha_1,\alpha_2=1 \atop \alpha_1 \neq \alpha_2}}^{c-1}
   \sum_{\beta_1=c}^{n}
    {\spa{\alpha_1}.{\alpha_2} \spa{\beta_1-1}.{\beta_1}
      \over \spa{\alpha_1}.{\beta_1} \spa{\alpha_2}.{\beta_1-1}}
   \ F( k_{\alpha_1} , P_{\alpha_1+1,\alpha_2-1},
     k_{\alpha_2} ,  P_{\alpha_2+1,\alpha_1-1}+P_{\{\beta\}})
    \ ,  \cr }
\eqn\GandS
$$
and
$$
  P_{\{\alpha\}}\ \equiv\ \sum_{\alpha_i\in\{\alpha\}} k_{\alpha_i}\ ,
  \qquad
  P_{\{\beta\}}\ \equiv\ \sum_{\beta_i\in\{\beta\}} k_{\beta_i}\ ;
\eqn\totPalphadef
$$
We define $F(k_{i_1},P_1,k_{i_2},P_2)$ to vanish if either $P_1^\mu=0$
or $P_2^\mu=0$.
We also set
$P_{\alpha_1+1,\alpha_1} \equiv P_{\beta_1+1,\beta_1} \equiv  0$.

For $c=2$, we have, with $\{\alpha\}=\{1\}$:
$$
\eqalign{
  A_{n;2}\ =\ \cg\  (\mu^2)^\e i \,
   { {\spa{j}.{k}}^4 \over \spa2.3\cdots\spa{n}.{2} }
   \sum_{{\beta_1,\beta_2=2 \atop \beta_1 \neq \beta_2}}^{n}
  \Bigl[ & \Soft_1(\beta_1,\beta_2)
   \ F( k_{\beta_1} , P_{\beta_1+1,\beta_2-1},
     k_{\beta_2} , P_{\beta_2+1,\beta_1-1}+k_1 ) \cr
       - & \Soft_1(\beta_1-1,\beta_1)
   \ F( k_1 , P_{\beta_1,\beta_2-1},
     k_{\beta_2} , P_{\beta_2+1,\beta_1-1} ) \Bigr]
 \ .  \cr }
\eqn\Antwoansatz
$$

We have also verified directly using the Cutkosky rules that the
expression in eq.~(\use\Ancansatz) has all the correct cuts.


\listrefs

\break

\centerline{\bf Figure Captions }
\vskip 0.5 truecm

\noindent
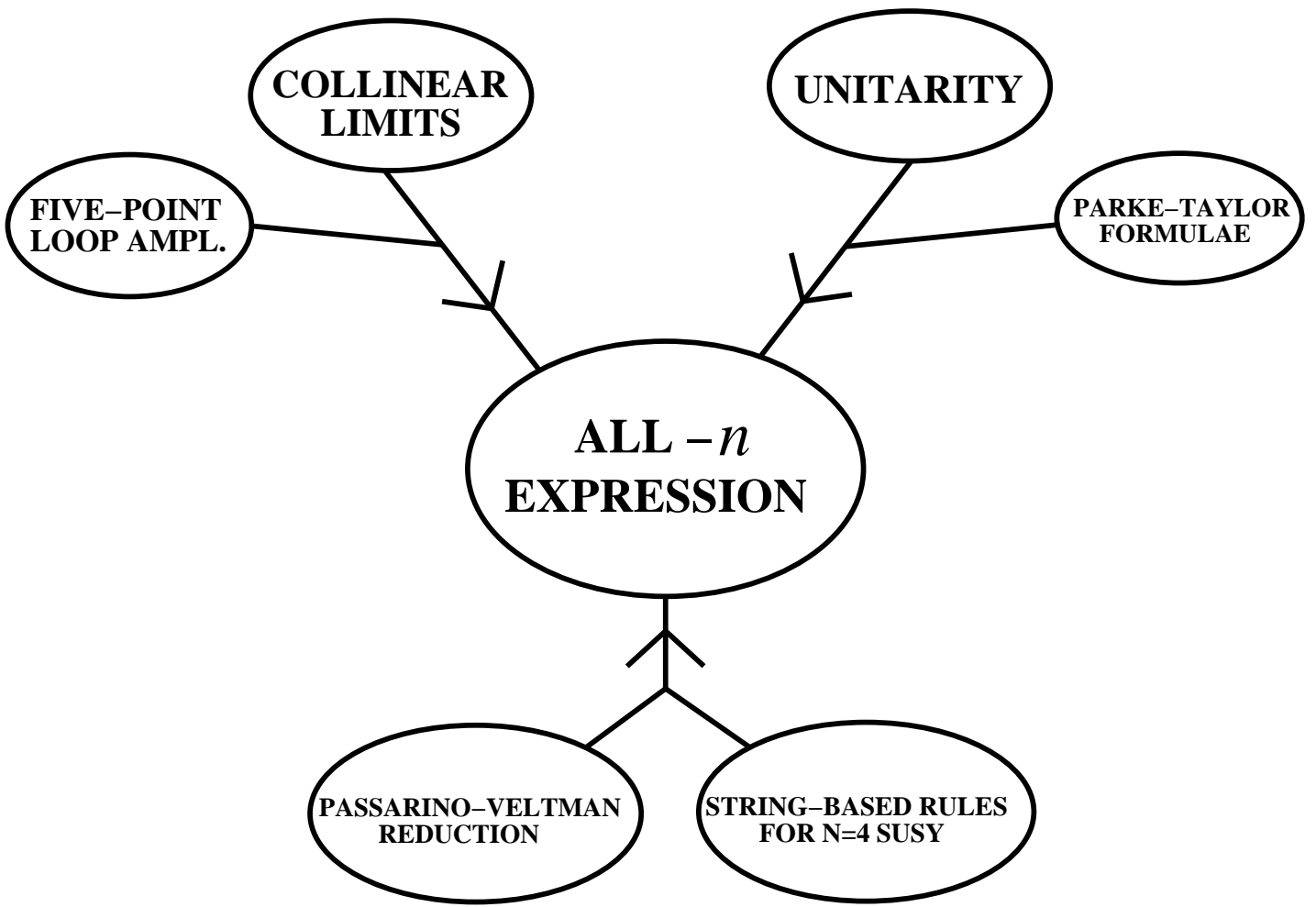
{\bf Fig.~1:}
To obtain all-$n$ expressions we impose a variety of
constraints summarized here.
\vskip 0.4 truecm

\noindent
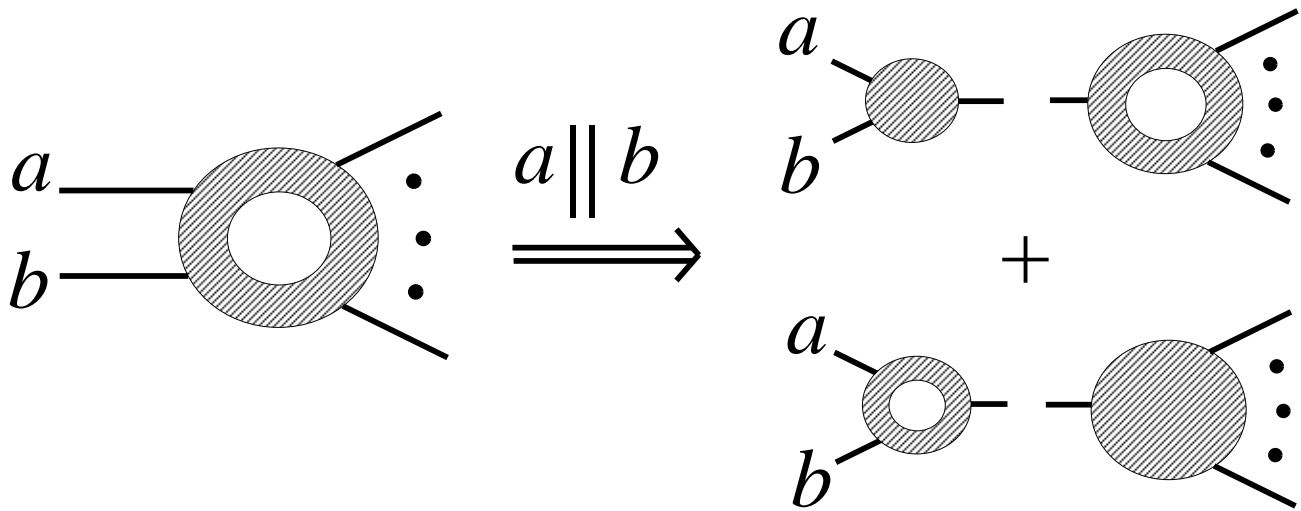
{\bf Fig.~2:} In the collinear limit of a one-loop amplitude we obtain two type
of terms: tree splitting amplitudes
multiplying one-loop amplitudes and
loop  splitting amplitudes multiplying tree amplitudes.
\vskip 0.4 truecm

\noindent
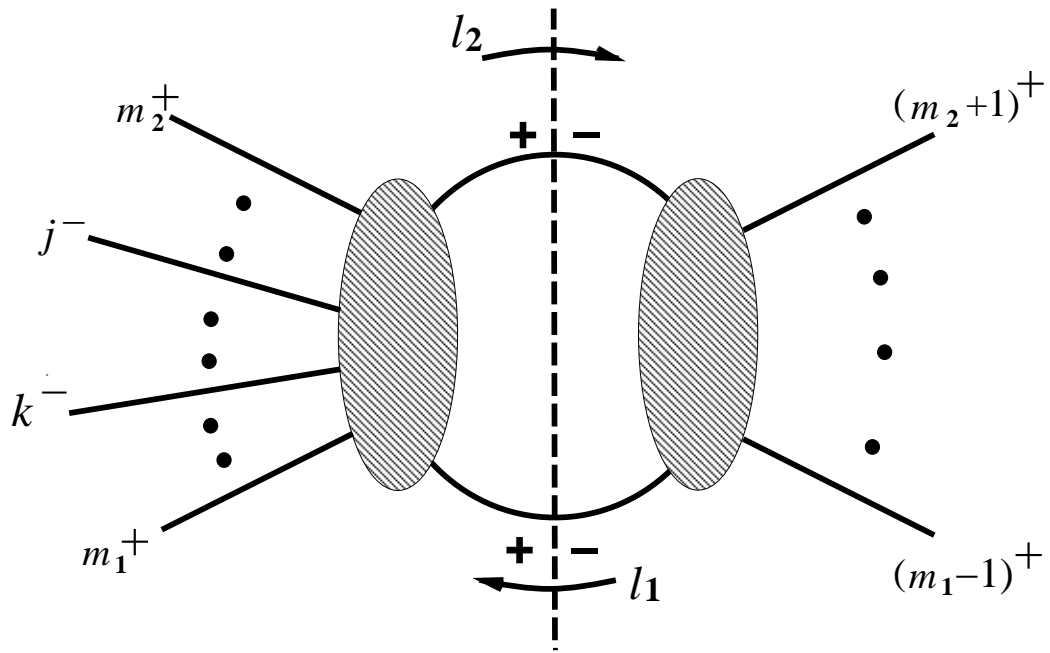
{\bf Fig.~3:} The possible intermediate helicities when both
negative helicity gluons lie on the same side of the cut.

\noindent
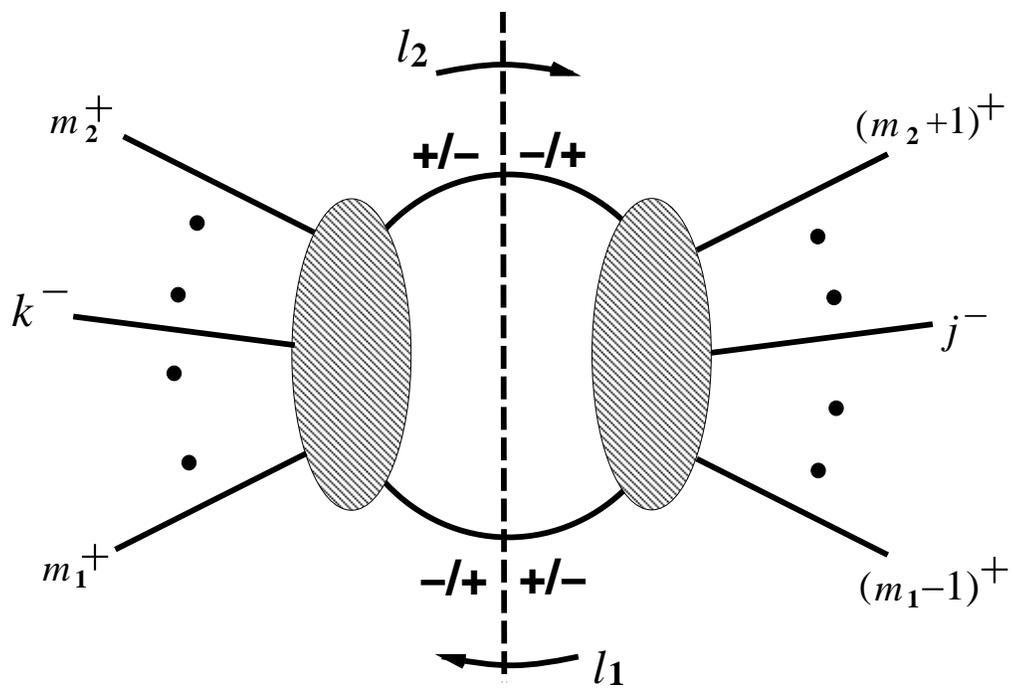
{\bf Fig.~4:} The two possible intermediate helicities when
the negative helicity gluons lie on opposite sides of the cut.
\vskip 0.4 truecm

\noindent
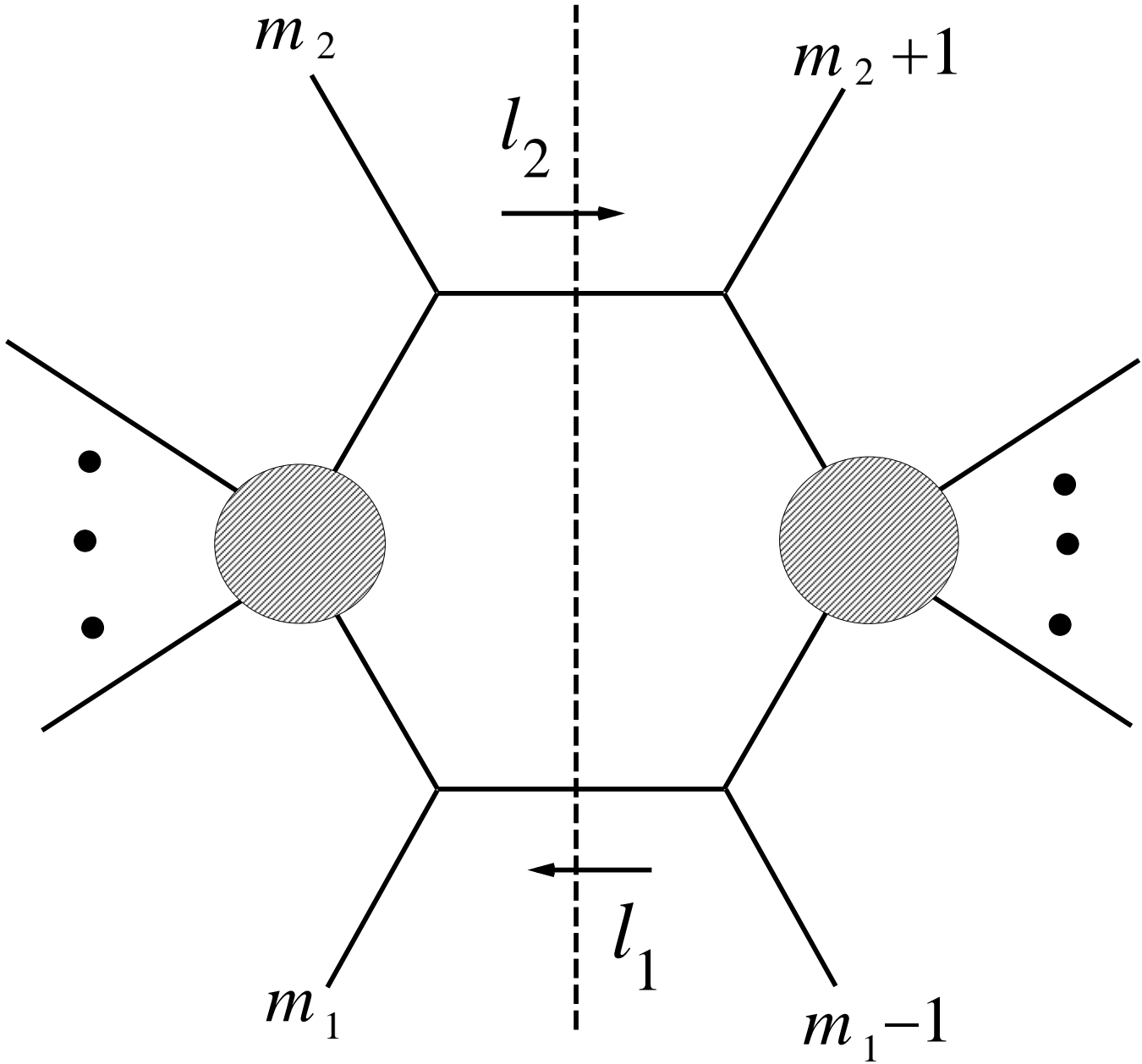
{\bf Fig.~5:} After sewing the MHV tree amplitudes
together, the cut can be rearranged to be the cut of the hexagon
integral shown.
\vskip 0.4 truecm

\noindent
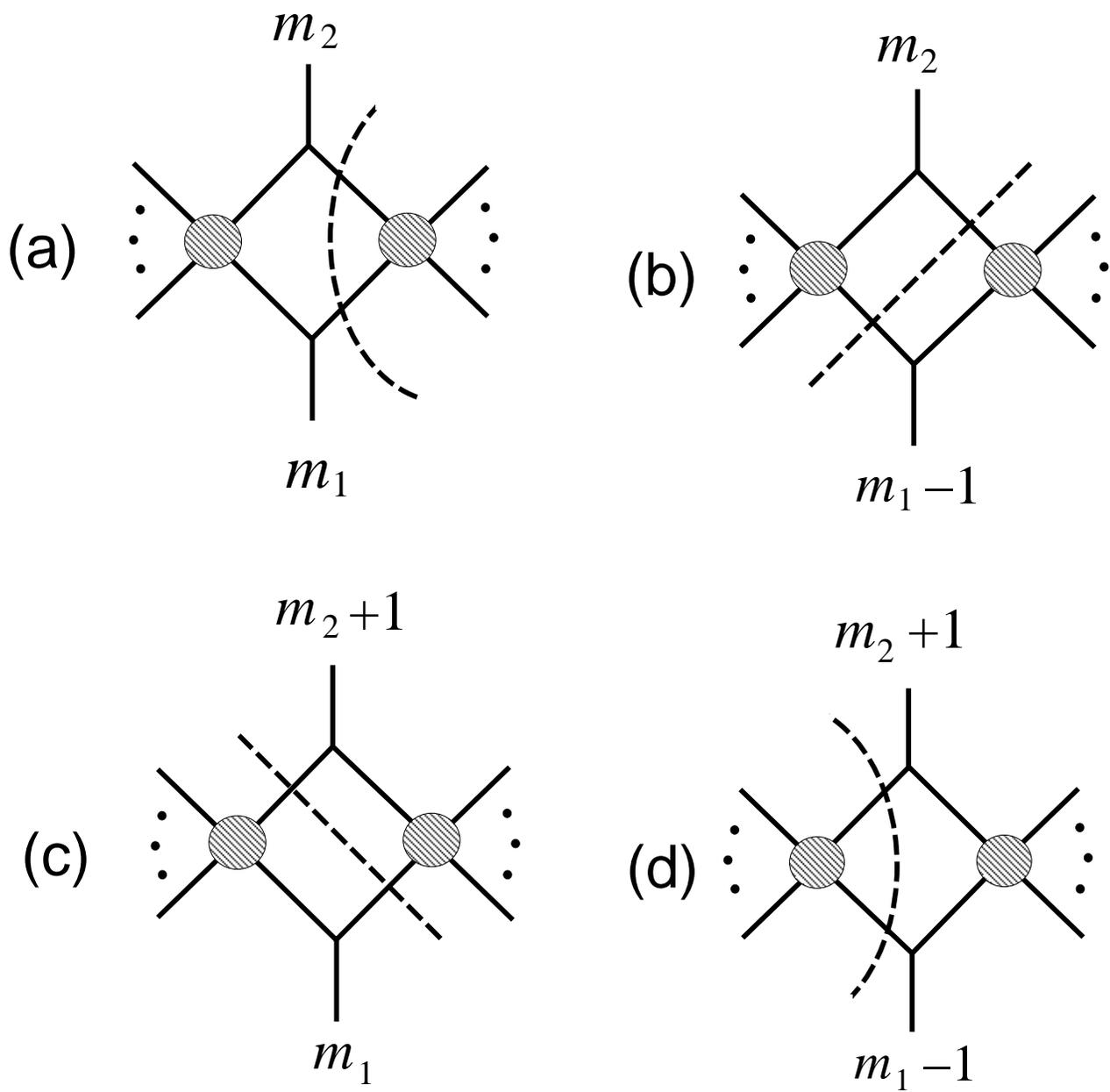
{\bf Fig.~6:} The cut hexagon integral can be reduced to the
sum of the four cut box integrals shown.
\vskip 0.4 truecm

\noindent
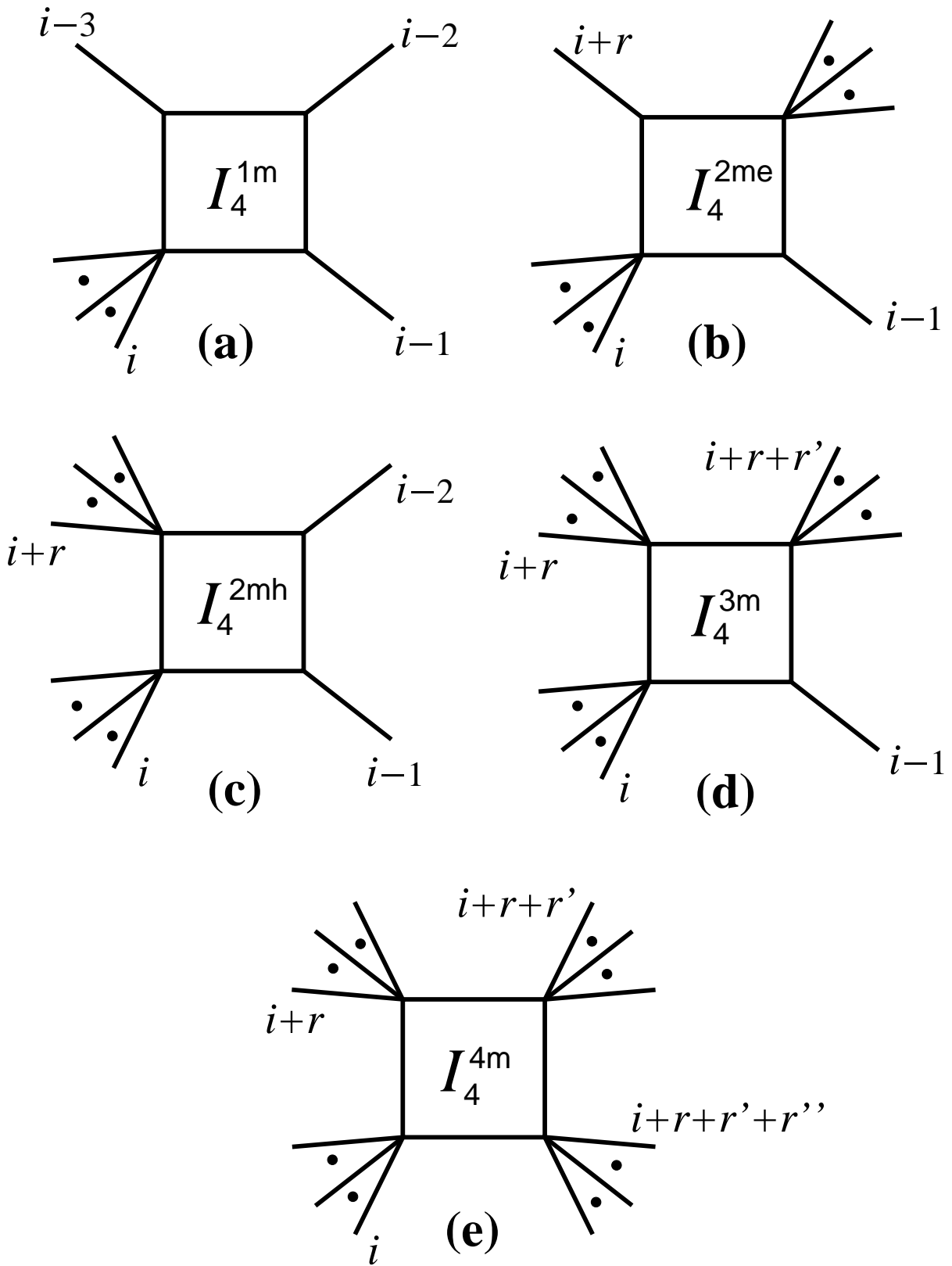
{\bf Fig.~7:} The different types of box integrals
given in eq.~(\use\boxes).
\vskip 0.4 truecm

\noindent
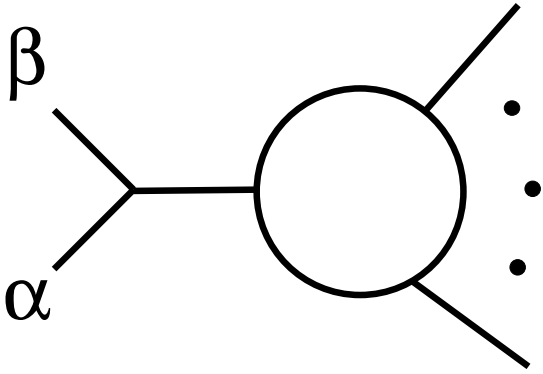
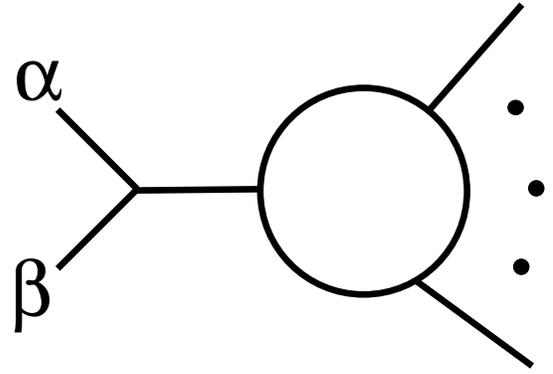
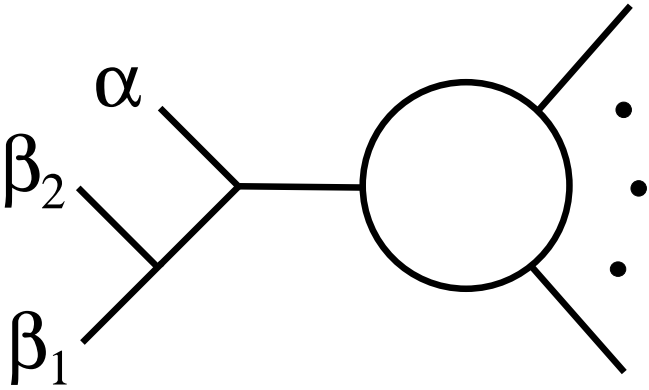
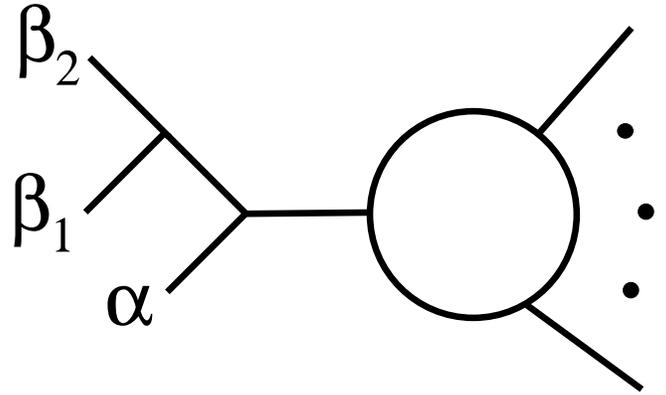
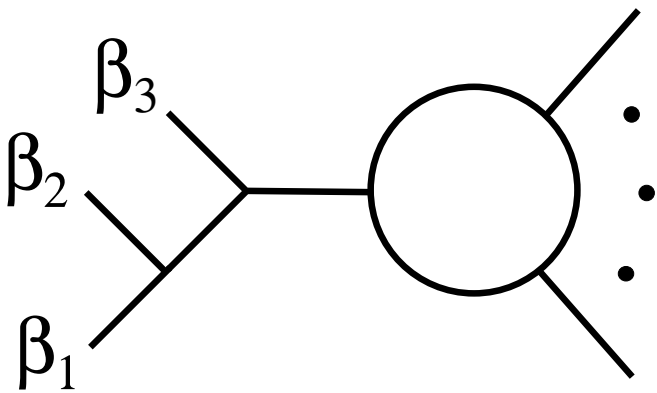
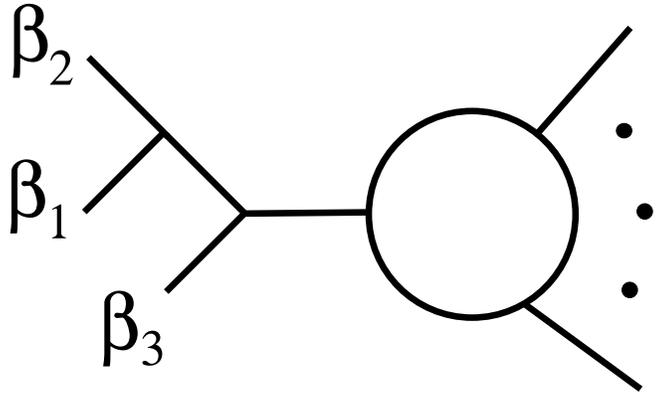
{\bf Fig.~8:} These pairs of diagrams cancel by the antisymmetry
of the three-point vertex.
\vskip 0.4 truecm

\noindent
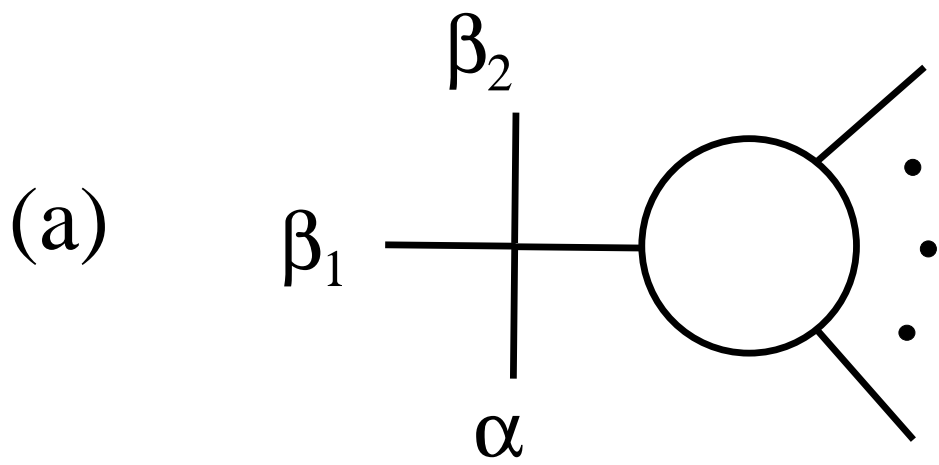
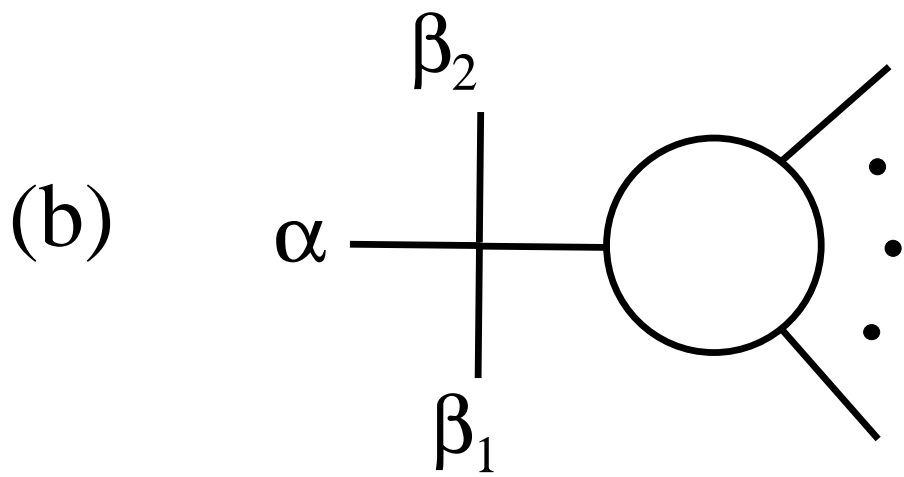
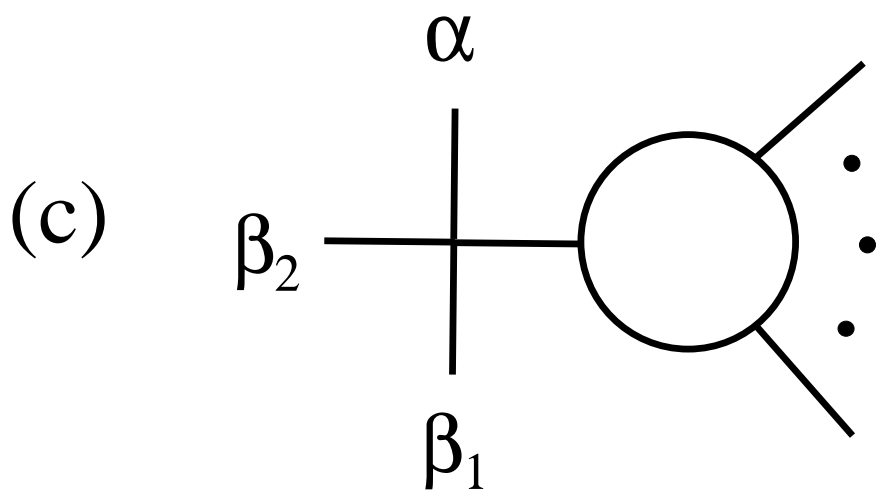
{\bf Fig.~9:} Diagrams with four-point vertices cancel in triplets.
\vskip 0.4 truecm

\end